\documentclass[preprint,12pt]{elsarticle}

\usepackage{amssymb}
\usepackage{graphicx}
\usepackage{amsmath}
\usepackage{caption}
\usepackage{subcaption}
\usepackage{upgreek}
\usepackage{stackengine}

 \usepackage{lineno}

\journal{Physica D}

\begin{document}

\begin{frontmatter}



\title{The influence of initial perturbation power spectra on the growth of a turbulent mixing layer induced by Richtmyer--Meshkov instability}


\author[label1]{M. Groom\corref{cor1}}
\author[label1]{B. Thornber}

\address[label1]{School of Aerospace, Mechanical and Mechatronic Engineering, The University of Sydney, Sydney, Australia}
\cortext[cor1]{michael.groom@sydney.edu.au}

\begin{abstract}
This paper investigates the influence of different broadband perturbations on the evolution of a Richtmyer--Meshkov turbulent mixing layer initiated by a Mach 1.84 shock traversing a perturbed interface separating gases with a density ratio of 3:1. Both the bandwidth of modes in the interface perturbation, as well as their relative amplitudes, are varied in a series of carefully designed numerical simulations at grid resolutions up to $3.2\times10^9$ cells. Three different perturbations are considered, characterised by a power spectrum of the form $P(k)\propto k^m$ where $m=-1$, $-2$ and $-3$. The growth of the mixing layer is shown to strongly depend on the initial conditions, with the growth rate exponent $\theta$ found to be $0.5$, $0.63$ and $0.75$ for each value of $m$ at the highest grid resolution. The asymptotic values of the molecular mixing fraction $\Theta$ are also shown to vary significantly with $m$; at the latest time considered $\Theta$ is $0.56$, $0.39$ and $0.20$ respectively. Turbulent kinetic energy (TKE) is also analysed in both the temporal and spectral domains. The temporal decay rate of TKE is found not to match the predicted value of $n=2-3\theta$, which is shown to be due to a time-varying {normalised dissipation rate $C_\epsilon$}. In spectral space, the data follow the theoretical scaling of $k^{(m+2)/2}$ at low wavenumbers and tend towards $k^{-3/2}$ and $k^{-5/3}$ scalings at high wavenumbers for the spectra of transverse and normal velocity components respectively. The results represent a significant extension of previous work on the Richtmyer--Meshkov instability evolving from broadband initial perturbations and provide useful benchmarks for future research.
\end{abstract}




\end{frontmatter}


\section{Introduction}
\label{sec:intro}
The Richtmyer--Meshkov instability (RMI) occurs when an interface separating two materials of differing densities is accelerated impulsively, usually by an incident shock wave \cite{Richtmyer1960,Meshkov1969}. The instability evolves due to the misalignment of density gradients across the interface and pressure gradients across the shock (typically due to surface perturbations on the interface or inclination of the shock wave), referred to as the deposition of baroclinic vorticity. This deposition leads to a net growth of the interface and the development of secondary Kelvin-Helmholtz instabilities, which drive the transition to a turbulent mixing layer. Unlike the closely related Rayleigh–Taylor instability (RTI), RMI can be induced for both light-heavy and heavy-light configurations, and in both cases the initial growth of the interface is linear and can be described analytically. However, as the perturbation amplitudes become large with respect to the wavelength, the layer growth enters the nonlinear regime, whereby numerical simulation is required to calculate the subsequent evolution. For a comprehensive and up-to-date review of the literature on RMI, the reader is referred to Zhou \cite{Zhou2017a,Zhou2017b}.

Once the instability is initiated and has passed the initial linear growth regime, it will evolve into a nonlinear state characterised by mushroom shaped bubbles (lighter fluid penetrating into heavier fluid) and spikes (heavier fluid penetrating into lighter fluid). A key area of interest in the study of RMI is the degree to which memory of the initial conditions is retained and how this affects the statistics of the flow at late time. Thornber et al. \cite{Thornber2010} investigated RMI induced by two different, multimode initial conditions using large eddy simulation (LES). The first of these was a narrowband perturbation, consisting of a narrow range of high wavenumber modes {$k_{min}$ to $k_{max}$ where $k_{min}=k_{max}/2$}, whose amplitudes are given by a constant power spectrum. This case was designed to give growth purely due to mode coupling/backscatter of the energetic scales, representing a lower bound on the expected growth rate due to pure RMI. Variations of this narrowband initial condition have appeared in subsequent studies \cite{Thornber2012,Thornber2012b,Thornber2016}, most notably the $\theta$-group collaboration \cite{Thornber2017}, which featured eight independent numerical algorithms and gave results for key integral properties of the layer at late time. In the turbulent regime, the width of the mixing layer grows as $h\propto t^{\theta}$. For narrowband initial conditions, experiments and simulations indicate $0.25<\theta<0.33$ \cite{Dimonte2000,Prasad2000,Thornber2010,Thornber2017}, while a theoretical growth rate exponent of $\theta=1/3$ is possible once a sufficient number of mode coupling generations have occurred \cite{Elbaz2018}. 

The second initial condition used in \cite{Thornber2010} was a broadband perturbation, consisting of a wide range of modes {($k_{max}/k_{min}\gg2$)} whose amplitudes satisfy a power spectrum $P(k)\propto k^{-2}$. If the initial conditions are forgotten then the late time statistics for this case should be the same as for the narrowband case. However, as proposed by Youngs \cite{Youngs2004}, it is possible that the linear growth of the largest wavelength modes is faster than that due to mode coupling, thus dictating the overall growth rate of the mixing layer. This was found to indeed be the case in the broadband simulations conducted by Thornber et al. \cite{Thornber2010}, which obtained a growth rate exponent of up to $\theta=0.62$ at the highest grid resolution considered, which is tending towards the theoretical prediction of $\theta=2/3$ given by just-saturated mode analysis \cite{Youngs2004}. This initial condition has also been studied after reshock  \cite{Thornber2011a}, where it was found to give a higher post-reshock growth rate than the narrowband case, as well as in two dimensions \cite{Thornber2015}. Simulations of turbulent mixing due to RMI in spherical implosions have also been performed using a $k^{-2}$ broadband perturbation \cite{Youngs2008,Boureima2018}, which is considered representative of the measured surface roughness power spectra of an inertial confinement fusion capsule \cite{Barnes2002}. 

Multiple experiments have also been performed that contain broadband initial perturbations. Weber et al. \cite{Weber2012} performed shock tube experiments using helium (seeded with acetone) and argon. A broadband initial condition was created by first forming a stagnation plane between the two gases and then injecting streams of argon and helium above and below this stagnation plane. These two streams interact due to buoyancy and shear to generate a statistically steady and repeatable broadband initial perturbation. {Due to this method of perturbing the interface, some non-linearity is already present prior to the arrival of the shock wave.} The initial condition was characterised as consisting of three distinct spectral ranges that scale as $k^{-1}$, $k^{-3}$ and $k^{-5}$ respectively. {The ratio of largest to smallest initial wavelengths in the perturbation was approximately 100.} The width of the layer, based on average mole fraction profiles, was found to have a growth rate exponent of $\theta=0.58$. This is considerably higher than that due to high wavenumber narrowband perturbations, indicating that the overall growth of the mixing layer is being dominated by long wavelength modes that have a slower but more persistent growth rate. Subsequent experiments using the same facility gave $\theta=0.43\pm0.01$ when data from two different shock Mach numbers was used \cite{Weber2014}, while a value of $\theta=0.34\pm0.01$ was obtained when the mole fraction field was adjusted to remove large-scale structures from the mixing layer prior to spanwise averaging \cite{Reese2018}. 

Mohaghar et al. \cite{Mohaghar2017} performed shock tube experiments between nitrogen (seeded with acetone) and carbon dioxide using two different initial conditions. The first of these was a predominantly single-mode interface, created by inclining the shock tube by $20^\circ$, while the second was a broadband interface that is created by injecting heavy/light gas above/below the stagnation plane between the two gases, which is inclined at $20^\circ$ as in the single-mode case. {In both cases the inclination results in an amplitude to wavelength ratio of 0.088 for the large-scale single mode. As with the experiments in Weber et al. \cite{Weber2012}, it is expected that some non-linearity is present in the broadband initial perturbation due to how it is formed.} Characterisation of the broadband initial condition was performed by computing the density power spectra, showing that the perturbation consisted of three distinct ranges that follow $k^{-0.1}$, $k^{-0.8}$ and $k^{-1.8}$ scalings. {For this initial condition, the ratio of largest to smallest wavelengths was 67.} The experimental dataset was extended in Mohaghar et al. \cite{Mohaghar2019} to include data at higher shock Mach number. Comparisons between the single mode and broadband cases showed that although the layer width $h$ is very similar, the mixed-mass thickness $\updelta$ is substantially larger prior to reshock, indicating greater mixing due to the presence of more fine-scale structure. Finally, Krivets et al. \cite{Krivets2017} performed shock tube experiments using air and sulphur hexafluoride, with smoke used to seed either the light or heavy gas. The initial perturbation was created by oscillating the shock tube with loudspeakers to produce Faraday waves at the interface between the two gases. Both the bubble and spike integral widths $W_b$ and $W_s$ were obtained, with the corresponding growth rate exponents varying over a wide range from $\theta=0.18$ to $\theta=0.57$. This suggests the possibility that low amplitude, long wavelength modes are present in the initial perturbation, which dominate the growth rate at later times. The results were reported for a small number of experiments however ($n=5$), hence there may be significant sample size effects in the data.

This evidence of enhanced growth rates in experiments where broadband perturbations are present, as well as the simulations and theory given in Youngs \cite{Youngs2004} and Thornber et al. \cite{Thornber2010}, motivates a thorough study of RMI evolving from well-characterised broadband initial conditions. The present work generalises the $k^{-2}$ broadband perturbation used in \cite{Thornber2010} to a class of perturbations with power spectra $P(k)\propto k^m$ for integer exponents $m$. Three different values of $m$ are considered, $m=-1$, $m=-2$ and $m=-3$, which represent idealised versions of the majority of initial conditions found in experiments and applications. A computational approach similar to that in \cite{Thornber2010} is used to study these perturbations, {where the ratio $k_{max}/k_{min}$, referred to as the bandwidth of the perturbation,} increases as the grid resolution is increased. {The aim of the set of simulations reported here is to validate theoretical predictions of various quantities in the self-similar regime for varying $m$ \cite{Youngs2004,Thornber2010}, as well as examine how these are affected by finite bandwidth. The results represent a significant extension of these previous studies, for example the highest bandwidth $m=-2$ case is very similar to the one presented in Thornber et al. \cite{Thornber2010} but was run to $10\times$ later dimensionless time.}

The paper is structured as follows; Sec. \ref{sec:description} describes the governing equations solved in the simulations as well the numerical method used to solve them. The specifics of the broadband initial perturbations are given in detail, along with the theoretical predictions for the growth rate given by just-saturated mode analysis. Sec. \ref{sec:results} presents a discussion of the results for the nine simulations performed, while conclusions are given in Sec. \ref{sec:conclusion}.

\section{Problem description}
\label{sec:description}
\subsection{Governing equations}
\label{subsec:governing}

The governing equations for binary mixtures of ideal gases with linear constitutive relations are given in strong conservation form by
\begin{subequations}
	\begin{align}
	\frac{\partial \rho}{\partial t}+\boldsymbol{\nabla}\cdot(\rho\boldsymbol{u}) & = 0, \label{subeqn:compressible1} \\
	\frac{\partial \rho \boldsymbol{u}}{\partial t}+\boldsymbol{\nabla}\cdot(\rho \boldsymbol{u}\boldsymbol{u}^t+p{\boldsymbol{I}}) & = \boldsymbol{\nabla}\cdot\boldsymbol{\tau}, \label{subeqn:compressible2}\\
	\frac{\partial \rho E}{\partial t}+\boldsymbol{\nabla}\cdot\left(\left[\rho E+p\right]\boldsymbol{u}\right) & =\boldsymbol{\nabla}\cdot\left(\boldsymbol{\tau}\cdot\boldsymbol{u}-\boldsymbol{q}_c-\boldsymbol{q}_d\right), \label{subeqn:compressible3}  \\
	\frac{\partial \rho Y_1}{\partial t}+\boldsymbol{\nabla}\cdot(\rho Y_1\boldsymbol{u}) & = \boldsymbol{\nabla}\cdot\boldsymbol{J}_1. \label{subeqn:compressible4}
	\end{align}
	\label{eqn:compressible}
\end{subequations}
In Eqn. \ref{eqn:compressible} above, (\ref{subeqn:compressible1}), (\ref{subeqn:compressible2}) and (\ref{subeqn:compressible3}) are the compressible Navier--Stokes equations, written in terms of the mass-weighted velocity of the mixture $\boldsymbol{u}$, while (\ref{subeqn:compressible4}) describes the conservation of mass for species 1. For a more thorough description and background, see Zhou et al. \cite{Zhou2019}. The total energy is given by $E=e+\frac{1}{2}\boldsymbol{u}\cdot\boldsymbol{u}$, where the internal energy $e$ is related to the density $\rho$ and pressure $p$ through the equation of state. For ideal gases this relation is
\begin{equation}
\rho e = \frac{p}{\overline{\gamma}-1},
\label{eqn:eos}
\end{equation}
where $\overline{\gamma}$ is the ratio of mass-weighted specific heats, {equal to 5/3 for all cases presented here}. The viscous stress tensor $\boldsymbol{\tau}$ is given by Newton's law of viscosity,
\begin{equation}
\boldsymbol{\tau}=\mu\left(\boldsymbol{\nabla u}+(\boldsymbol{\nabla u})^t-\frac{2}{3}\left(\boldsymbol{\nabla}\cdot\boldsymbol{u}\right){\boldsymbol{I}}\right),
\label{eqn:tau}
\end{equation}  
noting that Stokes' hypothesis of zero bulk viscosity is invoked. The heat flux vector is given by Fourier's law of conductivity to be
\begin{equation}
\boldsymbol{q}_c=-\kappa\boldsymbol{\nabla}T.
\end{equation}
The mass flux for species 1 is given by Fick's law of diffusion,
\begin{equation}
\boldsymbol{J}_1=\rho D_{12} \boldsymbol{\nabla}Y_1,
\end{equation}
where $D_{12}$ is the binary diffusion coefficient. Whenever required, the mass flux for species 2 is given by $\boldsymbol{J}_2=-\boldsymbol{J}_1$. Finally, changes in mixture composition due to species diffusion give rise to changes in energy which must be accounted for. This is done via the enthalpy diffusion flux, defined as
\begin{equation}
{\boldsymbol{q}_d=\sum_{l=1}^2 h_l\boldsymbol{J}_l},
\end{equation}
{where $h_l=e_l+p/\rho_l$ is the enthalpy of species $l$}. 

\subsection{Computational approach}
\label{subsec:numerical}
The governing equations presented in Sec. \ref{subsec:governing} are solved using the University of Sydney code Flamenco, which employs a method of lines discretisation approach in a structured multiblock framework. Spatial discretisation is performed using a Godunov-type finite-volume method, which is integrated in time via a second order TVD Runge-Kutta method \cite{Spiteri2002}. Spatial reconstruction of the inviscid terms is done using a fifth order MUSCL scheme \cite{Kim2005}, which is augmented by a modification to the reconstruction procedure to ensure the correct scaling of pressure and velocity in the low Mach number limit \cite{Thornber2007d,Thornber2008b}. The inviscid flux component is calculated using the HLLC Riemann solver \cite{Toro1994}, while the viscous and diffusive fluxes are calculated using second order central differences. This numerical algorithm has been extensively demonstrated to be an effective approach for solving shock-induced turbulent mixing problems \cite{Thornber2010,Thornber2011a,Thornber2012,Groom2019}.

In this paper, implicit large eddy simulation (ILES) is used to explore the high Reynolds number limit of key integral quantities in the regime of self-similar growth. In the ILES approach, it is assumed that the growth of the integral length scales are independent of the exact dissipation mechanism and that the species are intimately mixed within each computational cell, such that scalar dissipation rates are well represented and are insensitive to the actual values of viscosity and diffusivity. {The first of these assumptions is addressed in the design of the problem such that the integral length scales are large with respect to the grid scale, while the second assumption is considered to be valid provided the Reynolds number is sufficiently high. Since the aim is to explore the high Reynolds number limit,} the simulations are nominally inviscid and therefore all of the right hand side terms in Eqn. \ref{eqn:compressible} are zero. However, numerical dissipation in the spatial reconstruction and Riemann solver still acts to remove kinetic energy from the flow, which is used in lieu of an explicit subgrid model. For more details on the use of ILES for shock-induced turbulent mixing see \cite{Zhou2019,Groom2019,Grinstein2013}, as well as \cite{Thornber2019} for a quantification of the numerical dissipation. 

\subsection{Initial conditions}
\label{subsec:initial} 

The initial conditions used in the present simulations closely follow those used in previous fundamental studies of RMI turbulence, for example the $\theta$-group collaboration \cite{Thornber2017}. The setup consists of two quiescent gases separated by a perturbed material interface and with a shock wave initialised in the heavy gas travelling towards the interface. The interface is initially diffuse, with the profile given by an error function with characteristic initial thickness $\delta$. The volume fractions $f_1$ and $f_2=1-f_1$ of the two gases are computed as
\begin{equation}
f_1(x,y,z)=\frac{1}{2}\textrm{erfc}\left\{\frac{\sqrt{\pi}\left[x-S(y,z)\right]}{\delta}\right\},
\label{eqn:volume-fraction}
\end{equation}
where $S(y,z)=x_0+A(y,z)$ with $A(y,z)$ the amplitude perturbation of the interface and $x_0$ the mean position. A Cartesian domain of dimensions {$x\times y\times z=L_x\times L\times L$ where $L=2\pi$} is used for all simulations presented here. The extent of the domain in the $x$ direction, $L_x$, varies depending on the cross-sectional resolution and will be detailed {in Sec. \ref{subsec:simulations}}. Periodic boundary conditions are used in the $y$ and $z$ directions, while in the $x$ direction outflow boundary conditions are imposed very far away from the test section so as to minimise spurious reflections from outgoing waves impacting the flow field.  The initial mean positions of the shock wave and the interface are $x_s=3.0$ and $x_0=3.5$ respectively and the initial pressure of both (unshocked) fluids is $p=1.0\times10^5$. The shock Mach number is $M=1.8439$, equivalent to a four-fold pressure increase, the initial densities of the heavy and light fluids are $\rho_1=3.0$ and $\rho_2=1.0$ and the post-shock densities are $\rho_1^+=5.22$ and $\rho_2^+=1.80$ respectively. This gives a post-shock Atwood number of $A^+=0.487$. The variation in density $\rho$ and mass fraction $Y_1$ across the interface is computed using $\rho=\rho_1f_1+\rho_2(1-f_1)$ and $\rho Y_1=\rho_1f_1$ with $f_1$ given by Eqn. \ref{eqn:volume-fraction}. The evolution of the interface is solved in the post-shock frame of reference by applying a factor of $\Delta u=-291.575$ to the initial velocities of the shocked and unshocked fluids.

\subsubsection{Surface perturbation}
The surface perturbation of the material interface is defined in Fourier space as a power spectrum of the form
\begin{equation}
P(k) = \left\{
\begin{array}{ll}
Ck^m, & k_{min}<k<k_{max}, \\
0, & \textrm{otherwise},
\end{array} \right.
\label{eqn:power-spectrum}
\end{equation}
where $k=\sqrt{k_y^2+k_z^2}$ is the radial wavenumber of the perturbation and $m\le0$. {This form is chosen as it allows for a theoretical analysis of the perturbation growth, along the lines of \cite{Youngs2004,Thornber2010}.} In the majority of previous studies using initial conditions of this form, a narrowband surface perturbation was used, with $k_{min}=k_{max}/2$ and $m=0$. {In the present study, $m=-1, -2$ and $-3$ and the bandwidth $R=k_{max}/k_{min}$ of the initial perturbation is sought to be maximised. For $R>2$, the perturbation defined by $P(k)$ is referred to as a broadband perturbation. The particular choice of $k_{min}$ and $k_{max}$ (and therefore $R$) will be detailed in Sec. \ref{subsec:simulations} below.}


The derivation of the surface perturbation for each value of the exponent $m$ will now be given. For a power spectrum of the form given in Eqn. \ref{eqn:power-spectrum}, taking the inverse Fourier transform and simplifying using the Euler formula gives the perturbation amplitude in real space, 
{
\begin{multline}
A(y,z) = \sum_{p,q=0}^{N_k}  \big[ a_{pq}\cos(pk_0y)\cos(qk_0z)+b_{pq}\cos(pk_0y)\sin(qk_0z) \\
+ c_{pq}\sin(pk_0y)\cos(qk_0z) + d_{pq}\sin(pk_0y)\sin(qk_0z) \big],
\label{eqn:pert}
\end{multline}}
{where $k_0=2\pi/L$ and $N_k=k_{max}/k_0$.}The coefficients {$a_{pq}\ldots d_{pq}$} are chosen as (using {$a_{pq}$} as an example)
{
\begin{equation}
a_{pq}=\mathcal{R}S(p)S(q)\sigma_{pq},
\end{equation}
}
where {$S(p)=1/\sqrt{2}$ if $p=0$} and 1 otherwise and $\mathcal{R}$ is a random number taken from a Gaussian distribution. Unlike previous broadband simulations in \cite{Thornber2010,Thornber2011a}, the random numbers are generated using a Mersenne Twister algorithm, which is deterministic. This allows for the same random numbers to be used across multiple perturbations, as is required for grid convergence studies to be performed. 

The mean standard deviation of {the amplitude of} each mode, {$\sigma_{pq}$}, is given by
{
\begin{equation}
\sigma_{pq}^2=\frac{1}{4}\left(\overline{a_{pq}^2}+\overline{b_{pq}^2}+\overline{c_{pq}^2}+\overline{d_{pq}^2}\right)=\frac{1}{2\pi}\frac{P(k_{pq})}{k_{pq}}\Delta k_y\Delta k_z,
\end{equation}}
{where $\Delta k_y=\Delta k_z=2\pi/L$}, while the total standard deviation is given by
\begin{equation}
\sigma^2=\int_{0}^{\infty}P(k)\:\mathrm{d}k = \int_{-\infty}^{\infty}\int_{-\infty}^{\infty}\frac{1}{2\pi}\frac{P(k)}{k}\:\mathrm{d}k_y\:\mathrm{d}k_z.
\end{equation}
For each case, the total standard deviation is
\begin{equation}
\sigma = \displaystyle \left\{
\begin{array}{ll}
\displaystyle \sqrt{C\log(R)}, & m=-1, \vspace{1em}\\
\displaystyle \sqrt{\frac{Ck^{m+1}}{m+1}\left(1-1/R^{m+1}\right)}, & m<-1.
\end{array} \right.
\label{eqn:stddev}
\end{equation}
The mean standard deviation of each mode can be related to the total standard deviation of the perturbation by 
{
\begin{equation}
\sigma_{pq} = \displaystyle \left\{
\begin{array}{ll}
\displaystyle \frac{2\pi\sigma}{L\sqrt{2\pi\log(R)}}\sqrt{k_{pq}^{m-1}}, & m=-1, \vspace{1em}\\
\displaystyle \frac{2\pi\sigma}{L}\sqrt{\frac{m+1}{2\pi k_{max}^{m+1}(1-1/R^{m+1})}}\sqrt{k_{pq}^{m-1}}, & m<-1.
\end{array} \right.
\label{eqn:stddev-mode}
\end{equation}}
The derivation is completed by defining the constant of proportionality $C$ in Eqn. \ref{eqn:power-spectrum}. Two approaches will be discussed; ensuring that all modes are initially linear and fixing the total standard deviation of the perturbation so that it is the same for all values of $m$. 

\subsubsection{Ensuring linearity}
{It is often desirable in studies of RMI that the initial amplitudes of modes in the perturbation are small, so that linear theory accurately describes their early-time evolution.} Mode $k$ is assumed to be linear {(i.e. growing at the rate given by linear theory)} if $ka_k=1/2$, where
\begin{equation}
\left(\frac{a_k}{2}\right)^2=\int_{k/2}^{k}P(k')\:\mathrm{d}k'
\end{equation}
is the power in a band around wavenumber $k$ {\cite{Thornber2010}}. This gives the following expressions for the amplitude in the band around wavenumber $k$;
\begin{equation}
a_k = \displaystyle \left\{
\begin{array}{ll}
\displaystyle 2\sqrt{C\log(2)}, & m=-1, \vspace{1em}\\
\displaystyle 2\sqrt{\frac{Ck^{m+1}}{m+1}\left(1-1/2^{m+1}\right)}, & m<-1,
\end{array} \right.
\label{eqn:ak}
\end{equation}
By specifying that the highest wavenumber $k_{max}$ must be linear, that is $k_{max}a_{k_{max}}=1/2$, then the coefficient $C$ is determined to be
\begin{equation}
C = \displaystyle \left\{
\begin{array}{ll}
\displaystyle \frac{1}{16\log(2)k_{max}^2}, & m=-1, \vspace{1em}\\ 
\displaystyle \frac{m+1}{16k_{max}^{m+3}\left(1-1/2^{m+1}\right)}, & m<-1.
\end{array} \right.
\end{equation}
Therefore the standard deviation can be written as
\begin{equation}
\sigma = \displaystyle \left\{
\begin{array}{ll}
\displaystyle \frac{\lambda_{min}}{8\pi}\sqrt{\frac{\log(R)}{\log(2)}}, & m=-1, \vspace{1em}\\
\displaystyle \frac{\lambda_{min}}{8\pi}\sqrt{\frac{1-1/R^{m+1}}{1-1/2^{m+1}}}, & m<-1,
\end{array} \right.
\end{equation}
while the mean standard deviation of each mode that ensures linearity is determined to be
{
\begin{equation}
\sigma_{pq} = \displaystyle \left\{
\begin{array}{ll}
\displaystyle \frac{\lambda_{min}}{4L\sqrt{2\pi\log(2)}}\sqrt{k_{pq}^{m-1}}, & m=-1, \vspace{1em}\\
\displaystyle \frac{1}{4L}\sqrt{\frac{\lambda_{min}^{m+3}(m+1)}{(2\pi)^{m+2} (1-1/2^{m+1})}}\sqrt{k_{pq}^{m-1}}, & m<-1.
\end{array} \right.
\end{equation}}
{The description of the perturbation is completed by a suitable choice of $\lambda_{min}$, which is described in Sec. \ref{subsec:simulations}.}
\subsubsection{Fixed standard deviation}
{An alternative approach is to keep the total standard deviation constant across all perturbations of the same bandwidth, so that changing the exponent $m$ can be thought of as changing the relative distribution of mode amplitudes. If $\sigma$ is held constant for a given bandwidth $R$}, then $C$ is determined by Eqn. \ref{eqn:stddev} and {$\sigma_{pq}$} by Eqn. 
\ref{eqn:stddev-mode}. Writing the total standard deviation as $\sigma=\alpha\lambda_{min}$, the only remaining choice to make is a suitable value of the constant $\alpha$ {(assuming $\lambda_{min}$ has already been chosen)}. Here there is a trade-off that must be made between ensuring that the highest modes remain linear for the $m=-1$ perturbation while also retaining a large enough overall amplitude in the $m=-3$ perturbation such that it is computationally efficient to simulate. The standard deviation that ensures linearity of all modes for $R=128$ {(the highest bandwidth considered here)} is $\sigma=0.105\lambda_{min}, 0.448\lambda_{min}$ and $2.94\lambda_{min}$ for $m=-1,-2$ and $-3$ respectively. Therefore in this study, $\alpha$ is chosen to be 0.2 for all values of $m$. This choice results in $k_{max}a_{k_{max}}=0.950, 0.223$ and $1.32\times10^{-4}$, which means that the highest modes in the $m=-1$ perturbation are initially nonlinear (modes greater than $k=135$) while for the $m=-3$ perturbation a longer physical time is required to simulate to the same non-dimensional time. A full investigation into the advantages and disadvantages of this trade-off is outside the scope of the present study and will be performed in future work.

\begin{figure}
	\centering
	\begin{subfigure}{\textwidth}
		\includegraphics[width=\textwidth]{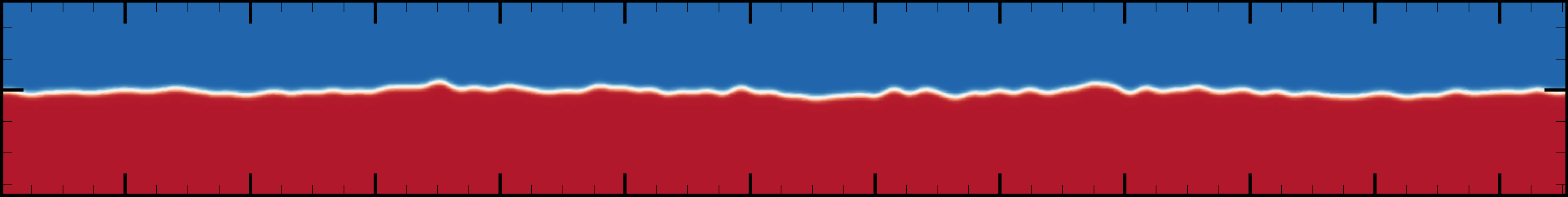}
		\subcaption{$m=-1$.}
	\end{subfigure}
	\begin{subfigure}{\textwidth}
		\includegraphics[width=\textwidth]{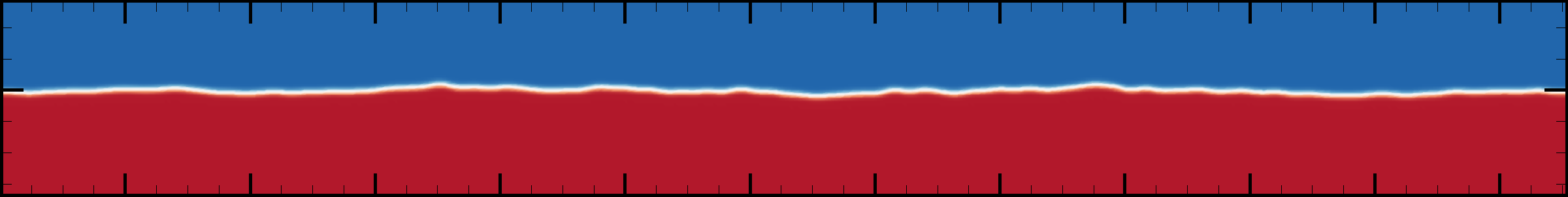}
		\subcaption{$m=-2$.}
	\end{subfigure}	
	\begin{subfigure}{\textwidth}
		\includegraphics[width=\textwidth]{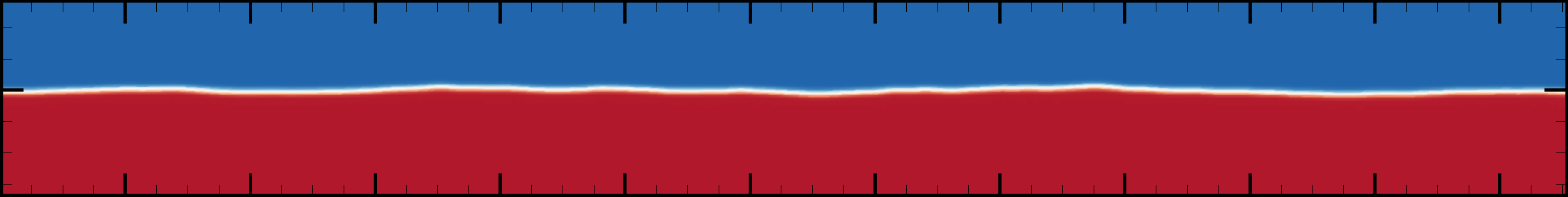}
		\subcaption{$m=-3$.}
	\end{subfigure}	
	\caption{\label{fig:IC} {Initial conditions for a bandwidth of $R=32$. The major ticks on both axes correspond to a grid spacing of $\Delta x=0.5$.}}
\end{figure}

\begin{figure}
	\centering
	\includegraphics[width=0.49\textwidth]{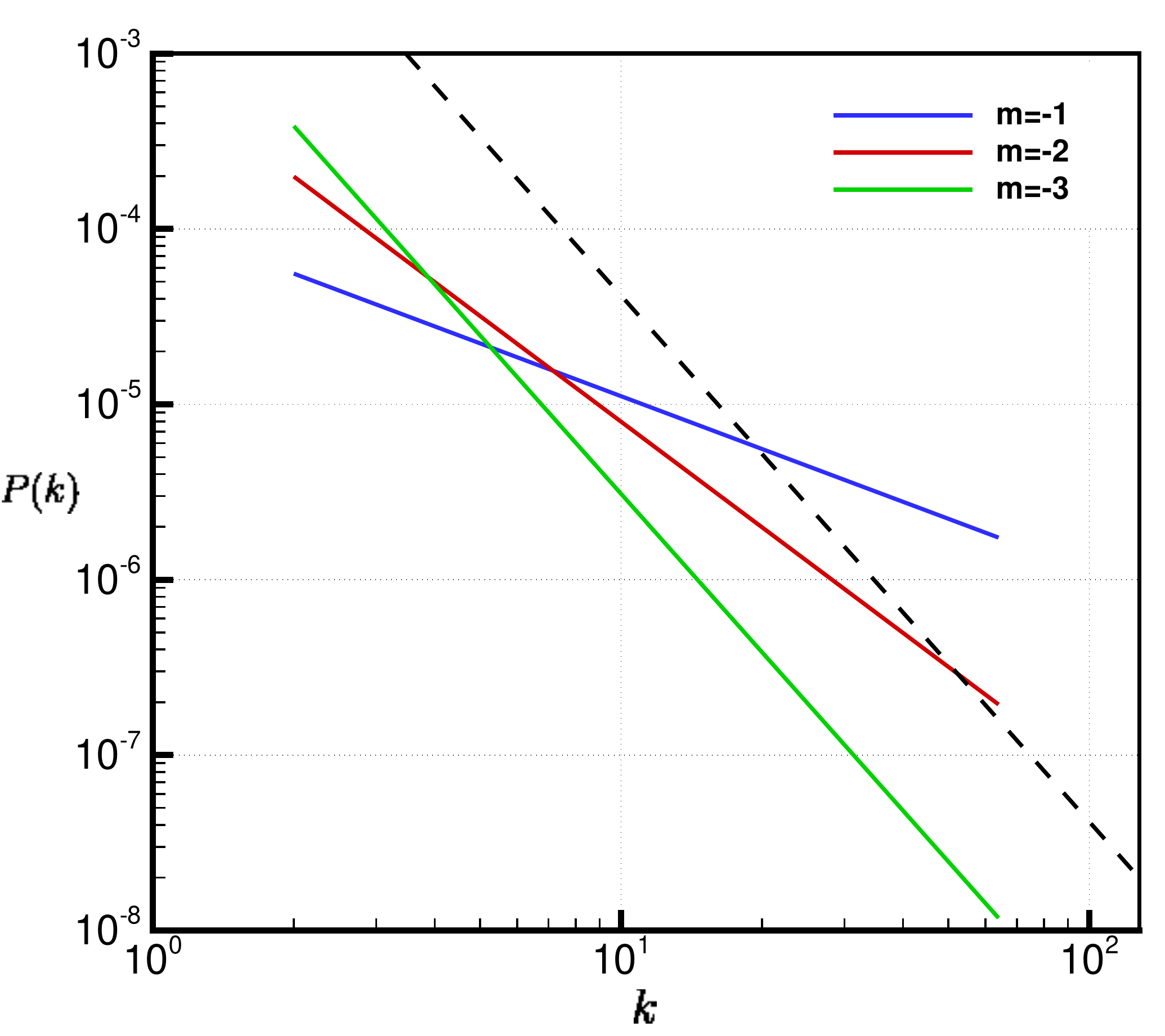}
	\includegraphics[width=0.49\textwidth]{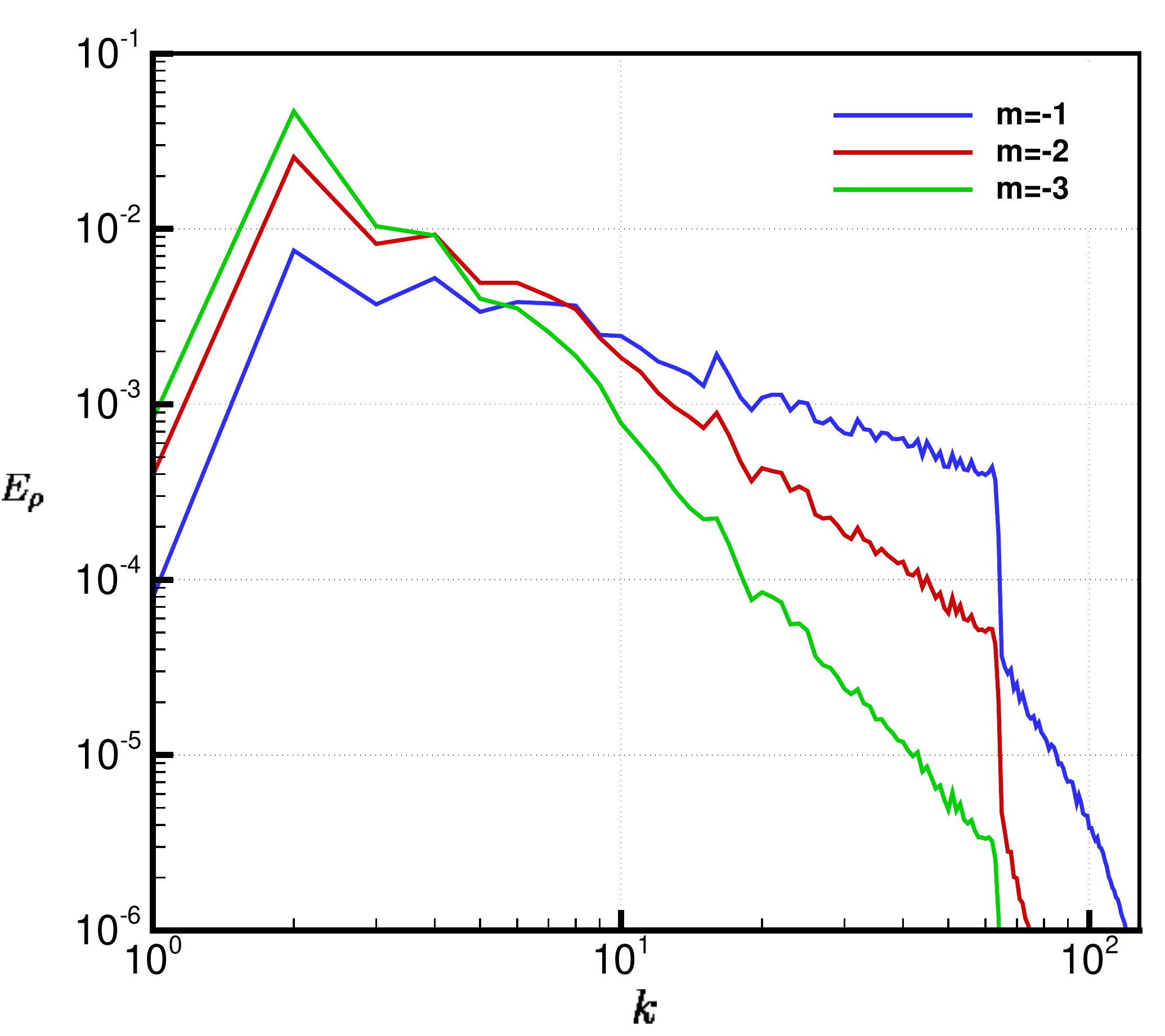}
	\caption{\label{fig:Pk-K} {Surface perturbation power spectrum (left) and density variance spectrum at time $t=0$ (right) for the $R=32$ bandwidth cases. Also shown is the power spectrum that bisects linear and nonlinear modes for this particular bandwidth (black dashed line).}} 
\end{figure}

{Fig. \ref{fig:IC} shows a contour flood of the heavy fluid volume fraction for the three different initial conditions at a bandwidth of $R=32$}, {while Fig. \ref{fig:Pk-K} shows the theoretical surface power spectrum at this bandwidth as well as the initial density variance spectrum for the particular realisation used in this study}. {The different distribution of mode amplitudes for the same overall standard deviation can be discerned from both of these figures}. {Also shown in Fig. \ref{fig:Pk-K} is the line separating linear and nonlinear modes so that the proportion of modes that are nonlinear in the $m=-1$ spectrum can be observed. For this particular bandwidth there is also a small fraction of modes in the $m=-2$ spectrum that are slightly nonlinear, however for higher bandwidths all modes in the $m=-2$ spectrum are linear}.

To facilitate the use of these test cases in future studies, a Fortran 90 implementation of the initial conditions may be made available to other research groups by contacting the authors.

\subsection{Numerical simulations and just-saturated mode analysis}
\label{subsec:simulations}
During the early-time evolution of a broadband perturbation, shorter wavelengths will initially grow faster than longer wavelengths (except for $m=-3$ where they all grow at the same rate), since the initial growth rate of {a single} mode $k$ is given by Richtmyer's formula
\begin{equation}
\dot{a_k}=a_k^+A^+\Delta u k,
\label{eqn:richtmyer}
\end{equation}
where $a_k^+=(1-\Delta u/U_s)a_k^-$ {is the post-shock amplitude} and $a_k^-$ {is the pre-shock amplitude}, given {by} Eqn. \ref{eqn:ak}. {In Eqn. \ref{eqn:richtmyer} $\Delta u$ is the change in velocity induced by the shock wave and $U_s$ is the incident velocity of the shock wave}. Note that other expressions are available for $\dot{a_k}$, such as the Vandemboomgaerde formula \cite{VDM}, as there are certain cases where the initial {linear} growth rate is not well described by Eqn. \ref{eqn:richtmyer}. {As the amplitude of mode $k$ grows, its growth rate will eventually decrease due to nonlinear effects. This is referred to as saturation and typically becomes significant for an amplitude $a_k=0.1\lambda_k$ \cite{Brouillette2002,Chapman2006}. If the linear growth rate of longer wavelength modes that are yet to saturate is faster than growth due to mode coupling in the range $k/2$ to $k$, then these modes will begin to dominate the overall growth rate of the layer and the growth rate exponent $\theta$ will be dependent on the initial conditions \cite{Thornber2010}}. If the bandwidth of the initial perturbation is large {(i.e. $R\gg2$)} then this regime of self-similar growth can extend for a significant amount of time, until the longest wavelength saturates. 

In just-saturated mode analysis, the growth rate of the mixing layer at time $t$ is assumed to be dominated by the growth rate of the mode that is saturating at time $t$. The model was first proposed by Dimonte et al. \cite{Dimonte1995} and was extended to include the effects of initial conditions by Youngs \cite{Youngs2004}. For a broadband perturbation of the form given in Eqn. \ref{eqn:power-spectrum}, the linear growth rate {in a band around} mode $k$ is given by 
\begin{equation}
\dot{a_k}=B\sqrt{C}A^+\Delta u k^{\frac{m+3}{2}},
\end{equation}
where $C$ is determined as described in Sec. \ref{subsec:initial} and $B$ is given by
\begin{equation}
B = \displaystyle \left\{
\begin{array}{lll}
\displaystyle (1-\Delta u/U_s)2\sqrt{\log(2)}, & m=-1, \vspace{1em}\\ 
\displaystyle (1-\Delta u/U_s)2, & m=-2, \vspace{1em}\\
\displaystyle (1-\Delta u/U_s)2\sqrt{3/2}, & m=-3.
\end{array} \right.
\end{equation}
Following Thornber et al. \cite{Thornber2010}, a structure of size $1/k$ becomes nonlinear at time $t=1/(k\dot{a_k})$. Assuming that the growth of $W$ is dominated by the growth of mode $k$ at time $t$ gives
\begin{equation}
W\propto\frac{1}{k}=\left(B\sqrt{C}A^+\Delta u t\right)^{\frac{2}{m+5}},
\label{eqn:just-saturated}
\end{equation}
and therefore $W\propto t^\theta$ where $\theta=2/(m+5)$. 

The duration of time for which Eqn. \ref{eqn:just-saturated} is valid can be estimated by estimating the time at which the smallest and largest modes saturate. Assuming that mode $k$ saturates when $a_k=0.1{\lambda_k}$, the saturation time can be estimated as
\begin{equation}
t_{sat}=\frac{0.1{\lambda_k}-a_k^+}{\dot{a_k}},
\label{eqn:tsat}
\end{equation}
where the linear growth rate $\dot{a_k}$ is given by Eqn. \ref{eqn:richtmyer}. An additional consideration must be made for the initial inversion that occurs due to the heavy--light configuration. The time for this inversion to occur may be estimated as
\begin{equation}
t_{inv}=\frac{2a_k^+}{\dot{a_k}}=\frac{2}{A^+\Delta u k}.
\label{eqn:tinv}
\end{equation}
Therefore the time to saturation of the longest wavelength in the initial perturbation may be estimated as $t_{sat}(k_{min})+t_{inv}(k_{min})$.

{For each value of $m$, three different bandwidths are simulated for a total of nine cases. The choice of $k_{min}$ and $k_{max}$ is made so as to maximise the time during which the layer is growing self-similarly, while also keeping numerical errors below an acceptable level.} {It is important to clarify how convergence is defined in the present set of implicit large eddy simulations, as this determines the particular choice of $k_{min}$ and $k_{max}$. In fact, there are two different notions of convergence that are relevant here; convergence with respect to the initial impulse and convergence with respect to the infinite bandwidth limit. The first of these is straightforward to assess and has been performed in the previous study of broadband RMI by Thornber et al. \cite{Thornber2010}. In that study, a grid convergence analysis was performed using the CNS3D code, which uses very similar numerics to Flamenco, by varying the grid resolution for a fixed initial condition so that $\lambda_{min}=4\Delta x$, $8\Delta x$ and $16\Delta x$. By the end of the simulations, the difference in integral width when $\lambda_{min}=8\Delta x$ vs. $\lambda_{min}=16\Delta x$ was less than 1.5\%. A comparison of the kinetic energy spectra between these two grid resolutions also showed that the first 48 modes were well resolved (i.e. 75\% of the possible modes supported by the grid), which was deemed to be a sufficient level of convergence. Given that the present study uses a very similar computational setup, the restriction of the shortest wavelength to $\lambda_{min}=8\Delta x$ is also adopted here.}

{The second notion, that of convergence with respect to the infinite bandwidth limit, may be analysed by considering the results of simulations with successively increased bandwidths that have the same number of grid points per minimum wavelength. If the results for a given quantity at two different bandwidths are the same (when appropriately non-dimensionalised) then these results are considered to be representative of those that would be obtained in the limit of infinite bandwidth. This is expected to be true for as long as the mixing layer evolving from a given bandwidth is growing self-similarly, which according to just-saturated mode theory should be between the saturation times of the shortest and longest wavelengths in the perturbation. This therefore motivates choosing $\lambda_{max}$ to be as large as possible. In actual simulations there are other sources of error that may cause the results at a given bandwidth to depart from the infinite bandwidth limit earlier than this. Aside from errors due to insufficient resolution of the smallest scales, there will also be errors due to insufficient statistical sampling of the largest scales. In other words, the results may become sensitive to the particular choice of random numbers used to initialise these scales once they begin to dominate the growth of the layer. In Thornber et al. \cite{Thornber2010} it was concluded that simulations with $\lambda_{max}=L/2$ did not show any influence of the domain size over the time scales being considered and for that reason the same choice of maximum wavelength is also made here.}
 
{Therefore the choice of $k_{min}=2$ and $k_{max}=N/8$ determines the bandwidth $R$, which in turn is determined by the number of cells in the $y$ and $z$ directions $N=L/\Delta x$. The three different bandwidths simulated are $R=16$, $R=32$ and $R=64$ for the $m=-1$ case, and $R=32$, $R=64$ and $R=128$ for the $m=-2$ and $m=-3$ cases. The corresponding grid resolutions used are $384\times256^2$, $768\times512^2$, $768\times1024^2$ and $768\times2048^2$ in order of increasing bandwidth.} To minimise computational expense, the domain length of uniformly refined mesh in the $x$-direction $L_x$ is set to be $1.5L$ for the $256^2$ and $512^2$ grids, $0.75L$ for the $1024^2$ grid and $0.375L$ for the $2048^2$ grid, {such that} the grid spacing $\Delta x$ is the same as $\Delta y=\Delta z$. {Each simulation is run until the at least the saturation time of the longest wavelength, which is estimated using Eqn. \ref{eqn:tsat}, although some simulations were extended beyond this time to explore the behaviour of the layer once it had fully saturated. The estimated time to saturation of the longest wavelength in each initial perturbation is given in Table \ref{tab:sattimes}, while the total physical time of each simulation is given in Table \ref{tab:phystimes}.}

\begin{table*}[]
	\centering
	\caption{Saturation time of the longest wavelength for each simulation, {estimated using Eqn. \ref{eqn:tsat}}.}
	\label{tab:sattimes}
	\begin{tabular}{llll}
		\hline
		& $m=-1$  & $m=-2$ & $m=-3$  \\
		\hline
		{$R=16$}& 0.036 & - &  - \\
		{$R=32$}& 0.083 & 0.044 &  0.031 \\
		{$R=64$}& 0.184 & 0.090 &  0.064 \\
		{$R=128$}& - & 0.183 &  0.129 \\
		\hline
	\end{tabular}
\end{table*}

\begin{table*}[]
	\centering
	\caption{Total physical time of each simulation.}
	\label{tab:phystimes}
	\begin{tabular}{llll}
		\hline
		 & $m=-1$  & $m=-2$ & $m=-3$  \\
		\hline
		{$R=16$}& 0.5 & - &  - \\
		{$R=32$}& 0.25 & 0.15 &  0.1 \\
		{$R=64$}& 0.4 & 0.2 &  0.15 \\
		{$R=128$}& - & 0.185 &  0.13 \\
		\hline
	\end{tabular}
\end{table*}

\section{Results}
\label{sec:results}
\begin{figure}
	\centering
	\includegraphics[width=0.98\textwidth]{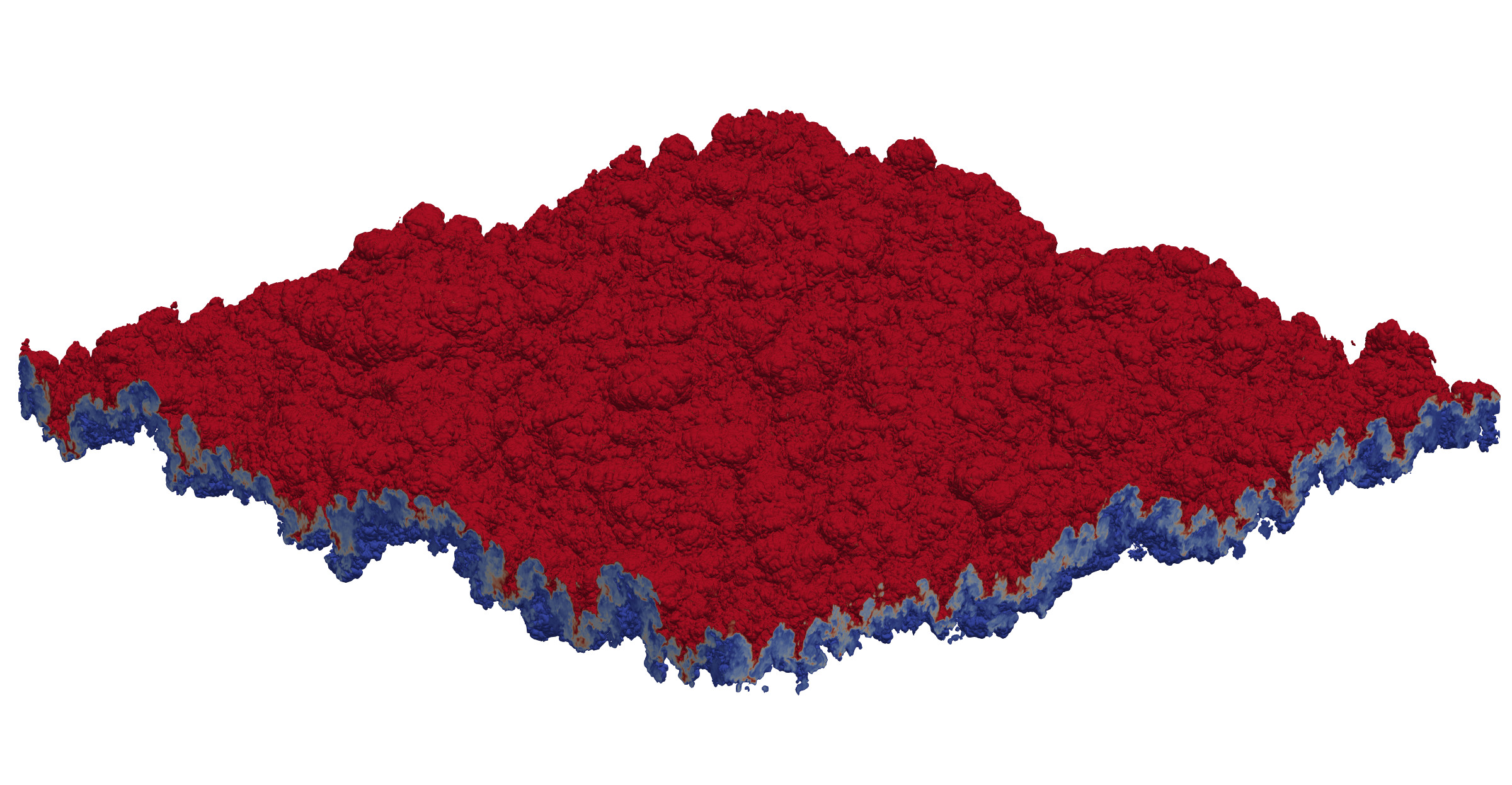}
	\caption{\label{fig:contour3D} Contours of heavy fluid volume fraction $f_1$ between the isosurfaces $f_1=0.01$ (blue) and $f_1=0.99$ (red), for the $m=-2$ case {with an initial bandwidth $R=128$} at time $t=0.185$.}
\end{figure}

Fig. \ref{fig:contour3D} gives a visualisation of the highest {bandwidth} $m=-2$ case, showing red bubbles rising into the heavy fluid and blue spikes penetrating into the light fluid. The data are plotted at the latest time in the simulation, just after the saturation of the longest wavelength in the initial perturbation. The long wavelength modes are still clearly visible at this time, with fine-scale turbulent structures due to the breakdown of shorter wavelength modes superimposed on top of them. At this point in the simulation, the width of the layer is still relatively narrow compared to the longest wavelength.

\begin{figure}
	\centering
	\includegraphics[width=0.98\textwidth]{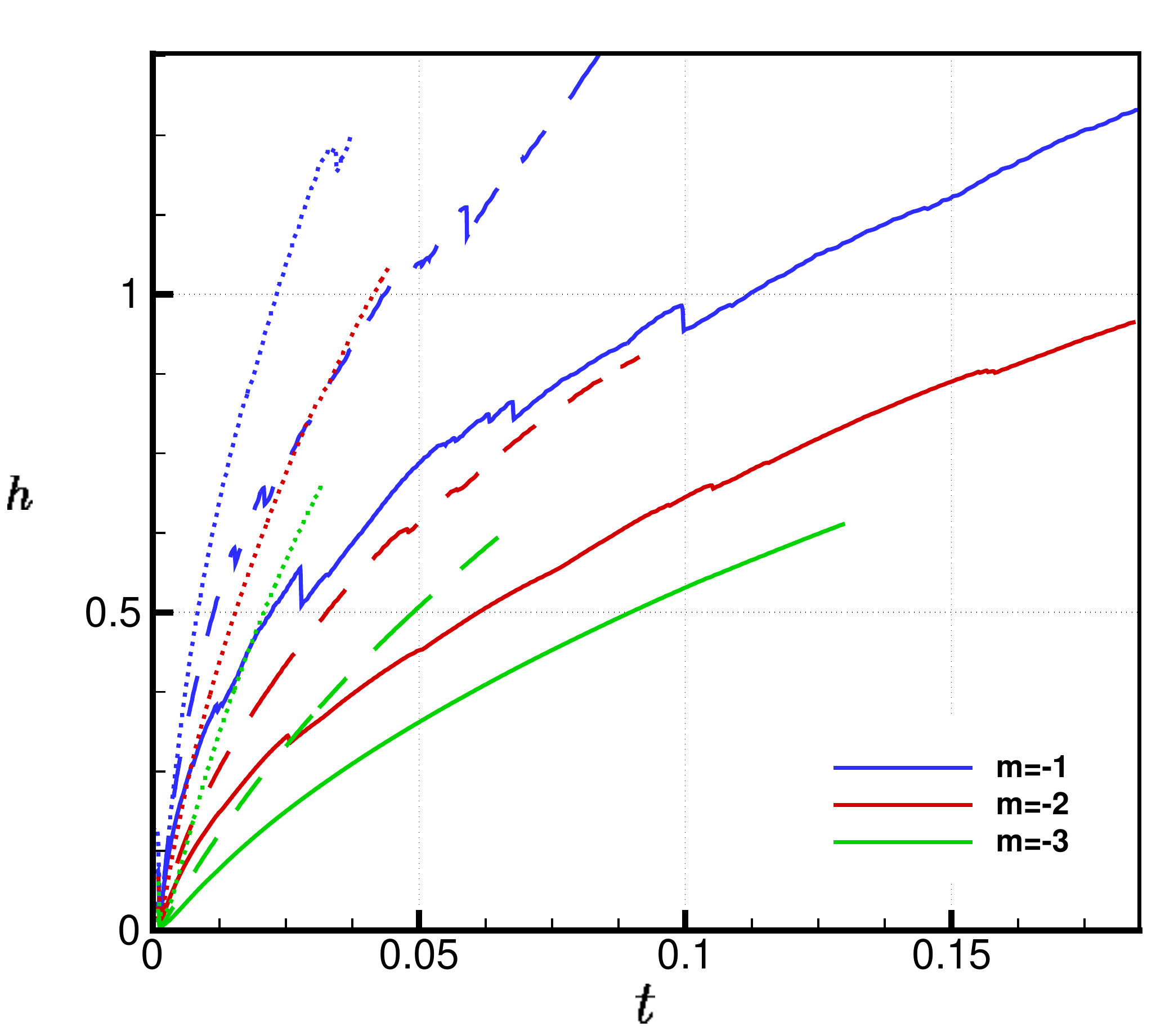}
	\caption{\label{fig:H-t} Peak-to-peak width of the mixing layer vs. physical time. Dotted lines represent the {smallest bandwidth}, dashed lines the medium {bandwidth} and solid lines the {largest bandwidth}. The data are plotted between $t=0$ and the saturation time of the longest wavelength.}
\end{figure}

The width of the mixing layer may be defined in a number of ways, for example the visual width $H$ based on the mean volume fraction profile $\langle f_1 \rangle$ \cite{Cabot2006}. An alternative definition is the peak-to-peak width $h$, taken as the distance between the minimum and maximum $x$ positions where the volume fraction of fluid 1 $f_1=0.5$. Note that both definitions are susceptible to fluctuations caused by turbulent breakup of the interface and acoustic waves, hence they are not well suited for estimating the growth rate exponent $\theta$. It is still useful to examine $h$ (or $H$) however as a way of comparing the growth of the layer for different {bandwidths} and values of $m$. It might be expected that the different cases will all have the same value of $h$ at the saturation time of the longest wavelength, since {$\lambda_{max}$} is the same for all cases.
This comes from observation of Fig. \ref{fig:contour3D}, where it is plausible that the fine scale turbulence superimposed on top of large scale coherent structures has a negligible impact on the overall width of the layer. However, as shown in Fig. \ref{fig:H-t}, {this expectation of the same value of $h$ at saturation time is only realised between cases with the same value of $m$}; as $m$ decreases the peak-to-peak width at saturation time also decreases.

\begin{figure}
	\centering
	\begin{subfigure}{\textwidth}
		\includegraphics[width=\textwidth]{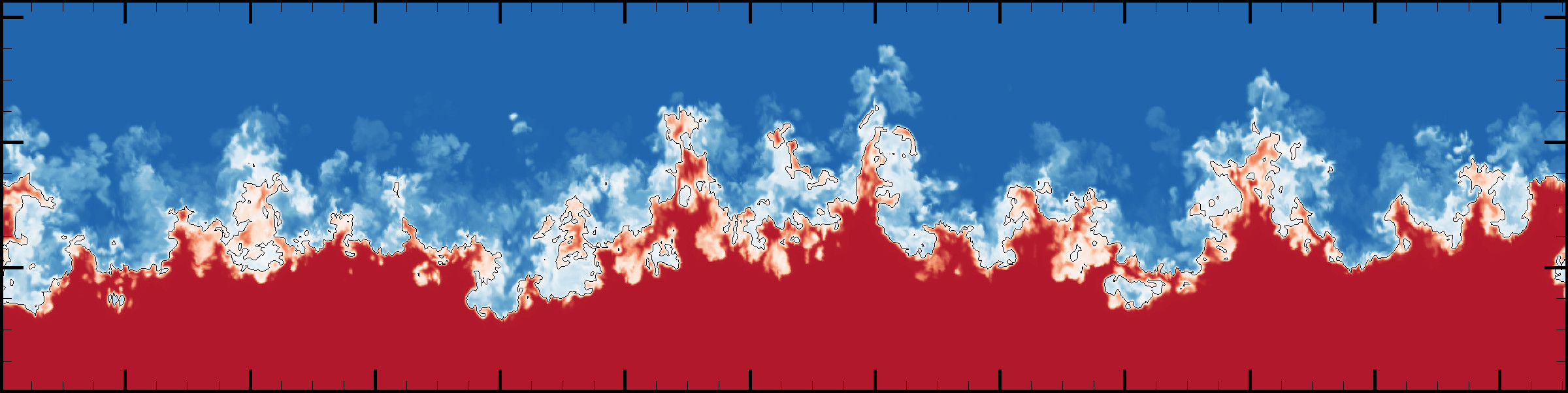}
		\subcaption{$m=-1$.}
	\end{subfigure}
	\begin{subfigure}{\textwidth}
		\includegraphics[width=\textwidth]{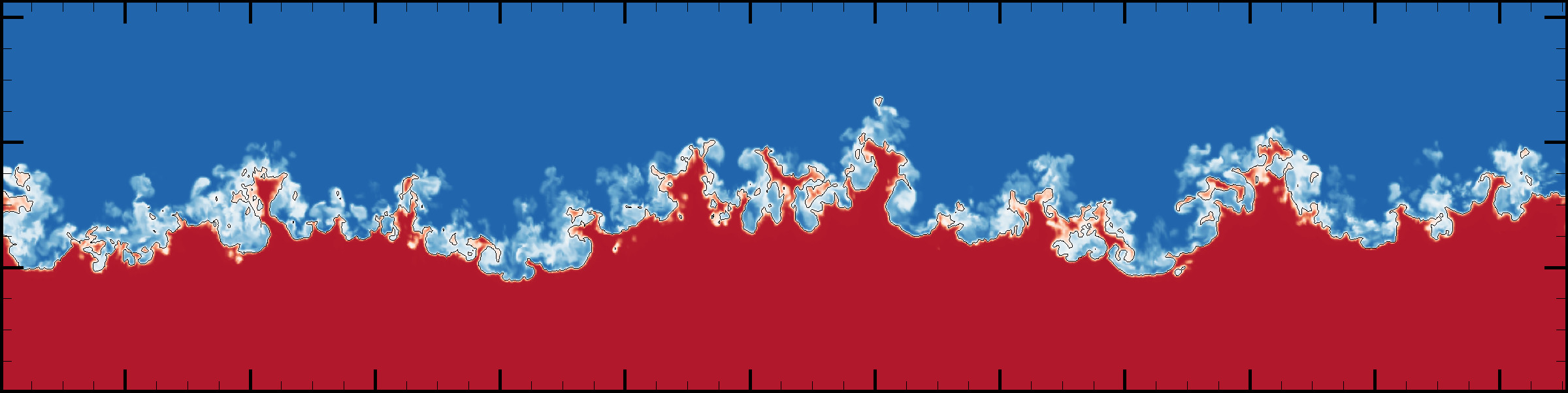}
		\subcaption{$m=-2$.}
	\end{subfigure}	
	\begin{subfigure}{\textwidth}
		\includegraphics[width=\textwidth]{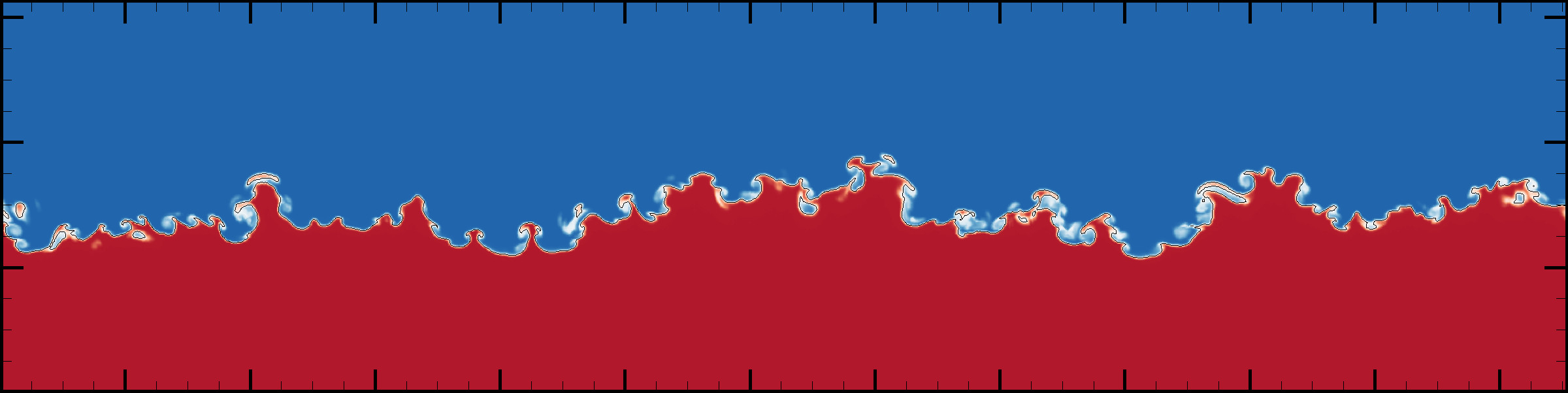}
		\subcaption{$m=-3$.}
	\end{subfigure}	
	\caption{\label{fig:contour1} {Contour flood of volume faction $f_1$ and contour line of $f_1=0.5$ at saturation time for the medium bandwidth cases. The major ticks on both axes correspond to a grid spacing of $\Delta x=0.5$.}}
\end{figure}

The differences in $h$ between cases with the same value of $m$, {which are small compared to the differences between cases with different $m$}, are most likely due to inaccuracies in estimating the true saturation time of the layer. To investigate the cause of the larger differences between cases with different $m$, 2D slices of the $f_1$ volume fraction field are plotted in Fig. \ref{fig:contour1}, as well as lines along the $f_1=0.5$ contour that is used to calculate the peak-to-peak width $h$. One immediate observation is that there is more fine-scale structure in the $m=-1$ simulation than in the $m=-2$ simulation and similarly in the $m=-2$ simulation compared to the $m=-3$ simulation. This is because the smallest wavelengths have had more time to become nonlinear and transition to turbulence in the $m=-1$ simulation due to the longer time to saturation of the longest wavelength. It is also because of this additional time to develop more fine-scale structure that the values of $h$ at saturation time differ between the different simulations. In the $m=-1$ simulation (and less so in the $m=-2$ simulation), the interface is less coherent and there are multiple regions in the flow where blobs of fluid with $f_1\ge0.5$ have separated from the main interface. In contrast to this, the interface at the end of the $m=-3$ simulation is for the most part still simply connected as the smallest wavelengths are still in a relatively early stage of nonlinear development. 

\begin{figure}
	\centering
	\includegraphics[width=0.98\textwidth]{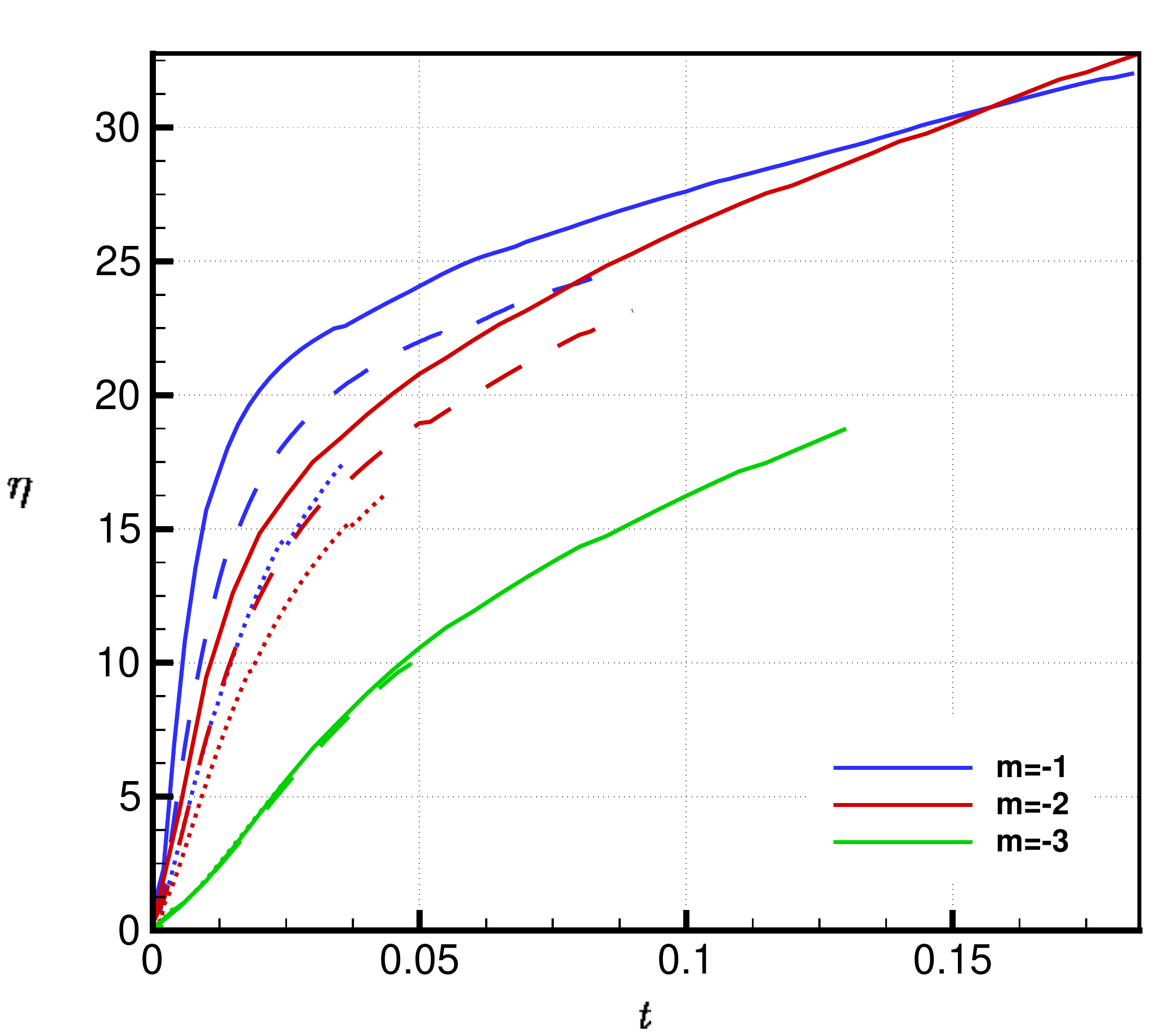}
	\caption{\label{fig:Eta-t} {Spectral bandwidth measure vs. physical time. Dotted lines represent the {smallest bandwidth}, dashed lines the medium {bandwidth} and solid lines the {largest bandwidth}. The data are plotted between $t=0$ and the saturation time of the longest wavelength.}}
\end{figure}

{These observations can be quantified more precisely by considering a measure of the spectral bandwidth, which is expected to rapidly increase during transition to turbulence. Following Gowardhan et al. \cite{Gowardhan2011}, the integral thickness $\updelta$ and mean zero-crossing frequency $\kappa$ are defined as}
{
\begin{subequations}
	\begin{align}
	\updelta & = \int_{-\infty}^{\infty}4\langle Y_1\rangle\langle Y_2\rangle\:\mathrm{d}x, \label{subeqn:delta} \\
	\kappa^2 & = \frac{\int_{0}^{\infty}k^2E_\rho(k)\:\mathrm{d}k}{\int_{0}^{\infty}E_\rho(k)\:\mathrm{d}k}, \label{subeqn:kappa}
	\end{align}
	\label{eqn:eta}
\end{subequations}}
{where $\langle\dots\rangle$ indicates a plane average taken over the statistically homogeneous directions and $E_\rho(k)$ is the density variance spectrum, calculated using the method given in Eqn. \ref{eqn:E2D}. A measure of the spectral bandwidth is then given by $\eta(t)=\updelta(t)\kappa(t)$. Fig. \ref{fig:Eta-t} gives the evolution of $\eta$ in time for each case, showing easily discernible differences between cases with different values of $m$. For a given initial bandwidth $R$, the increase in spectral bandwidth is most rapid for $m=-1$ and most slow for $m=-3$. This indicates that the $m=-1$ cases transition to turbulence first and have the most fine-scale structure present at saturation time, in accordance with the observations made in Fig. \ref{fig:contour1}.}

\begin{figure}
	\centering
	\begin{subfigure}{\textwidth}
		\includegraphics[width=\textwidth]{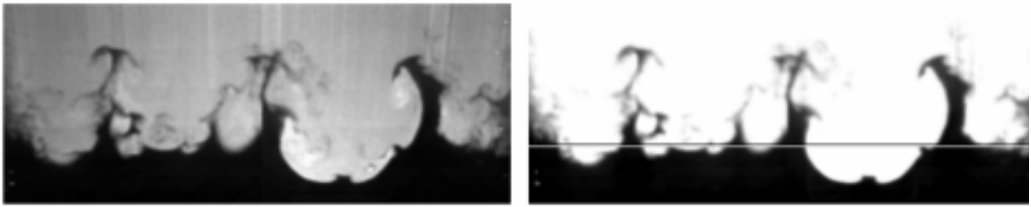}
		\subcaption{Original (left) and processed (right) images from Experiment 4.}
	\end{subfigure}
	\begin{subfigure}{\textwidth}
		\includegraphics[width=\textwidth]{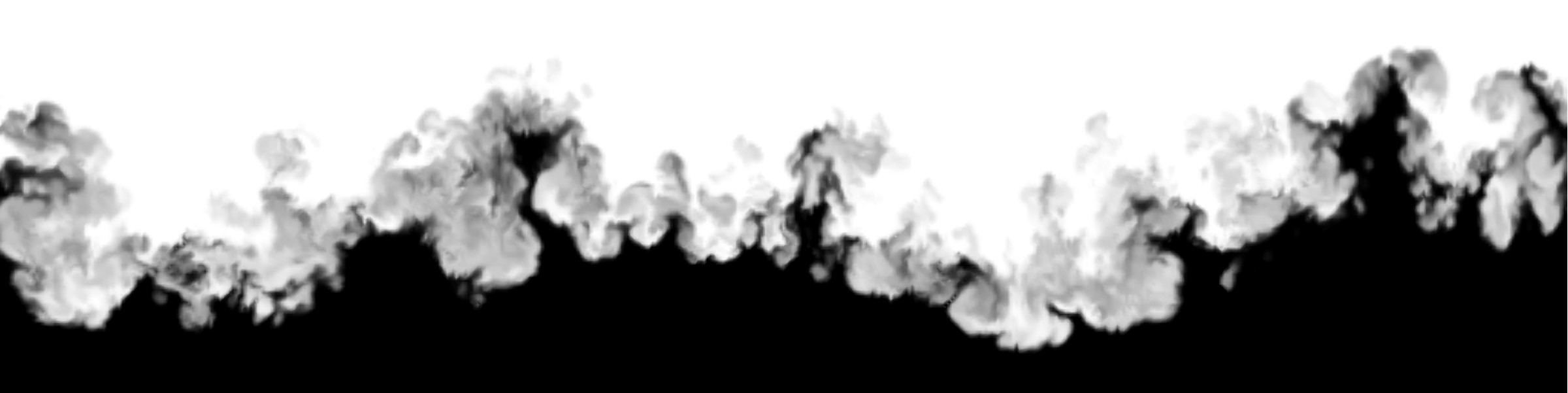}
		\subcaption{Contour flood of volume fraction from the $m=-2$, {$R=64$} simulation.}
	\end{subfigure}
	\caption{\label{fig:krivets} Comparison of (a) experimental images from Experiment 4 and (b) volume fraction contour flood from a section of the $m=-2$, {$R=64$} simulation at time $t=0.1$. \textit{Source}: From Fig. 1 of Krivets et al. \cite{Krivets2017}.}
\end{figure}

The contour plots in Fig. \ref{fig:contour1} also resemble quite closely the experimental images reported in Krivets et al. \cite{Krivets2017}. Those image sequences, particularly for Experiment 4, show a {transitional} mixing layer containing a broad range of modes, the smallest of which have become turbulent by the end of the experiment, while the largest modes are still mostly linear. Fig. \ref{fig:krivets} gives a comparison of the final images from Experiment 4 with an image taken from a section of the $m=-2$ simulation at {a bandwidth of $R=64$}, highlighting the very similar phenomenology. The majority of values reported for $\theta$ in \cite{Krivets2017} are also higher than the range of values typically reported for narrowband, short wavelength perturbations, suggesting that the growth rate of the layer is being {influenced} by linear growth of longer wavelengths. For example, in Experiment 4 the bubble and spike growth rates were $\theta_b=0.42$ and $\theta_s=0.51$ respectively, which suggests the perturbations used were broadband in nature. Relating those experiments to the present work, a perturbation with spectral exponent $-1\le m\le -2$ and relatively narrow bandwidth (i.e. $R=32$) would produce a similar growth rate, as will be shown in Sec. \ref{subsec:mix}. A more realistic representation would be a power spectrum consisting of multiple distinct ranges, each with a different exponent, such as in the initial conditions used in Weber et al. \cite{Weber2012}. This will be explored further in future work.

\subsection{Mixing Measures}
\label{subsec:mix}
In this section, various integral mixing measures based on plane-averaged volume fraction profiles are presented for each of the nine simulations. The most fundamental of these is the integral width, given by
\begin{equation}
W=\int_{-\infty}^{\infty}\langle f_1\rangle\langle f_2\rangle\:\mathrm{d}x.
\end{equation}
As with $h$, the integral width evolves as $W\propto t^\theta$ and is the more useful quantity for estimating $\theta$ as it is more robust to fluctuations. Bubble and spike integral widths may also be defined, following Krivets et al. \cite{Krivets2017}, as
\begin{eqnarray}
W_b & = & \int_{-\infty}^{x_c} \langle f_1\rangle\langle f_2\rangle\:\mathrm{d}x, \\
W_s & = & \int_{x_c}^{\infty} \langle f_1\rangle\langle f_2\rangle\:\mathrm{d}x,
\label{eqn:Wb-Ws}
\end{eqnarray}
where the mixing layer centre $x_c$ is defined as the $x$ position of equal mixed volumes \cite{Walchli2017}, given by
\begin{equation}
\int_{-\infty}^{x_c}\langle f_2\rangle\:\mathrm{d}x=\int_{x_c}^{\infty}\langle f_1\rangle\:\mathrm{d}x.
\label{eqn:xc}
\end{equation}
Another useful quantity, based on second-order moments, is the (global) molecular mixing fraction \cite{Youngs1991}, given by
\begin{equation}
\Theta=\frac{\int\langle f_1f_2\rangle\:\mathrm{d}x}{\int\langle f_1\rangle\langle f_2\rangle\:\mathrm{d}x}.
\end{equation}
$\Theta$ can take values anywhere between 0 and 1, with $\Theta=0$ corresponding to complete heterogeneity and $\Theta=1$ corresponding to complete homogeneity of mixing. A steady-state value of $\Theta$ is also an indication that the mixing layer is evolving in a self-similar fashion. 

In order to compare the evolution of $W$ across different {bandwidths} and values of $m$, a suitable non-dimensionalisation is sought, along the same lines of the analysis presented in Thornber et al. \cite{Thornber2017} for narrowband perturbations. For a multimode perturbation of the form given in Eqn. \ref{eqn:pert} with normally distributed coefficients, the initial growth rate of the integral width is
\begin{equation}
\dot{W_0}=0.564\sigma^+A^+\Delta u\bar{k},
\end{equation}
where $\sigma^+=(1-\Delta u/U_s)\sigma$ and $\bar{k}$ is a weighted average wavenumber of the perturbation, given by
\begin{equation}
\bar{k}=\sqrt{\frac{\displaystyle \int_{k_{min}}^{k_{max}}k^2P(k)\:\mathrm{d}k}{\displaystyle \int_{k_{min}}^{k_{max}}P(k)\:\mathrm{d}k}}.
\label{eqn:kbar}
\end{equation}
For each of the simulations $\sigma=0.2\lambda_{min}$ and $\bar{k}$ is
\begin{equation}
\bar{k} = \displaystyle \left\{
\begin{array}{lll}
k_{max}\sqrt{\frac{\displaystyle 1-1/R^2}{\displaystyle 2\log(R)}}, & m=-1, \vspace{1em}\\ 
k_{max}\sqrt{\frac{\displaystyle 1}{\displaystyle R}}, & m=-2, \vspace{1em}\\
k_{max}\sqrt{\frac{\displaystyle 2\log(R)}{\displaystyle R^2-1}}, & m=-3.
\end{array} \right.
\end{equation}
An additional correction factor must be included to account for the initial diffuse width of the interface. Following Duff et al. \cite{Duff1962} and Youngs \& Thornber \cite{Youngs2019}, the initial impulse is written as $\dot{W_0}=0.564\sigma^+A^+\Delta u\bar{k}/\psi$, where $\psi$ is given by
\begin{equation}
\psi=1+\sqrt{\frac{2}{\pi}}\bar{k}\epsilon.
\label{eqn:psi}
\end{equation}
In Eqn. \ref{eqn:psi}, $\epsilon=\delta^+/\sqrt{\pi}$ where $\delta^+=\overline{C}\delta$ is the post-shock characteristic thickness of the interface, $\delta=\lambda_{min}/4$ is the pre-shock characteristic thickness and $\overline{C}$ is the mean compression rate, given by
\begin{equation}
\overline{C}=\frac{\rho_1^-+\rho_2^-}{\rho_1^++\rho_2^+}.
\end{equation}
The initial growth rates of $W$ are tabulated in Table \ref{tab:W0dot}. In the following sections, all quantities are non-dimensionalised by $\dot{W_0}$ and $\lambda_{min}$, for example dimensionless time $\tau=t\dot{W_0}/\lambda_{min}$.

\begin{table*}[]
	\centering
	\caption{Initial growth rate $\dot{W_0}$ for each simulation.}
	\label{tab:W0dot}
	\begin{tabular}{llll}
		\hline
		& $m=-1$  & $m=-2$ & $m=-3$  \\
		\hline
		{$R=16$}& 11.99 & - &  - \\
		{$R=32$}& 10.91 & 5.466 &  2.639 \\
		{$R=64$}& 10.08 & 3.942 &  1.467 \\
		{$R=128$}& - & 2.827 &  0.7983 \\
		\hline
	\end{tabular}
\end{table*}

\begin{figure}
	\centering
	\includegraphics[width=0.98\textwidth]{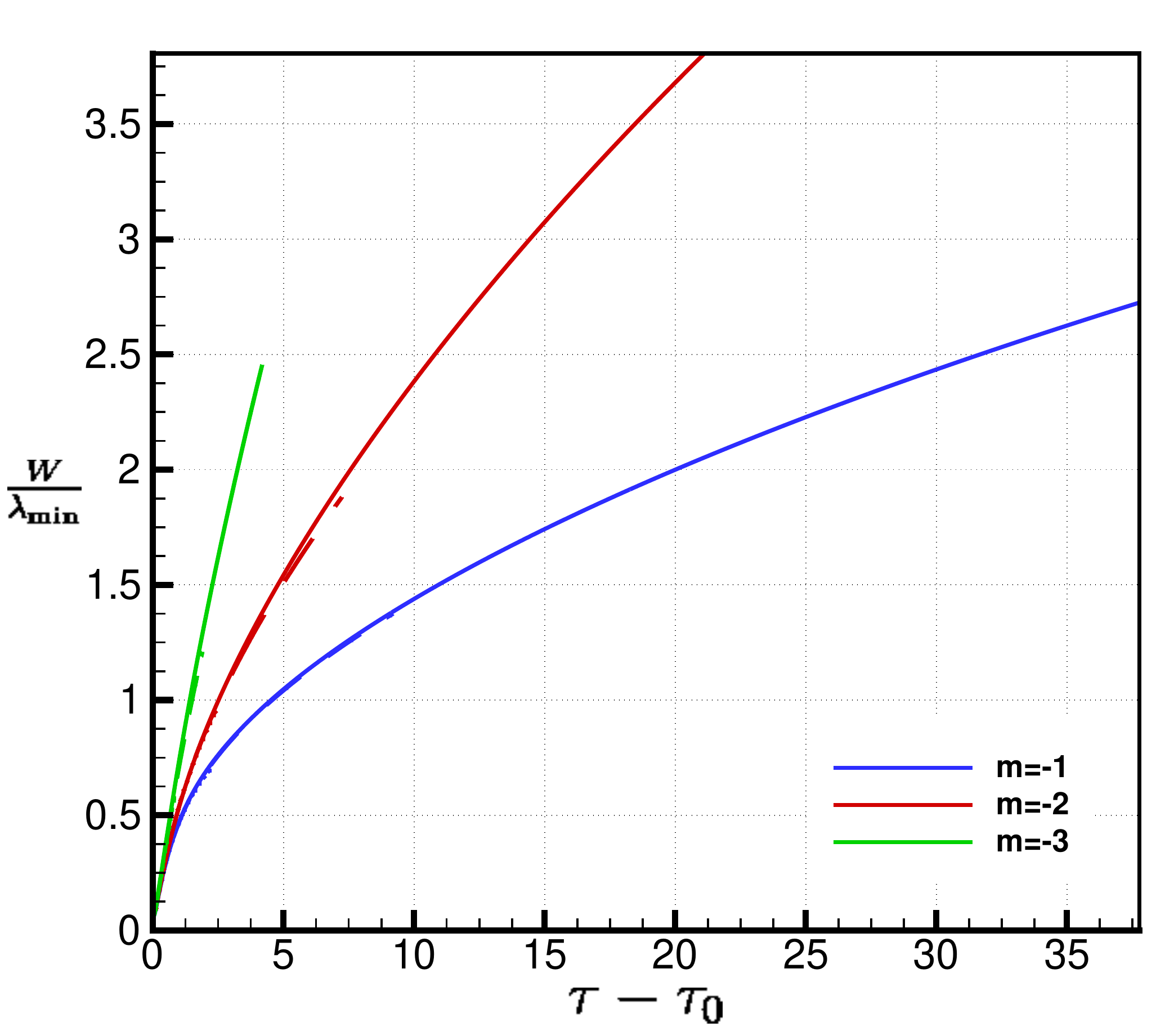}
	\caption{\label{fig:W-tau} Integral width vs. dimensionless time. Dotted lines represent the {smallest bandwidth}, dashed lines the medium {bandwidth} and solid lines the {largest bandwidth}.}
\end{figure}

The evolution of integral width in time is shown in Fig. \ref{fig:W-tau}. The data are plotted from the shock arrival time ($t_0=0.0011$) up to the saturation time of the largest wavelength. A good collapse of the data is observed at early time across all cases, as well as at late time between cases with the same value of $m$. This indicates that with this non-dimensionalisation, cases that are growing with a larger value of $\theta$ also have a larger value of $W/\lambda_{min}$ {for a given dimensionless time}. Fig. \ref{fig:W-tau} also shows how it becomes increasingly difficult to simulate to late dimensionless times with decreasing values of $m$ (prior to the largest wavelength saturating), which is in line with the qualitative observations made for Fig. \ref{fig:contour1}. In other words, to obtain the same dimensionless time at the point of saturation of the largest wavelength for the $m=-3$ case compared to the $m=-1$ case, a much larger bandwidth is required (or alternatively a larger impulse must be used). This is the reason why {smaller bandwidths} were used in the $m=-1$ cases compared to the other cases. 

\begin{figure}
	\centering
	\topinset{\includegraphics[width=0.2\textwidth]{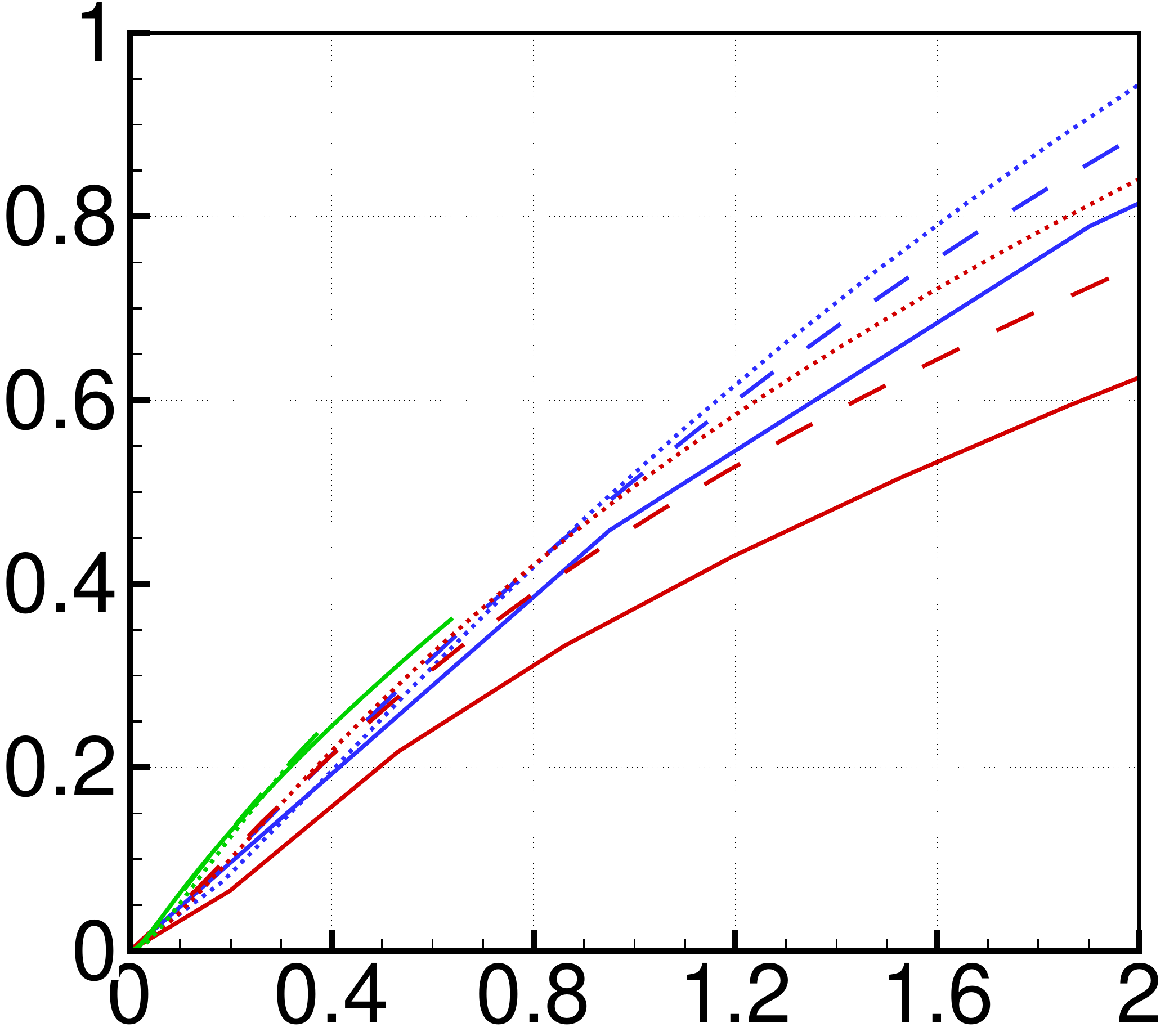}}{\includegraphics[width=0.49\textwidth]{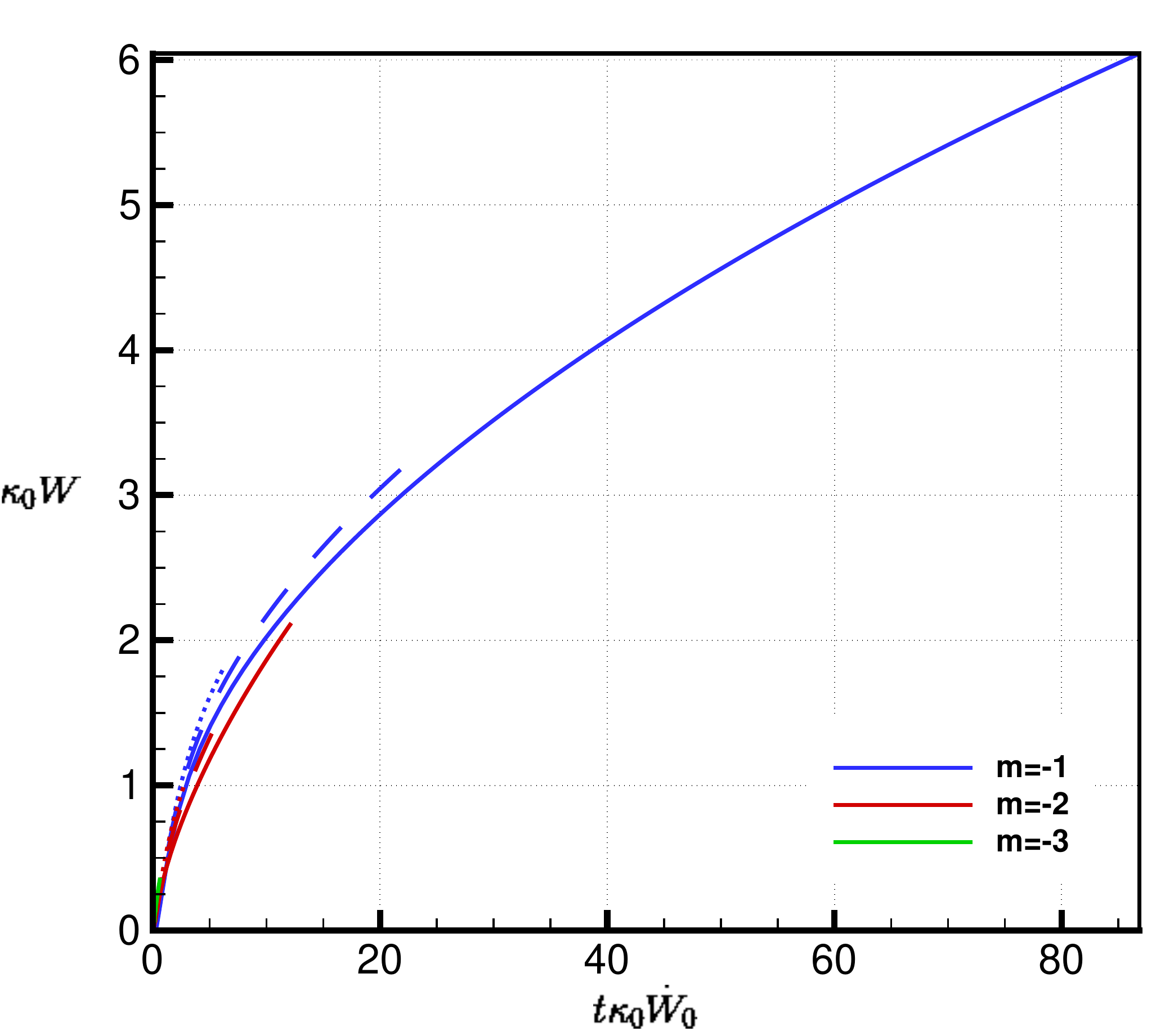}}{15pt}{10pt}
	\topinset{\includegraphics[width=0.2\textwidth]{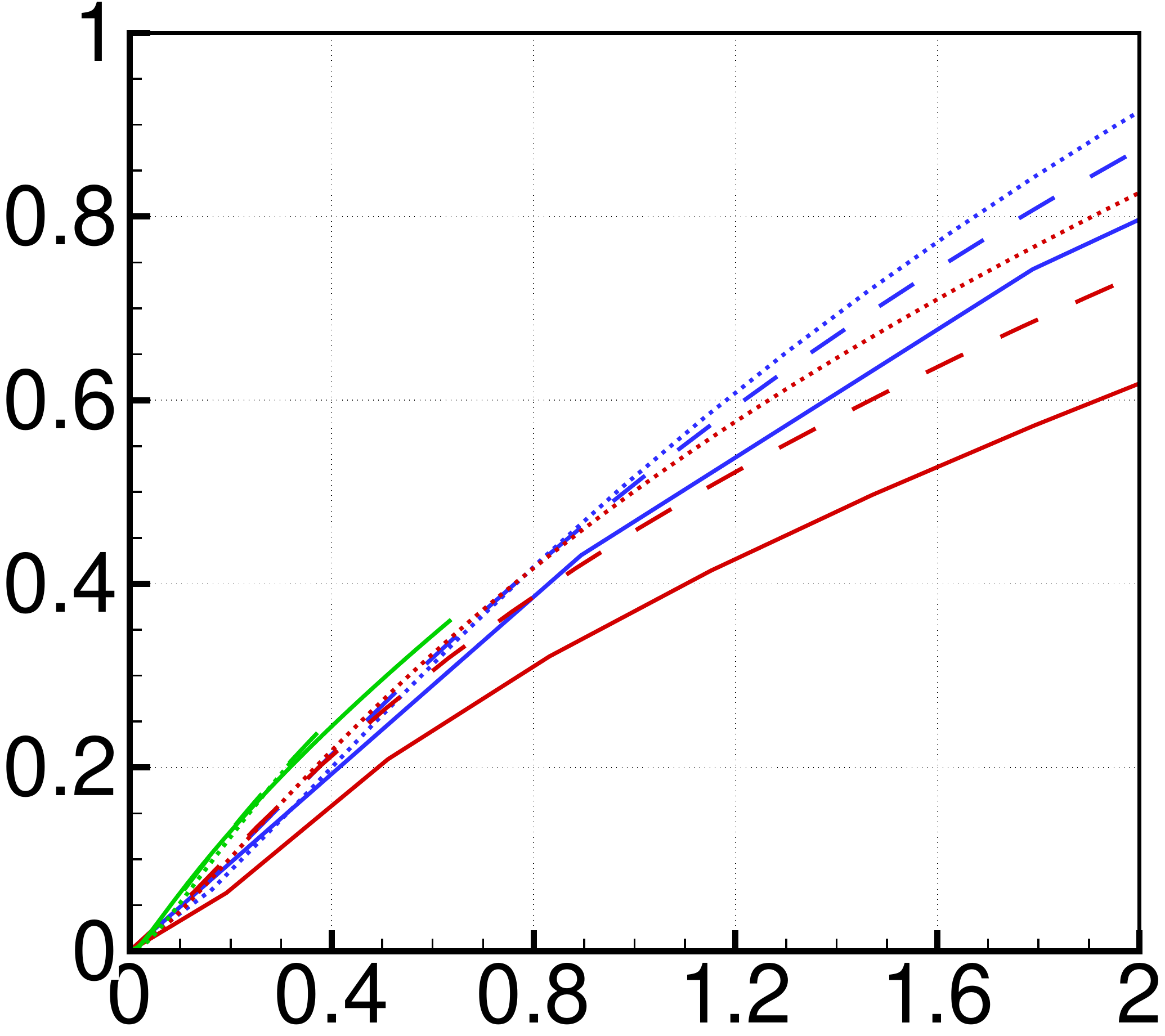}}{\includegraphics[width=0.49\textwidth]{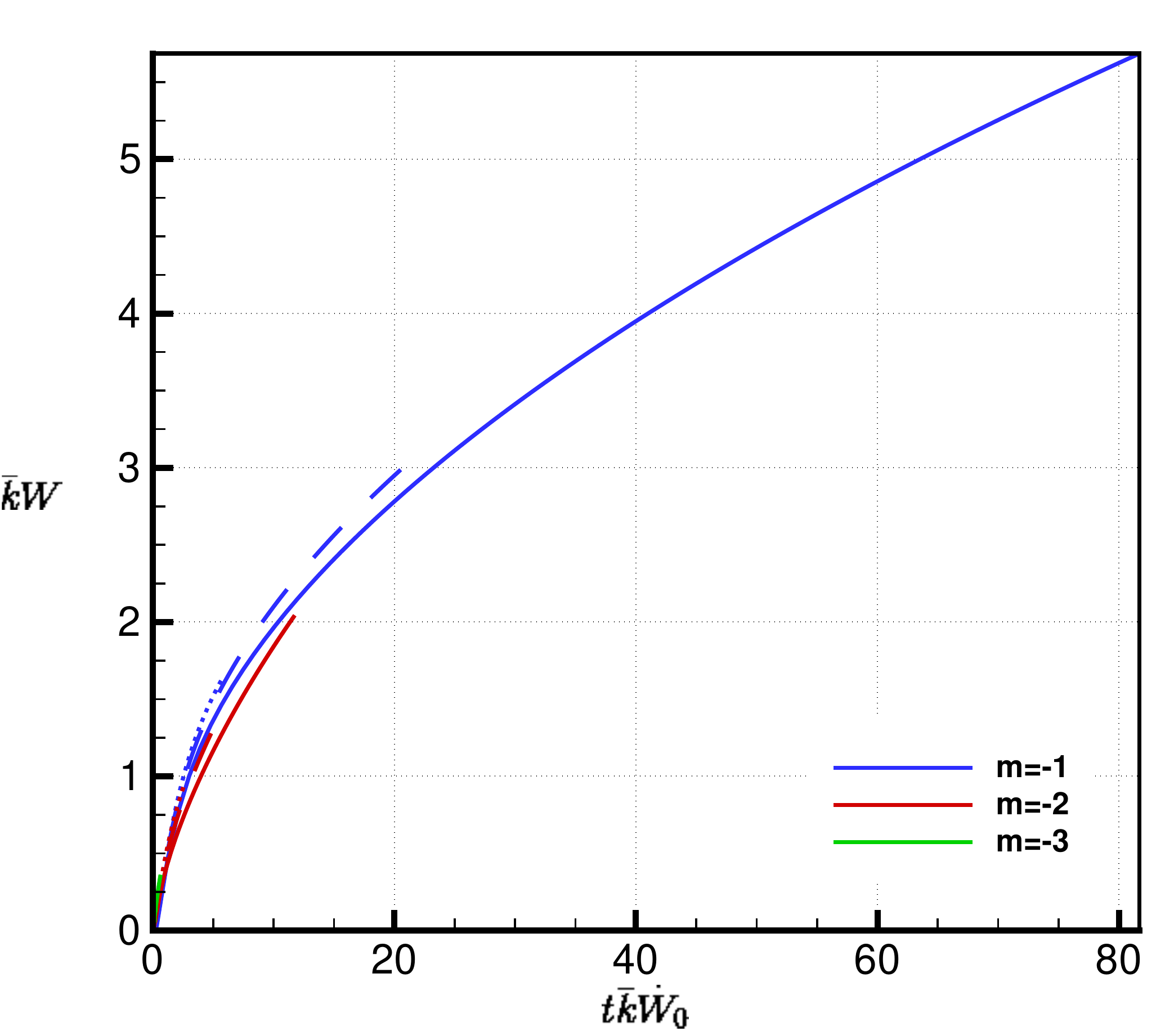}}{15pt}{10pt}
	\caption{\label{fig:Width-t} {Integral width vs. time using the non-dimensionalisation of Gowardhan et al. \cite{Gowardhan2011} (left) and an equivalent non-dimensionalisation based on $\bar{k}$ (right). Dotted lines represent the {smallest bandwidth}, dashed lines the medium {bandwidth} and solid lines the {largest bandwidth}.}}
\end{figure}

{As a comparison, the non-dimensionalisation presented in Gowardhan et al. \cite{Gowardhan2011} is also performed here. The integral width is scaled by $\kappa_0=\kappa(0)$ and time is scaled by $\kappa_0\dot{W_0}$, with the results shown in Fig. \ref{fig:Width-t}. Compared with the present non-dimensionalisation, the data are collapsed to approximately a single curve, although the collapse between cases with the same value of $m$ is not as good. The difference in dimensionless time between cases with different $m$ has also been substantially enhanced. It is interesting to note that $\kappa_0$ is very similar to the weighted average wavenumber $\bar{k}$ introduced in Eqn. \ref{eqn:kbar}. For example, for the $m=-2$ case at a bandwidth of $R=128$, $\kappa_0=23.46$ and $\bar{k}=22.62$. This suggests that a very similar collapse should occur if $\bar{k}$ is used instead to non-dimensionalise $W$ and $t$, with Fig. \ref{fig:Width-t} confirming that this is indeed the case. Since this scaling was introduced by Gowardhan et al. \cite{Gowardhan2011} to distinguish between cases with linear vs. nonlinear initial perturbations, it is not surprising that all of the present cases collapse to a single group. This can also be explained in terms of the initial impulse $\dot{W_0}$, which varies between cases with different $m$ due to varying $\bar{k}$. Given that the purpose here is to distinguish between cases with different $m$, the original non-dimensionalisation is used for the remainder of this study.} 

\def\stackalignment{r}

%

To estimate the growth rate exponent $\theta$ for each simulation, nonlinear regression was used to fit a model of the form $W=\beta(\tau-\tau_0)^\theta$. The interval over which this regression was performed was taken to be the period of time between the inversion time to the saturation time of the largest wavelength, as during this period the {results are expected to be representative of the infinite bandwidth limit}. The estimate of $\theta$ taken from the nonlinear regression for each simulation is given in Table \ref{tab:theta}. It can be seen that for the $m=-1$ and $m=-2$ cases $\theta$ is increasing with the increasing bandwidth of the perturbation and is approaching the theoretical value of $2/(m+5)$, particularly in the $m=-1$ simulations. In the $m=-3$ cases, $\theta$ is actually decreasing with increasing bandwidth. One possible explanation for this is that as the bandwidth increases, the shortest wavelengths are more nonlinear by the time the largest wavelength saturates and hence there is more dissipation of kinetic energy in the layer (for this specific value of $m$). {In all cases the error in the regression is very low; the coefficient of determination $R^2$ is at least 0.999 and the standard error is at most 0.096\%. This is not equivalent to the uncertainty in the value of $\theta$ but is merely a measure of how well the functional form $W=\beta(\tau-\tau_0)^\theta$ can explain the variation in the integral width data; the uncertainty in the data itself has not been taken into account. In order to obtain meaningful error bounds on $\theta$ multiple realisations would need to be run, or the same realisation simulated using multiple codes as in Thornber et al. \cite{Thornber2017}. However, by carefully designing the problem it is assumed these error bounds are small \cite{Thornber2016}.}

\begin{table*}[]
	\centering
	\caption{Growth rate exponent $\theta$ for each simulation.}
	\label{tab:theta}
	\begin{tabular}{llll}
		\hline
		& $m=-1$  & $m=-2$ & $m=-3$  \\
		\hline
		{$R=16$}& 0.426 & - &  - \\
		{$R=32$}& 0.432 & 0.571 &  0.872 \\
		{$R=64$}& 0.488 & 0.574 &  0.805 \\
		{$R=128$}& - & 0.622 &  0.790 \\
		\hline
	\end{tabular}
\end{table*}

\begin{figure}
	\centering
	\includegraphics[width=0.49\textwidth]{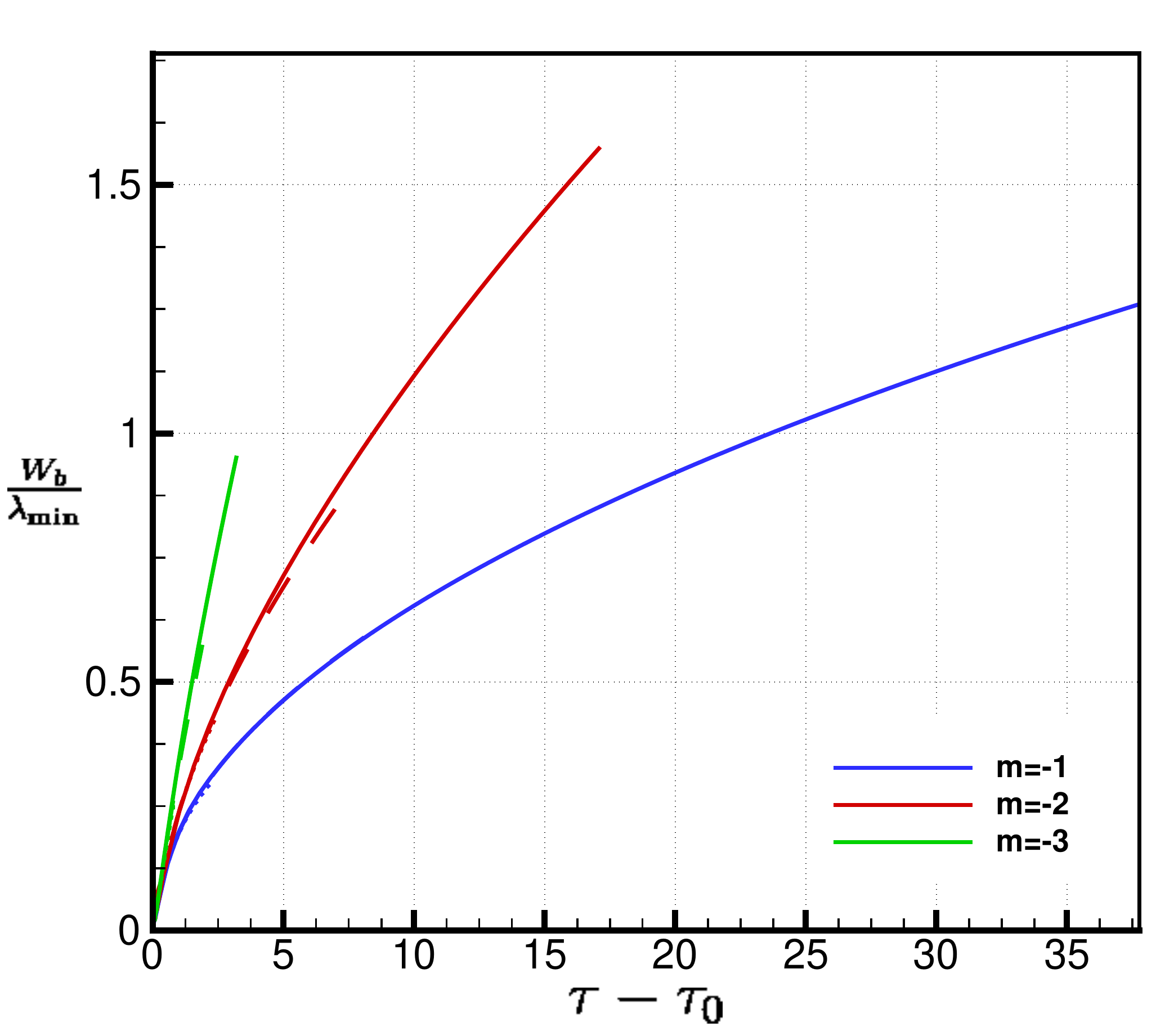}
	\includegraphics[width=0.49\textwidth]{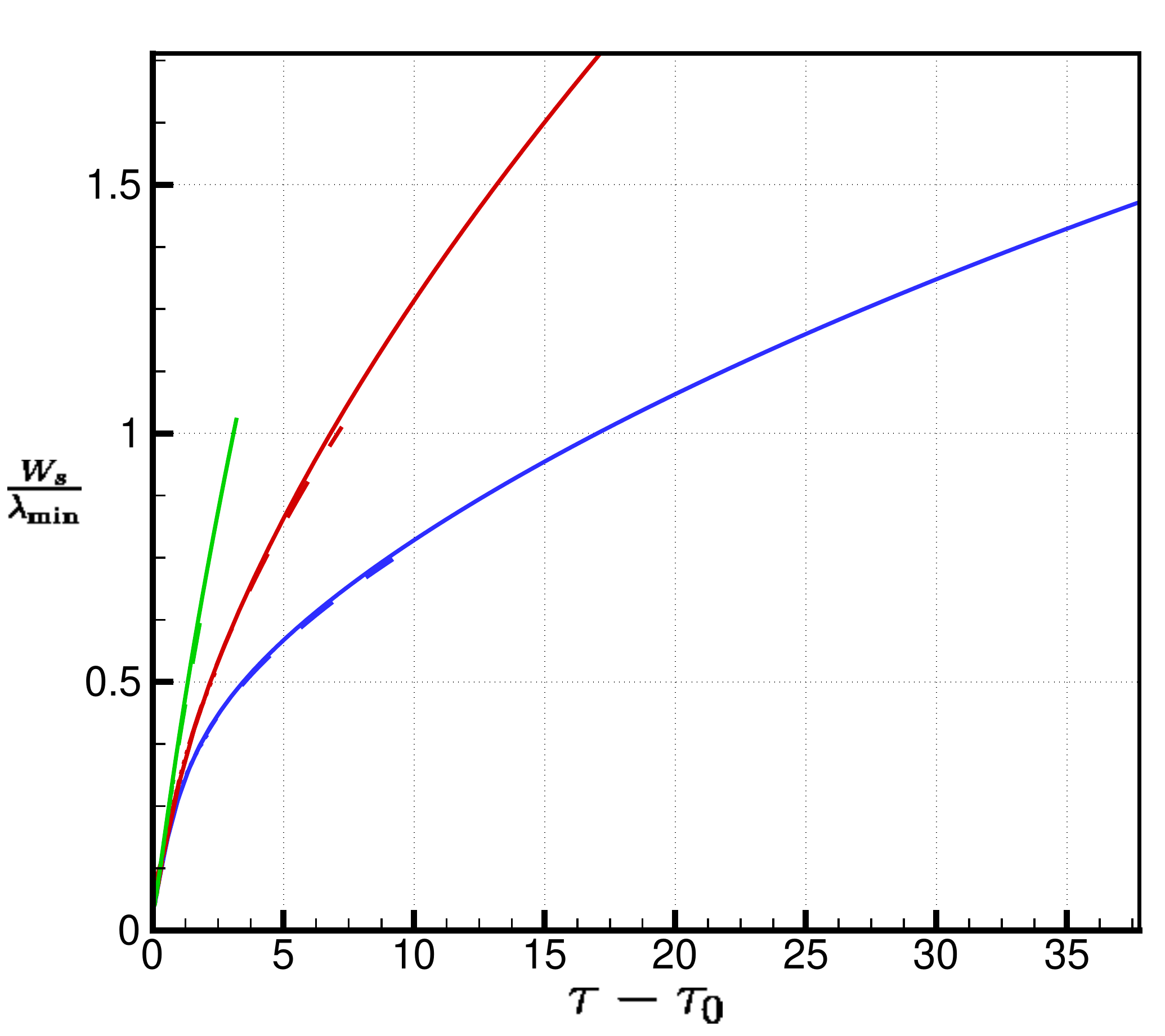}
	\caption{\label{fig:Wbs-tau} Bubble and spike integral widths vs. dimensionless time. Dotted lines represent the {smallest bandwidth}, dashed lines the medium {bandwidth} and solid lines the {largest bandwidth}.}
\end{figure}

\begin{figure} 
	\centering
	\begin{subfigure}{0.49\textwidth}
		\includegraphics[width=\textwidth]{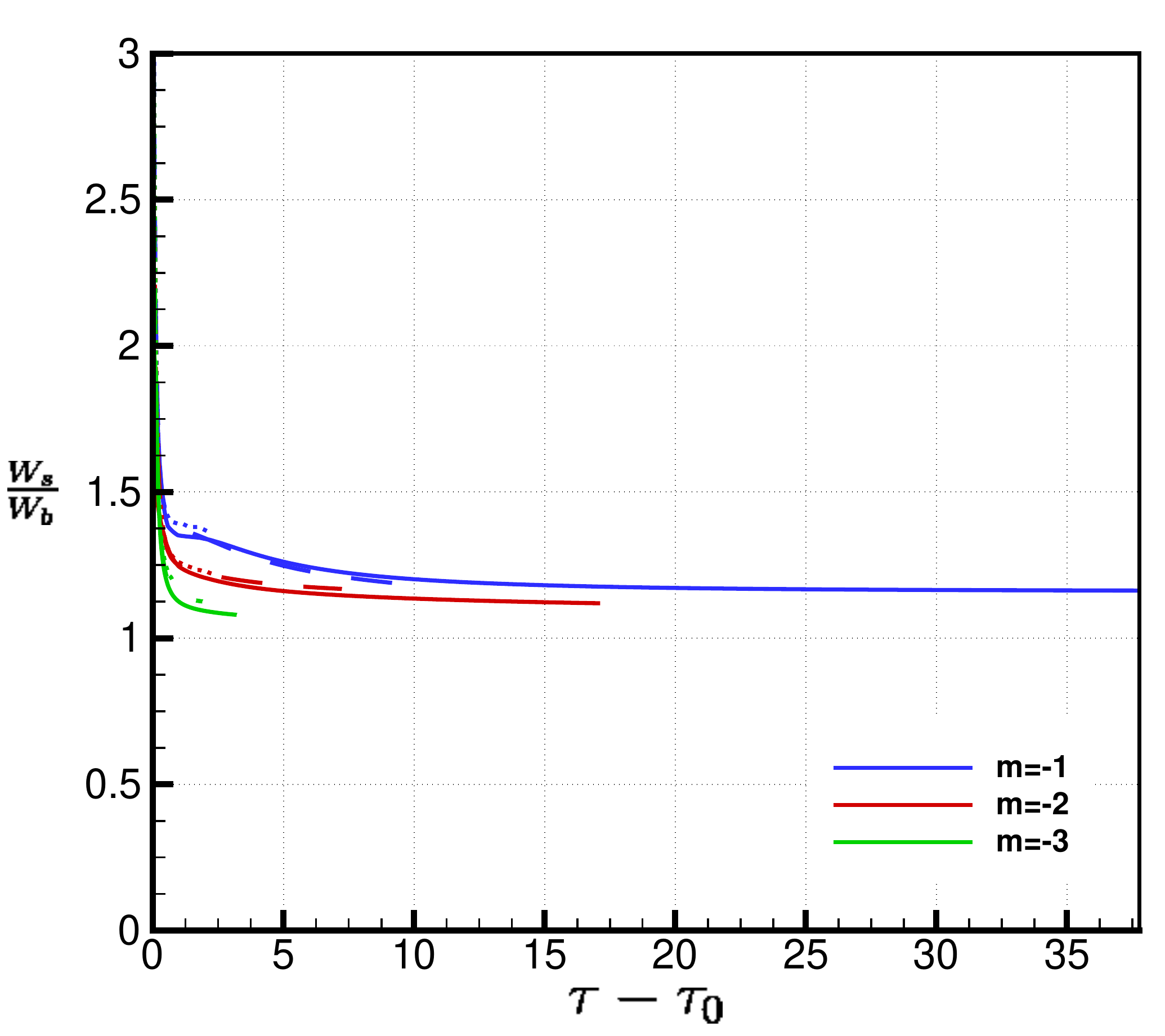}
		\subcaption{Saturation time.}
	\end{subfigure}
	\begin{subfigure}{0.49\textwidth}
		\includegraphics[width=\textwidth]{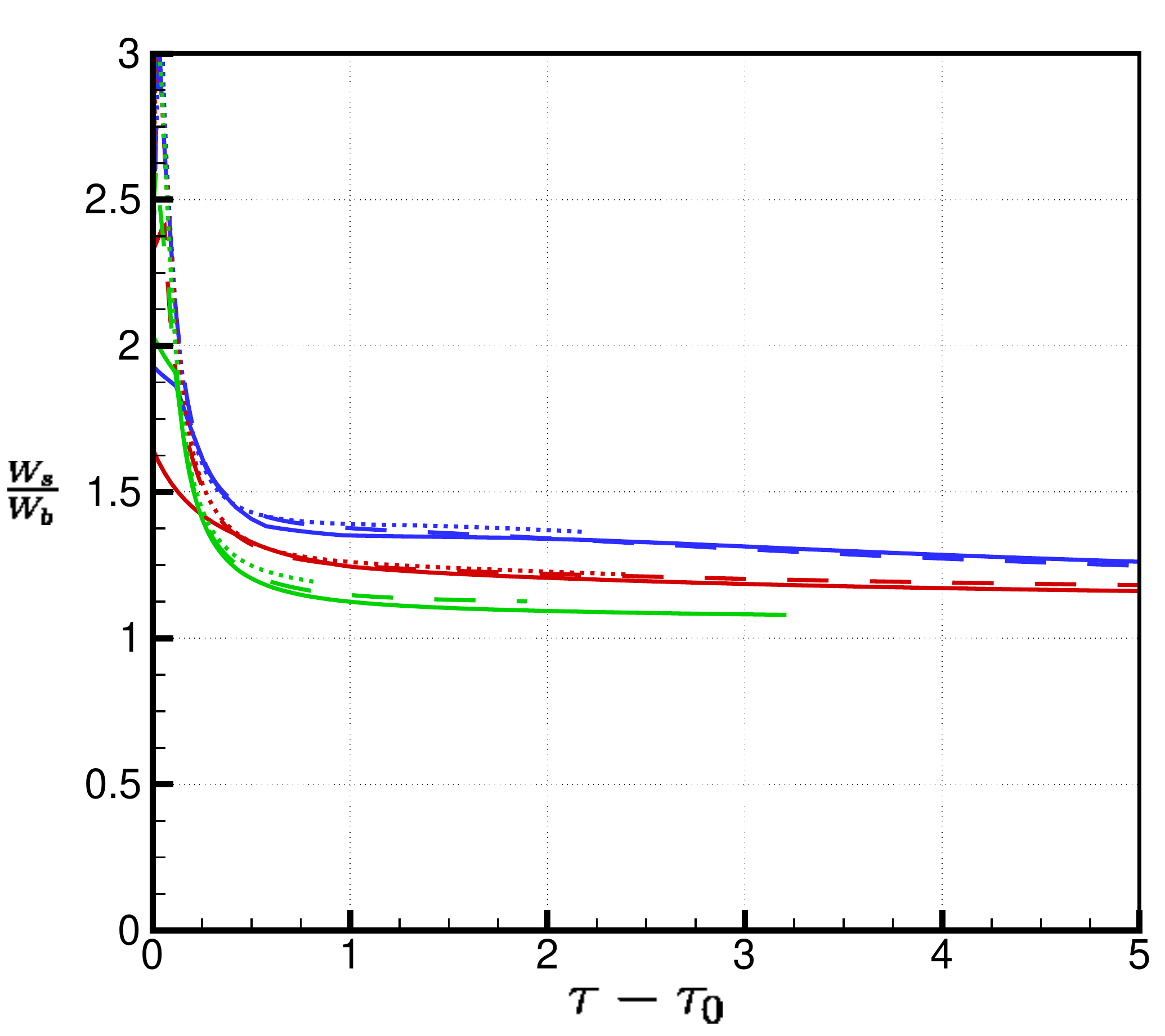}
		\subcaption{Early time.}
	\end{subfigure}
	\caption{\label{fig:ratio-tau} Ratio of bubble and spike integral widths. Dotted lines represent the {smallest bandwidth}, dashed lines the medium {bandwidth} and solid lines the {largest bandwidth}.}
\end{figure}

The evolution of the bubble and spike integral widths in time is shown in Fig. \ref{fig:Wbs-tau}. Qualitatively, both $W_b$ and $W_s$ evolve quite similarly, however $W_s$ is greater than $W_b$ for the entire duration of the simulation in all cases. To explore the relationship between $W_b$ and $W_s$ further, the evolution of the ratio $W_s/W_b$ in time is plotted in Fig. \ref{fig:ratio-tau}. In all cases this ratio is initially around 3 just after shock passage, but quickly reduces and asymptotes to a constant value. {The early-time variation in $W_s/W_b$ indicates that the initial impulsive growth rate of $W_s$ is greater than that of $W_b$. However, the fact that $W_s/W_b$ approaches a constant value at late time indicates that the bubbles and spikes eventually scale with the same exponent $\theta_b=\theta_s=\theta$. This is confirmed using nonlinear regression; for the highest bandwidth cases the difference between $\theta_b$ and $\theta_s$ when fit between the inversion and saturation times is 3.9\%, 4.3\% and 0.63\% for $m=-1$, $-2$ and $-3$ respectively.} This implies that at late time the {self-similar} evolution of the mixing layer can be described by a single length scale $W$.  The constant value is also not the same for different values of $m$; at the latest dimensionless time in each simulation the ratio $W_s/W_b$ is 1.16, 1.12 and 1.08 for decreasing $m$.

\begin{figure}
	\centering
	\begin{subfigure}{0.49\textwidth}
	\includegraphics[width=\textwidth]{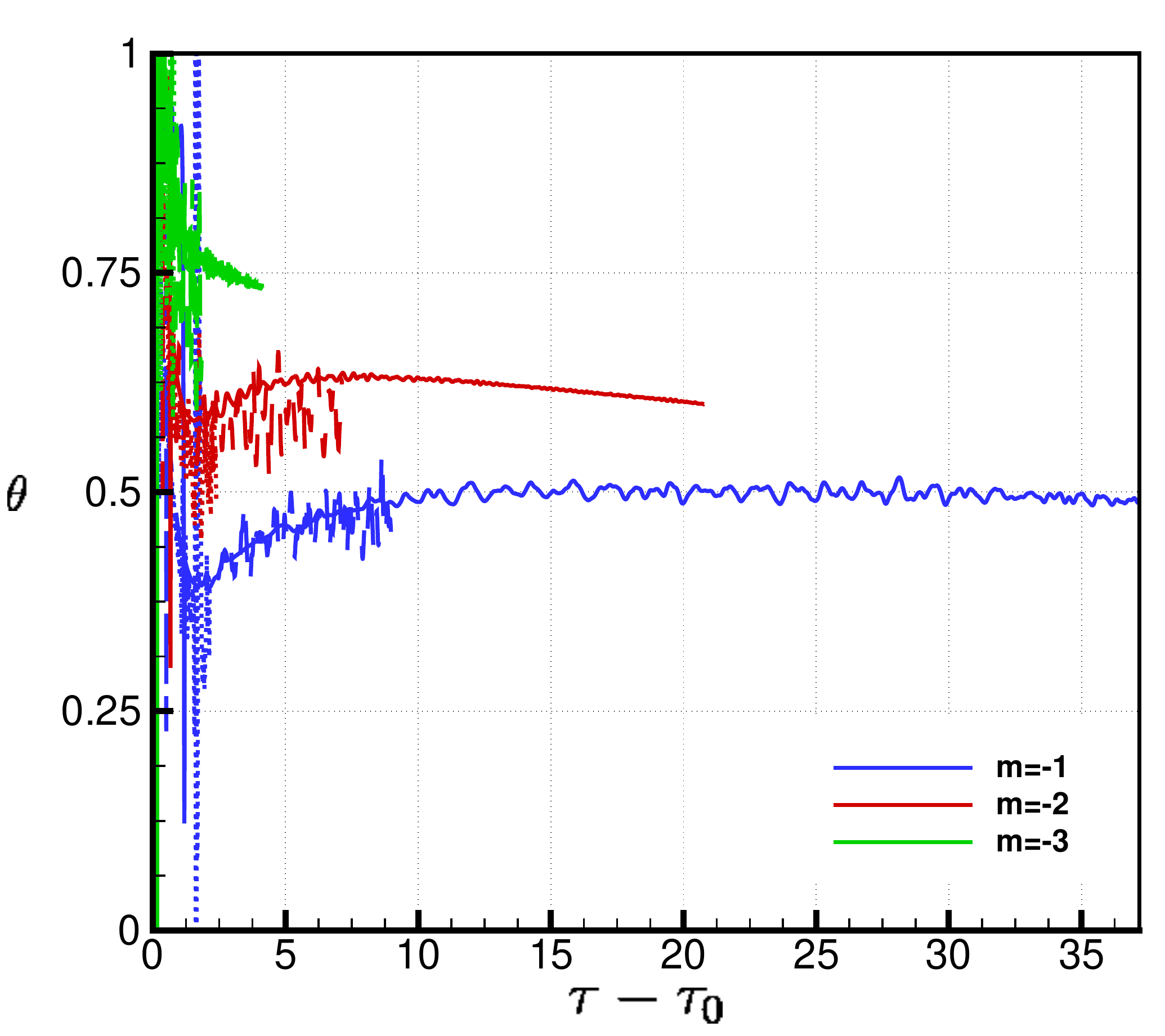}
	\subcaption{Saturation time.}
\end{subfigure}
\begin{subfigure}{0.49\textwidth}
	\includegraphics[width=\textwidth]{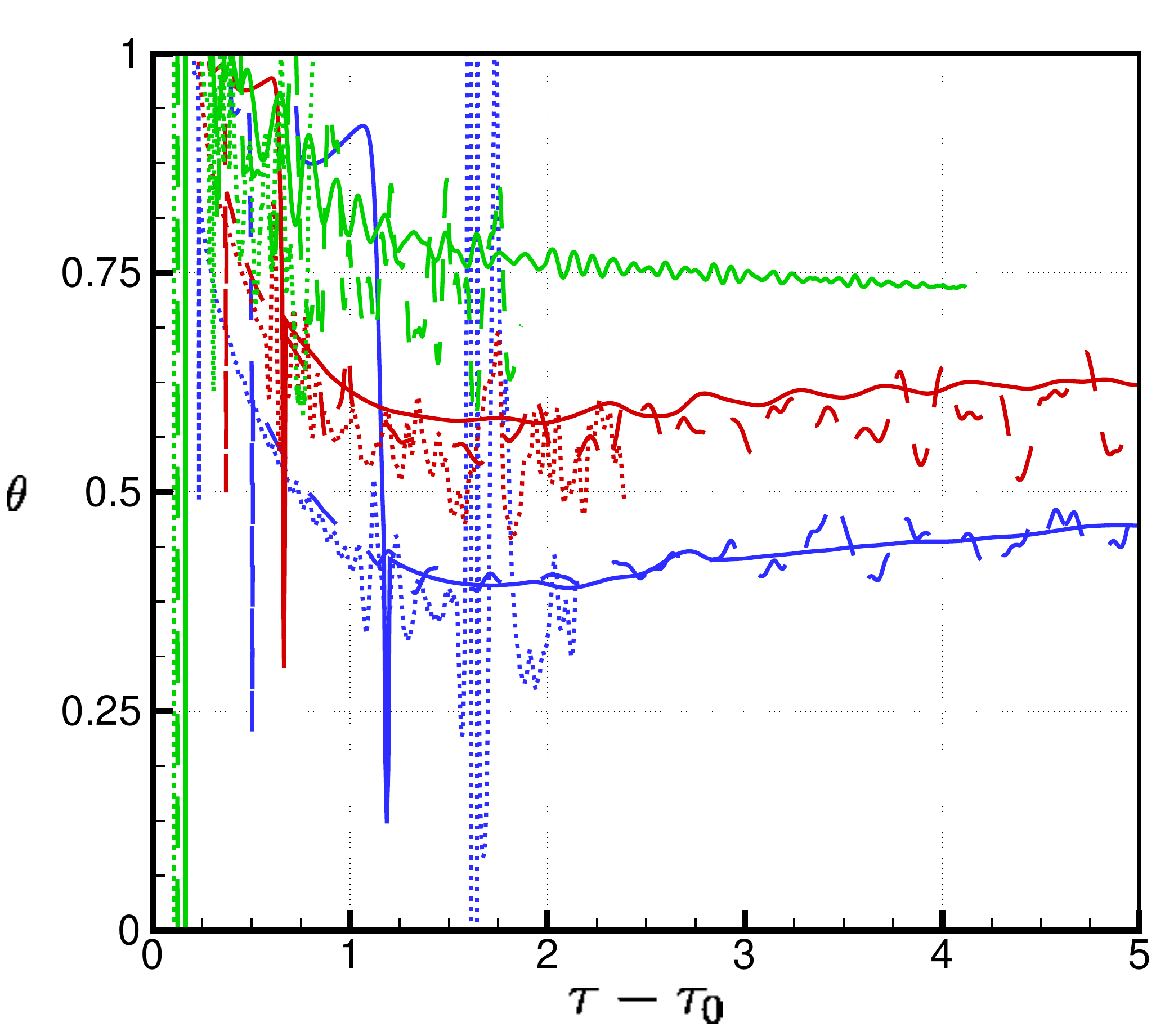}
	\subcaption{Early time.}
\end{subfigure}
	\caption{\label{fig:theta} Instantaneous growth rate exponent $\theta$ estimated from derivatives of $W$. Dotted lines represent the {smallest bandwidth}, dashed lines the medium {bandwidth} and solid lines the {largest bandwidth}.}
\end{figure}

The instantaneous value of $\theta$ can be computed by using a simple buoyancy-drag model (which also follows from self-similarity arguments), written as
\begin{equation}
\ddot{W}=-C_d\frac{\dot{W}^2}{W},
\label{eqn:bd}
\end{equation}
which has the solution $W=W_0(t-t_0)^\theta$ with $\theta=1/(1+C_d)$. This can be used to estimate $\theta$ by calculating the derivatives of $W$ with finite differences, the results of which are shown in Fig. \ref{fig:theta}. There is some noise in the data, mainly due to division by the numerical second derivative, however clear trends are still able to be determined. In the $m=-1$ case, {the $R=64$ bandwidth is sufficient} for the theoretical growth rate of $\theta=1/2$ to be obtained for a significant period of time; taking the average of instantaneous $\theta$ from $\tau=10$ to $\tau=30$ gives $\theta=0.50$. Performing the nonlinear regression over this same interval also yields $\theta=0.50$. For the $m=-2$ case, the {largest bandwidth} simulation comes close to obtaining the theoretical growth rate of $\theta=2/3$. If the average of instantaneous $\theta$ is taken over the interval from $\tau=5$ to $\tau=10$ this gives $\theta=0.63$, which is also the same value obtained from nonlinear regression performed over this interval. For all {bandwidths} in the $m=-3$ case, the instantaneous $\theta$ obtained from the buoyancy-drag model is substantially less than the theoretical value of $\theta=1$. For the {largest bandwidth}, taking the average of instantaneous $\theta$ from $\tau=2$ to $\tau=4$ gives $\theta=0.75$, which again matches the value obtained value obtained from nonlinear regression on this interval. 

\begin{figure}
	\centering
	\begin{subfigure}{0.49\textwidth}
	\includegraphics[width=\textwidth]{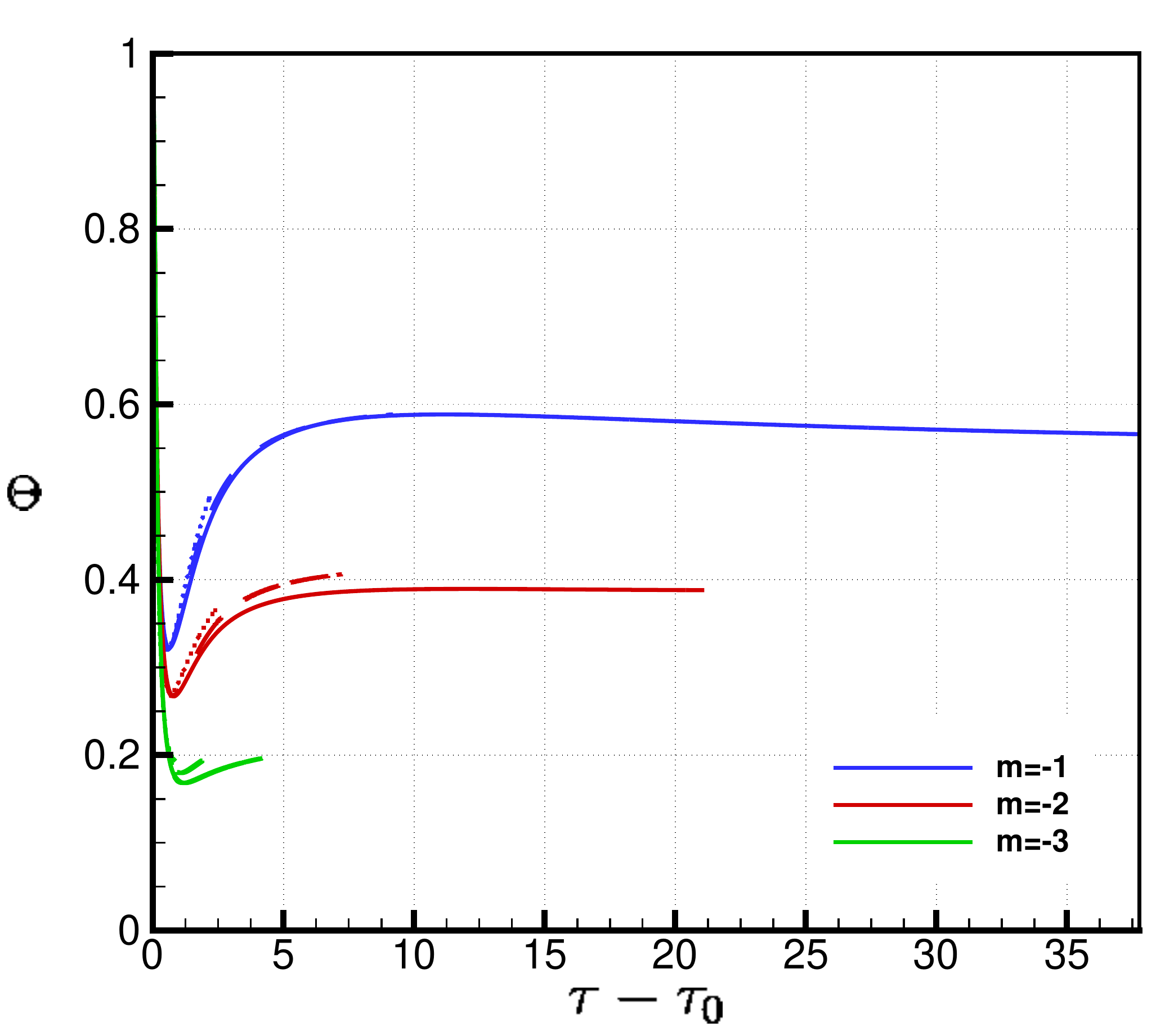}
	\subcaption{Saturation time.}
\end{subfigure}
\begin{subfigure}{0.49\textwidth}
	\includegraphics[width=\textwidth]{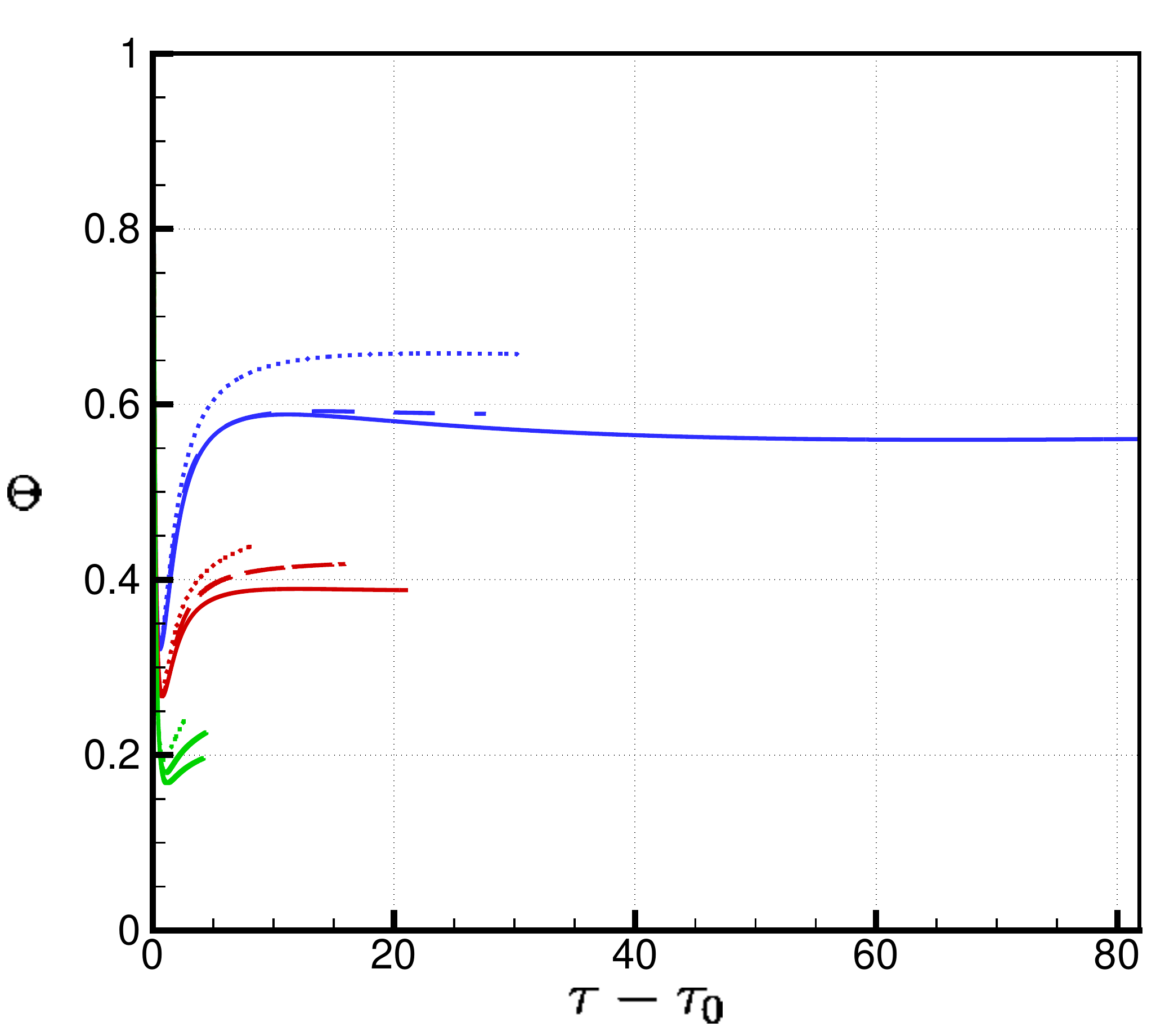}
	\subcaption{Extended time.}
\end{subfigure}
	\caption{\label{fig:Theta-tau} Molecular mixing fraction vs. dimensionless time. Dotted lines represent the {smallest bandwidth}, dashed lines the medium {bandwidth} and solid lines the {largest bandwidth}.}
\end{figure}

Finally, the evolution of the molecular mixing fraction $\Theta$ in time is plotted in Fig. \ref{fig:Theta-tau}. Shown are data for all simulations plotted until the saturation time of the longest wavelength, as well as data for some simulations that were extended beyond this saturation time to explore the late time behaviour. When the data are only plotted up to the saturation time, a good collapse is observed between cases with the same value of $m$. Beyond this point, the results for the {smaller bandwidth} cases begin to depart from those of the {largest bandwidth}, indicating that they are no longer representative of the infinite bandwidth layer. As stated earlier, a constant value of $\Theta$ is one measure of self-similarity of the mixing layer. However, over the interval where the instantaneous growth rate $\theta$ is approximately constant in each of the {largest bandwidth} simulations, $\Theta$ is either slowly decreasing in the $m=-1$ case or slowly increasing in the $m=-2$ and $m=-3$ cases. For the $m=-1$ case, a local maximum in $\Theta$ occurs after the initial (global) minimum, beyond which $\Theta$ gradually decays to an asymptotic value. Similar behaviour was also observed for the narrowband mixing layer in Thornber et al. \cite{Thornber2017}. Given that the {largest bandwidth} $m=-1$ case was shown previously to be growing self-similarly at the theoretically predicted rate over this interval, it is reasonable to assume that the {late time value of $\Theta=0.560$} is close to that which would be obtained in the limit of infinite bandwidth. For the $m=-2$ and $m=-3$ cases, $\Theta=0.388$ and 0.197 respectively, although this is not necessarily indicative of the infinite bandwidth asymptotic value (particularly in the $m=-3$ case). Simulations at higher bandwidths are likely required to determine whether $\Theta$ in the $m=-2$ and $m=-3$ cases behaves similarly to the $m=-1$ case. 

\subsubsection{Self-Similarity}
\label{subsec:similar}
\begin{figure}
	\centering
	\includegraphics[width=0.49\textwidth]{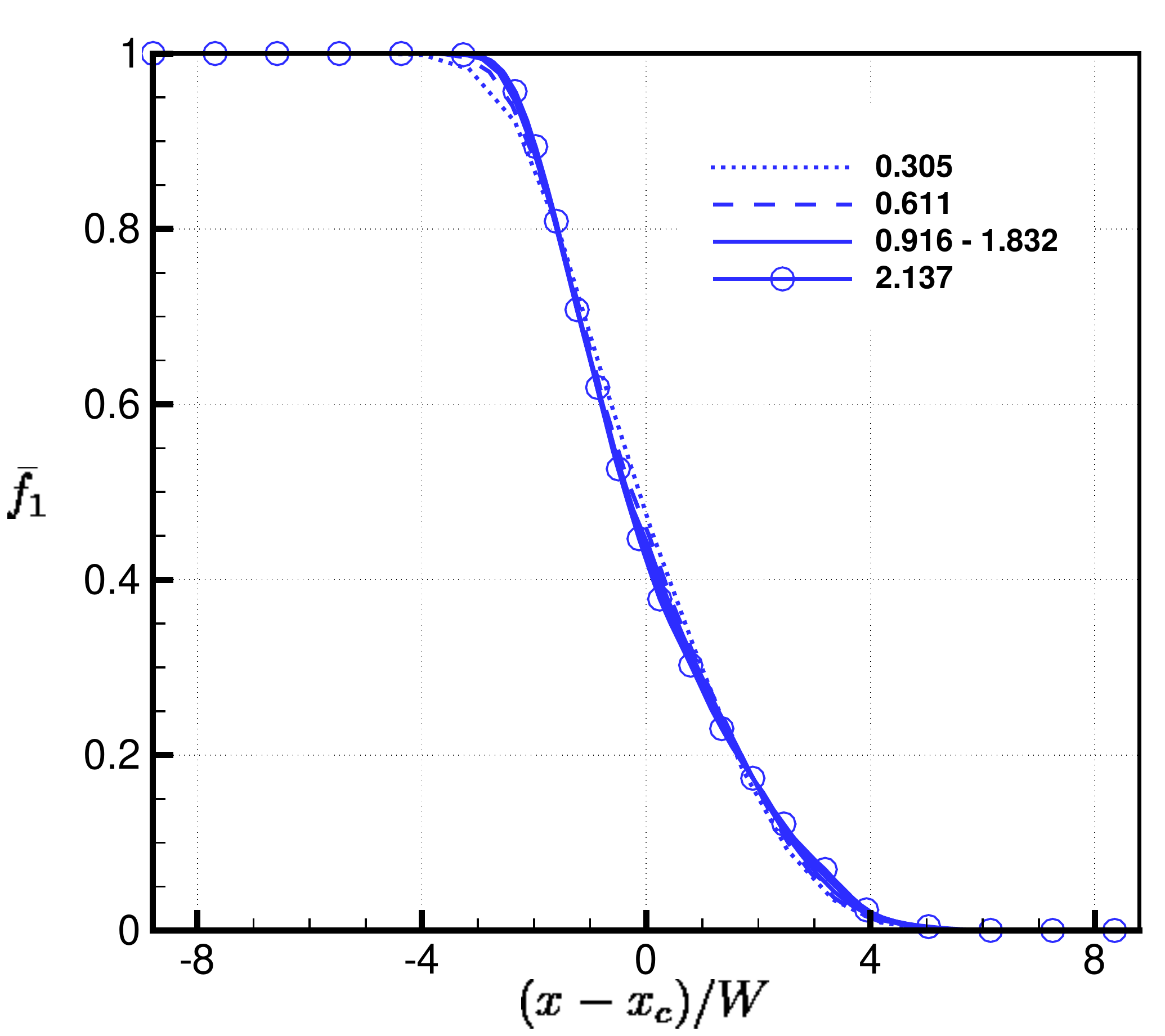}
	\includegraphics[width=0.49\textwidth]{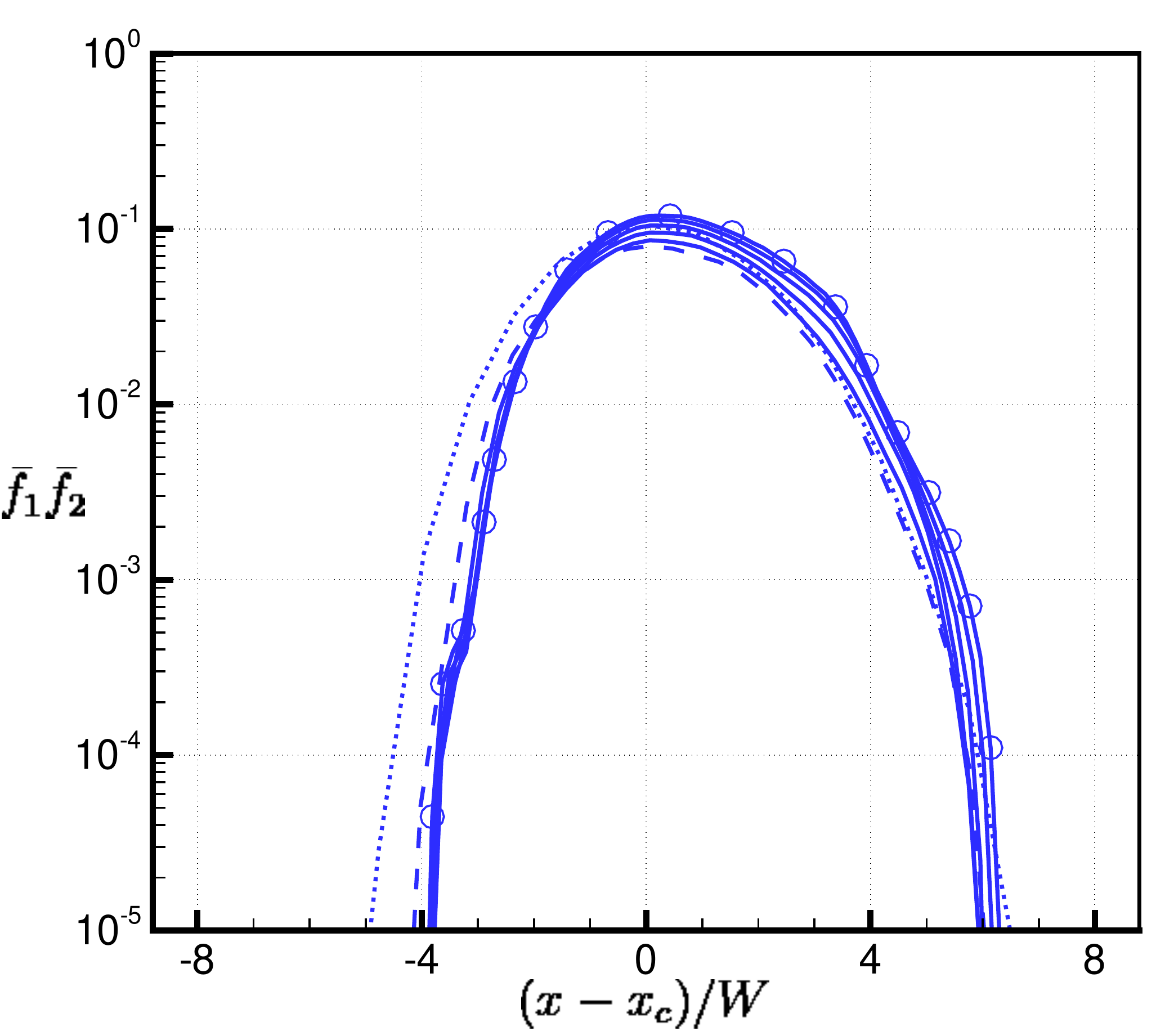}
	\includegraphics[width=0.49\textwidth]{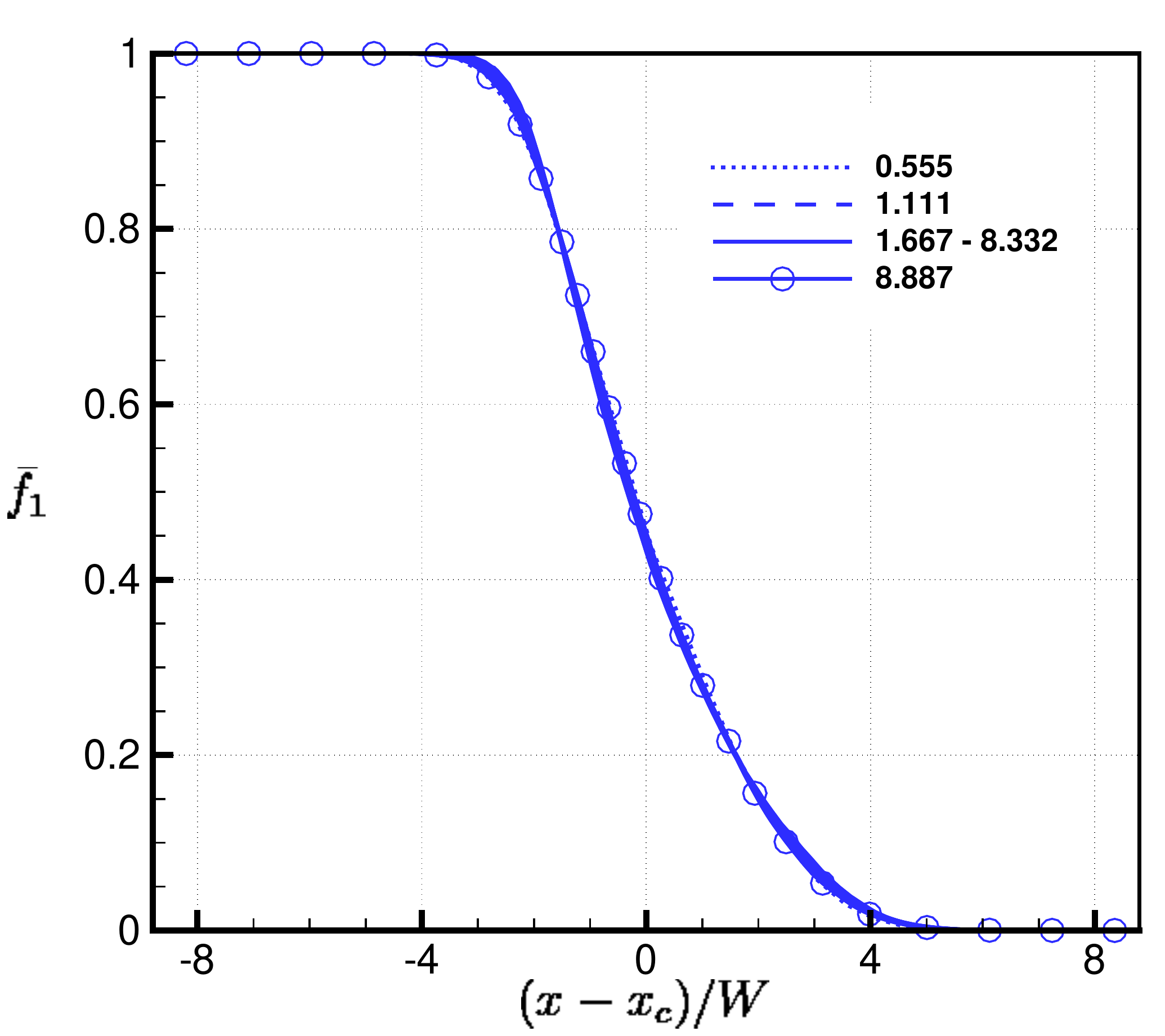}
	\includegraphics[width=0.49\textwidth]{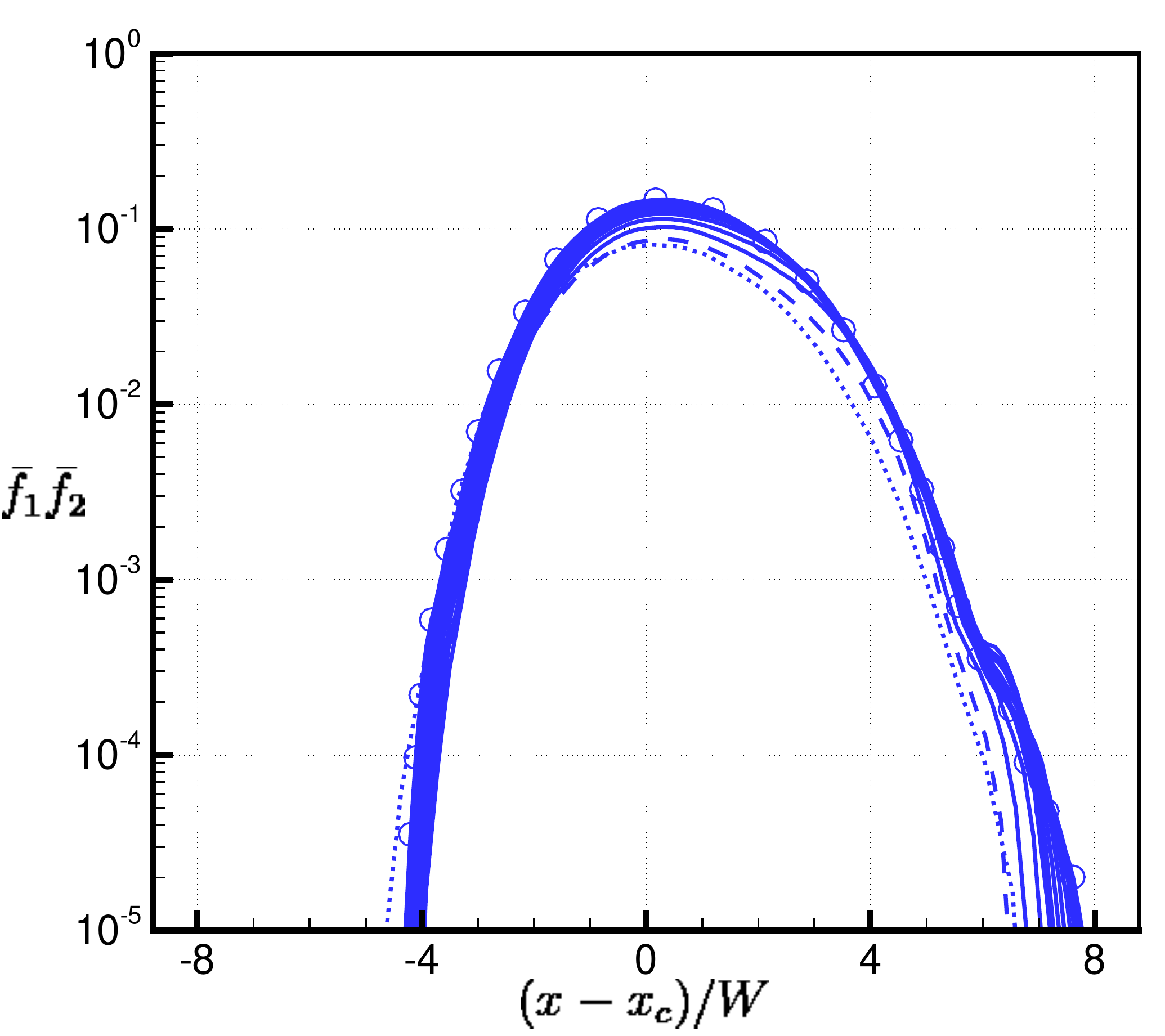}
	\includegraphics[width=0.49\textwidth]{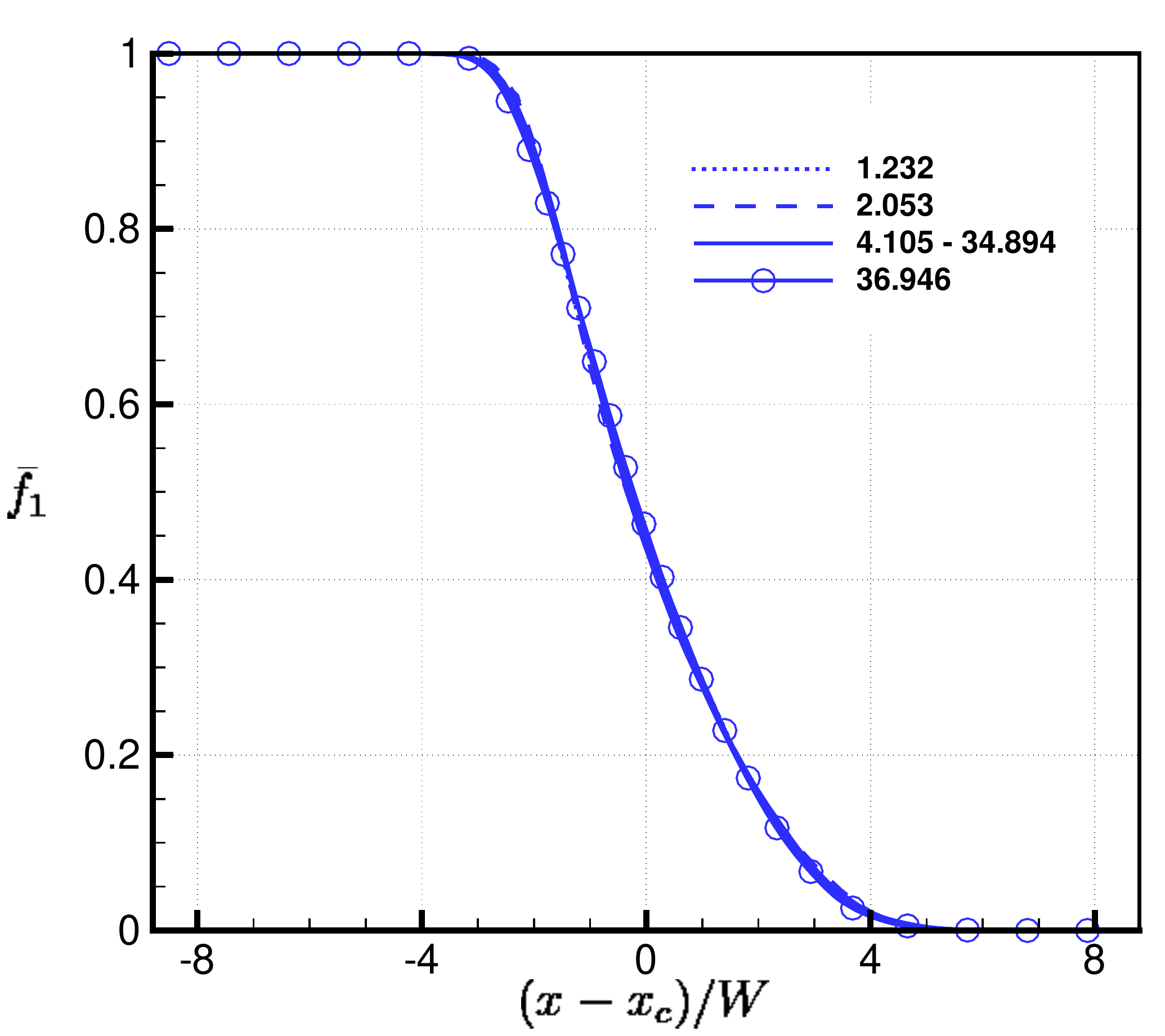}
	\includegraphics[width=0.49\textwidth]{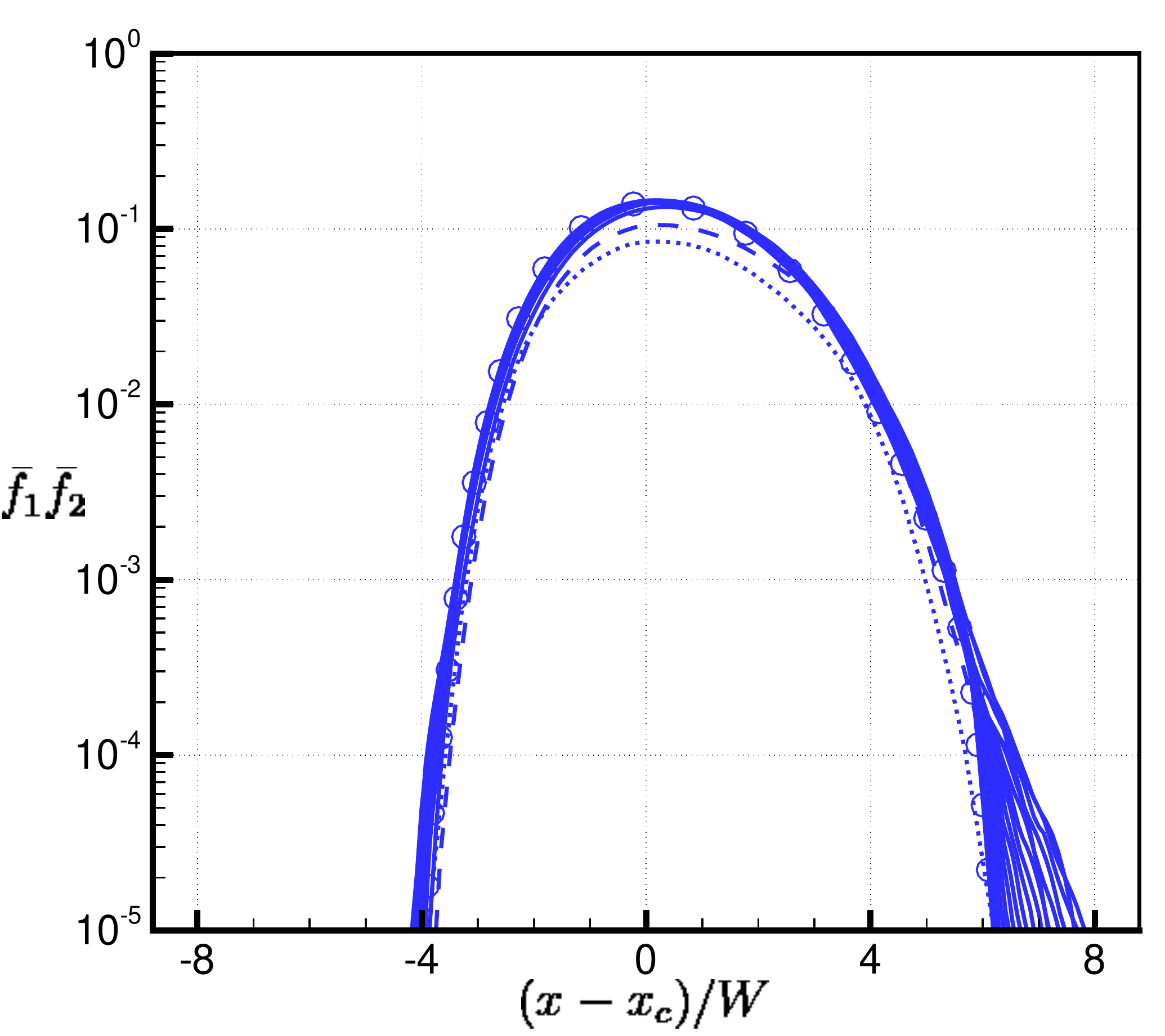}
	\caption{Plane averaged volume fraction profiles for $m=-1$. {Shown are data for $R=16$ (top), $R=32$ (middle) and $R=64$}. Dimensionless times are given in the legend.}
	\label{fig:km1}
\end{figure}

Another method for assessing the degree to which the layer is evolving self-similarly is to plot plane-averaged volume fraction profiles at different points in time, scaled by the integral width. If a perfect collapse is obtained then this shows that the evolution of the mixing layer can be completely described by a single length scale, in this case $W$, and is therefore self-similar. Fig. \ref{fig:km1} gives both the plane-averaged volume fraction $\langle f_1\rangle$ (denoted by $\bar{f_1}$ in the figures) as well as the product $\langle f_1\rangle\langle 1-f_1\rangle$ to highlight the variation at the extremes of the mixing layer for the $m=-1$ case. A good collapse is observed across all of the later times considered (solid lines), particularly for the {largest bandwidth}, showing that the layer is evolving self-similarly and can be appropriately scaled by $W$. There is a {narrowing} of the $\langle f_1\rangle\langle 1-f_1\rangle$ profile in time at the fringes of the spike side in the {larger bandwidth cases, which suggests that the integral width is becoming increasingly dominated by mixing in the core of the layer}. To give some more context to the figures, the 1\% bubble and spike heights $H_b$ and $H_s$ can be calculated, in a similar manner to the visual width, as 
\begin{subequations}
\begin{align}
H_b & = x_c - x(\langle f_1 \rangle =0.99), \\
H_s & = x(\langle f_1 \rangle =0.01) - x_c.
\end{align}
\end{subequations}
Based on these definitions, the ratios $H_b/W$ and $H_s/W$ are found to be 3.0 and 4.3 on average for the finest grid simulation. Therefore the departure from self-similarity observed on the spike side is occurring at the absolute fringes of the layer ($x/W\approx 6$), at a much greater distance from the layer centre than the 1\% spike height. 

\begin{figure}
	\centering
	\includegraphics[width=0.49\textwidth]{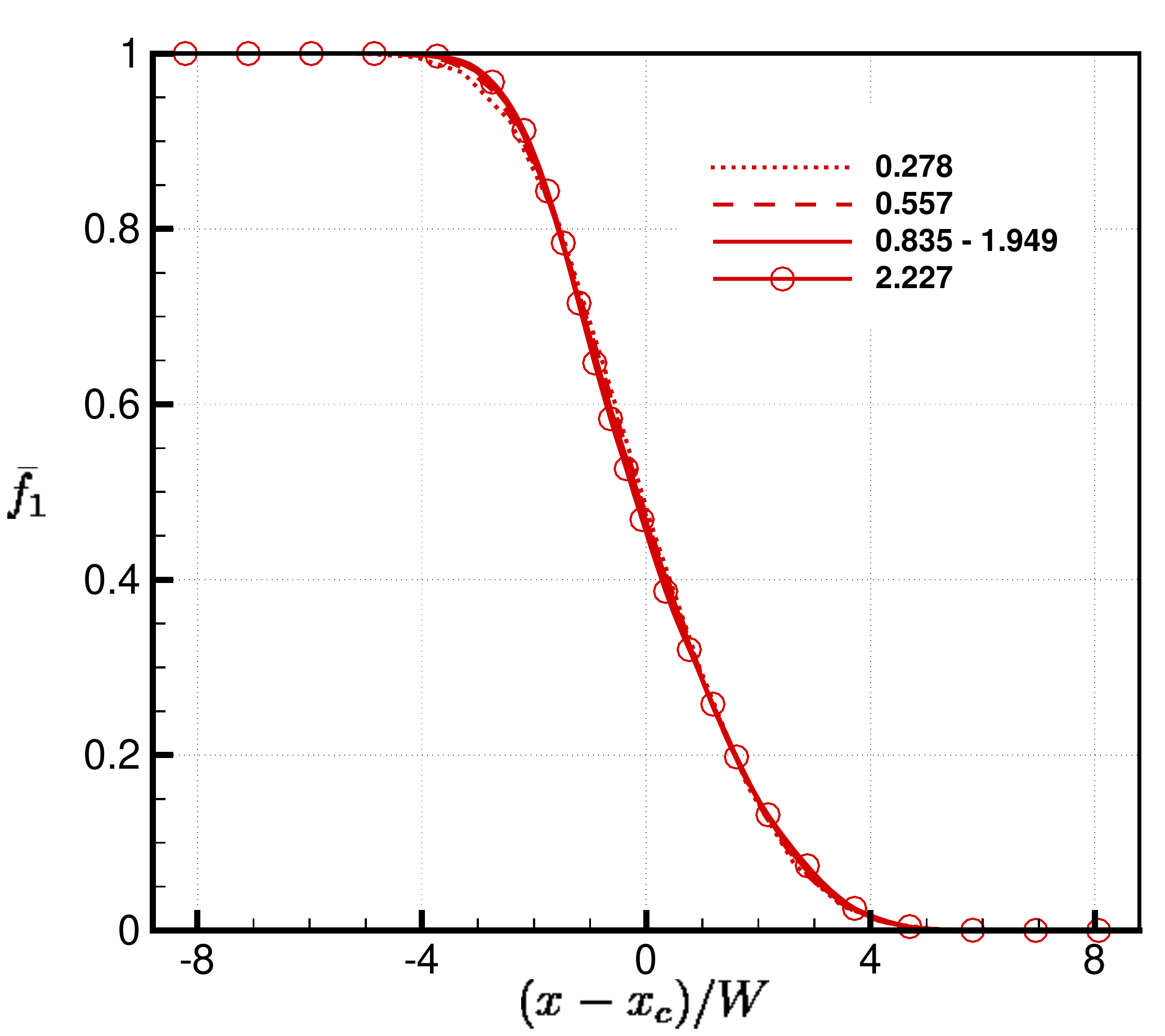}
	\includegraphics[width=0.49\textwidth]{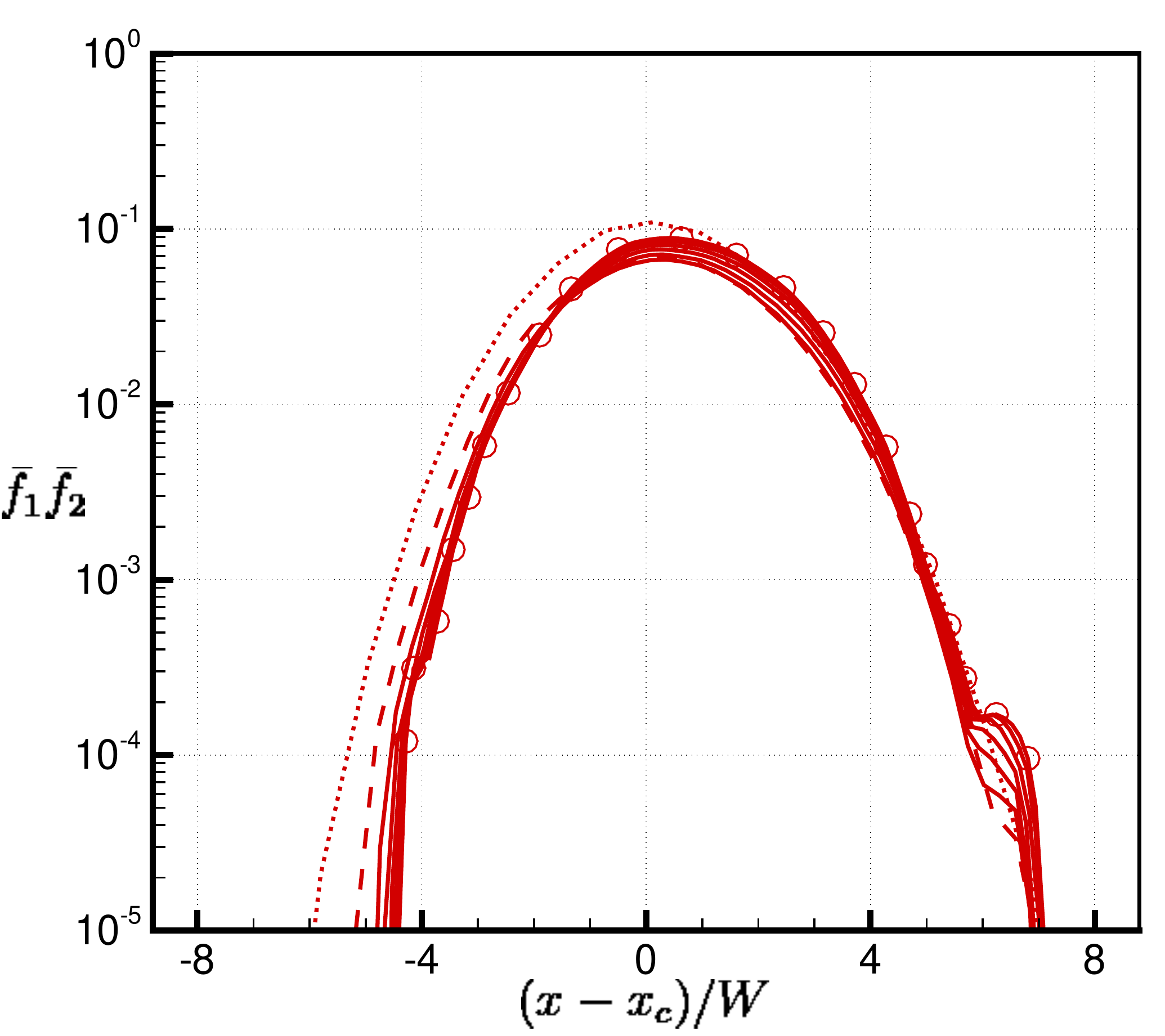}
	\includegraphics[width=0.49\textwidth]{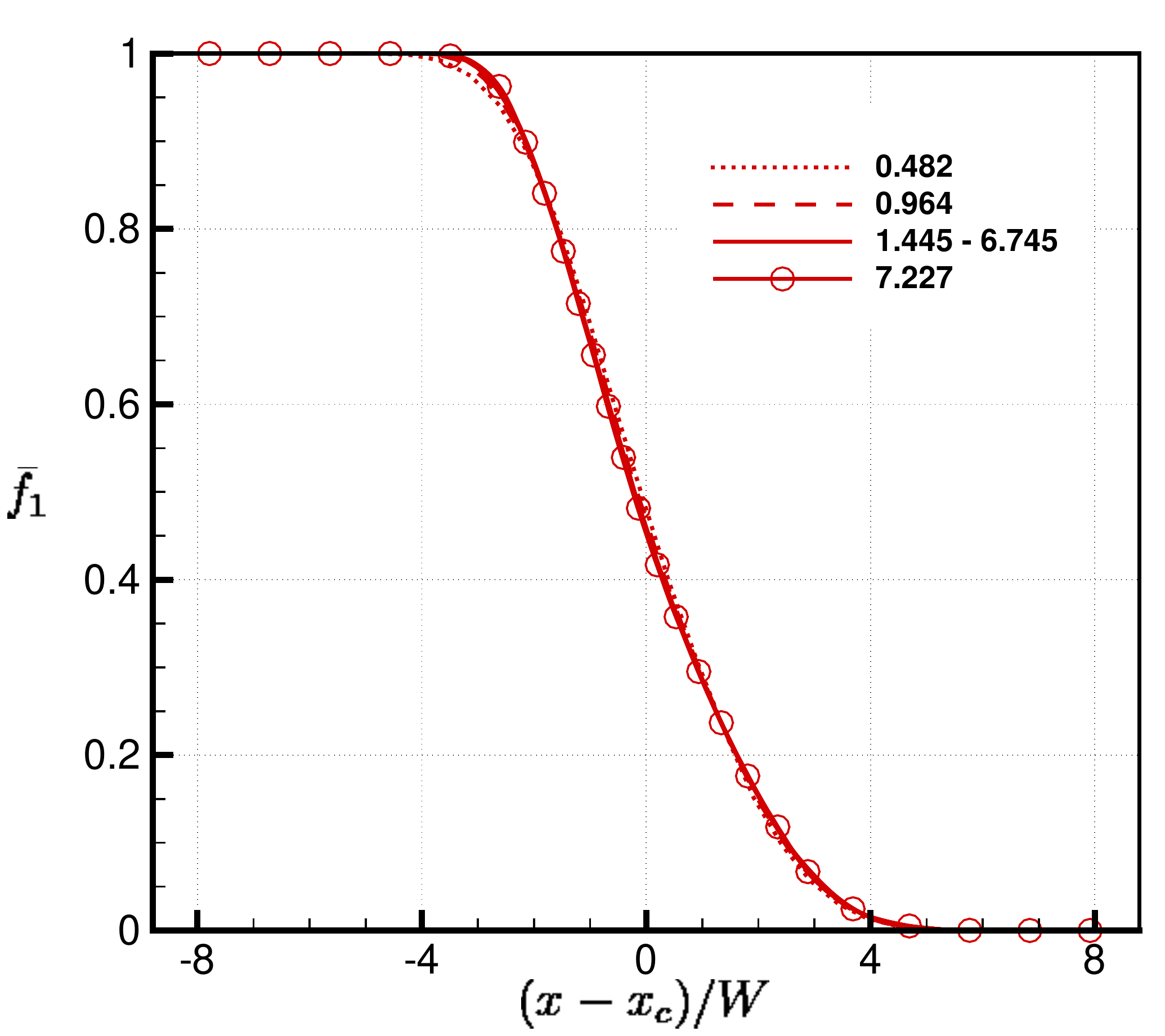}
	\includegraphics[width=0.49\textwidth]{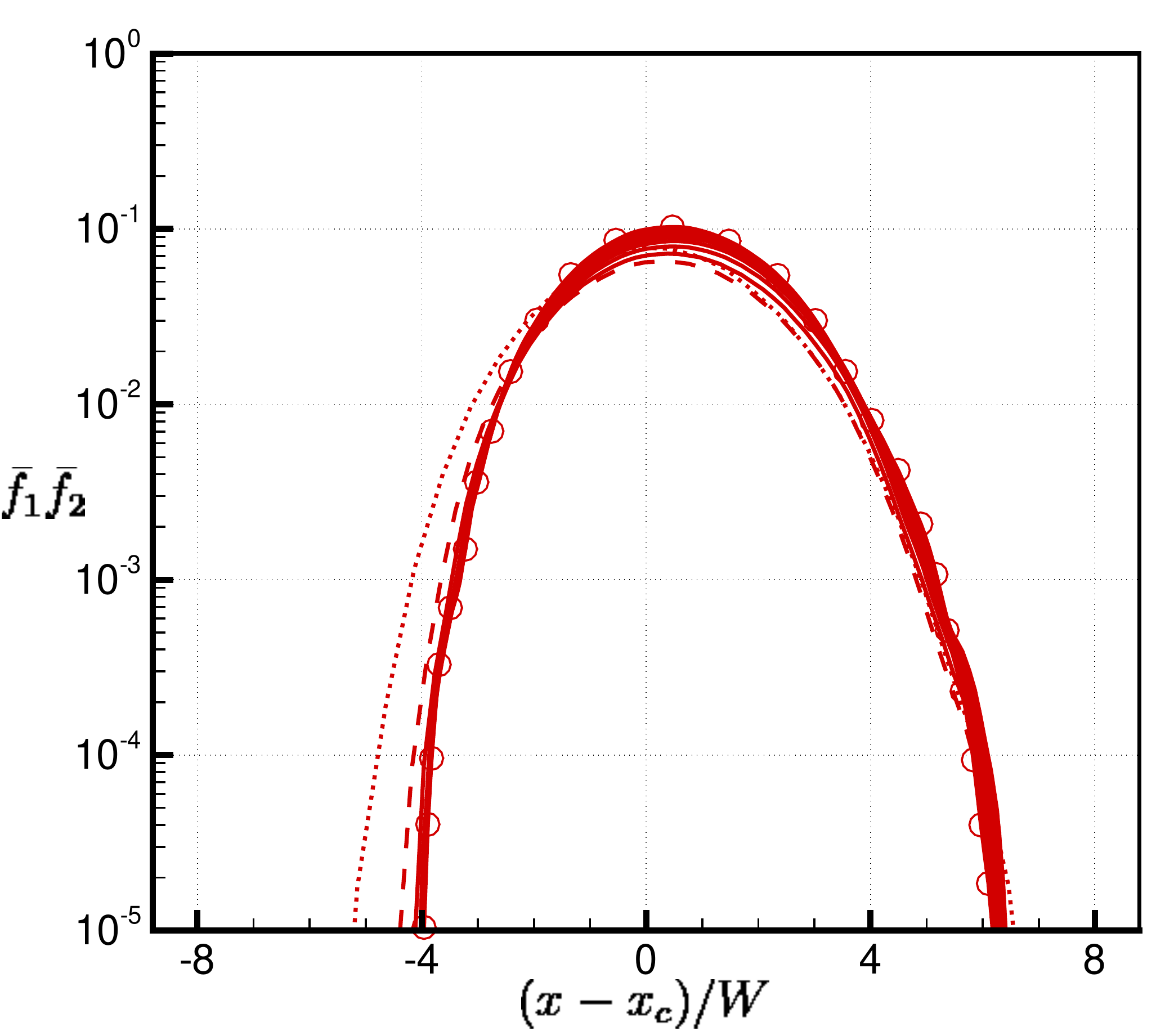}
	\includegraphics[width=0.49\textwidth]{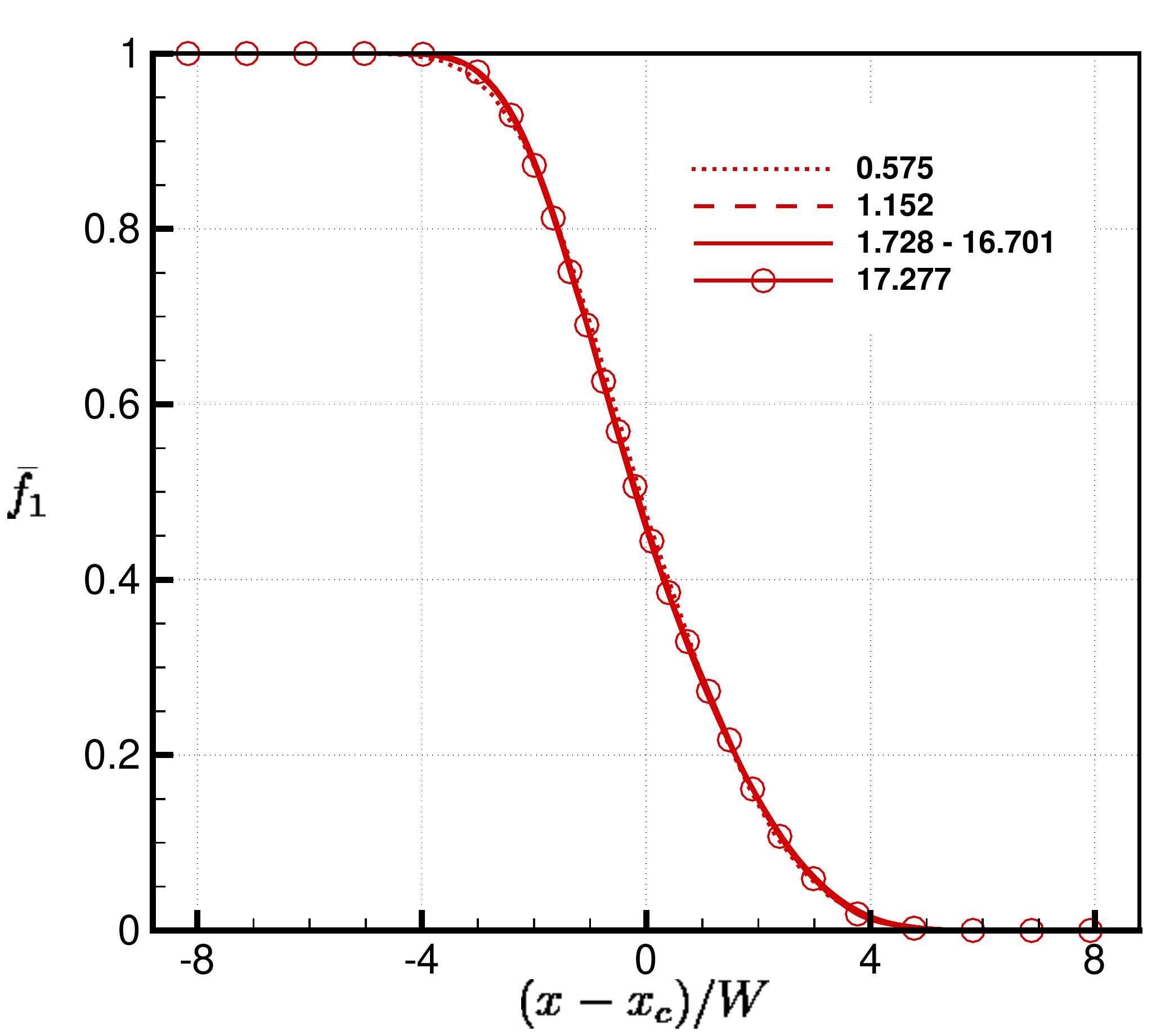}
	\includegraphics[width=0.49\textwidth]{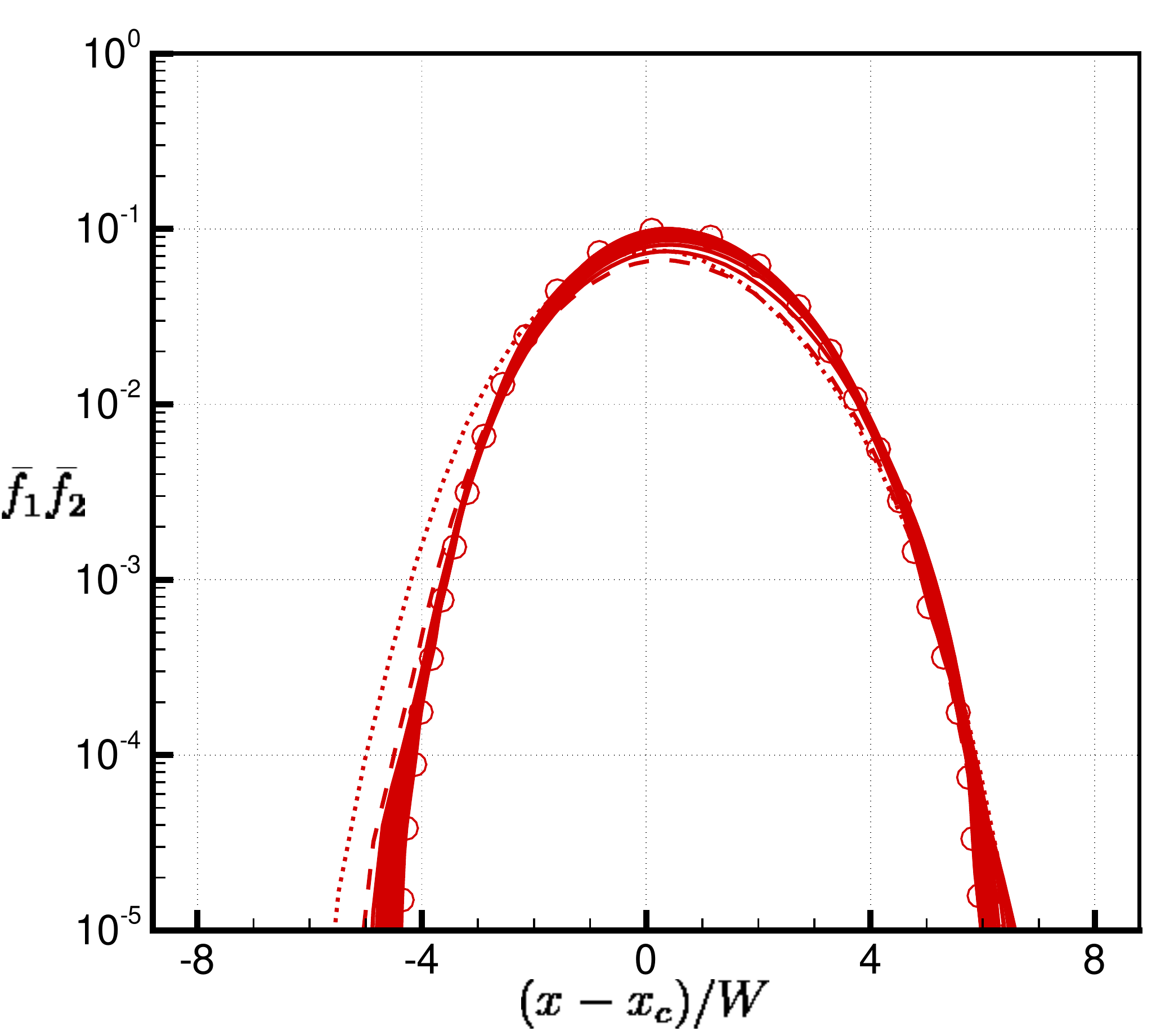}
	\caption{Plane averaged volume fraction profiles for $m=-2$. {Shown are data for $R=16$ (top), $R=32$ (middle) and $R=64$}. Dimensionless times are given in the legend.}
	\label{fig:km2}
\end{figure}

\begin{figure}
	\centering
	\includegraphics[width=0.49\textwidth]{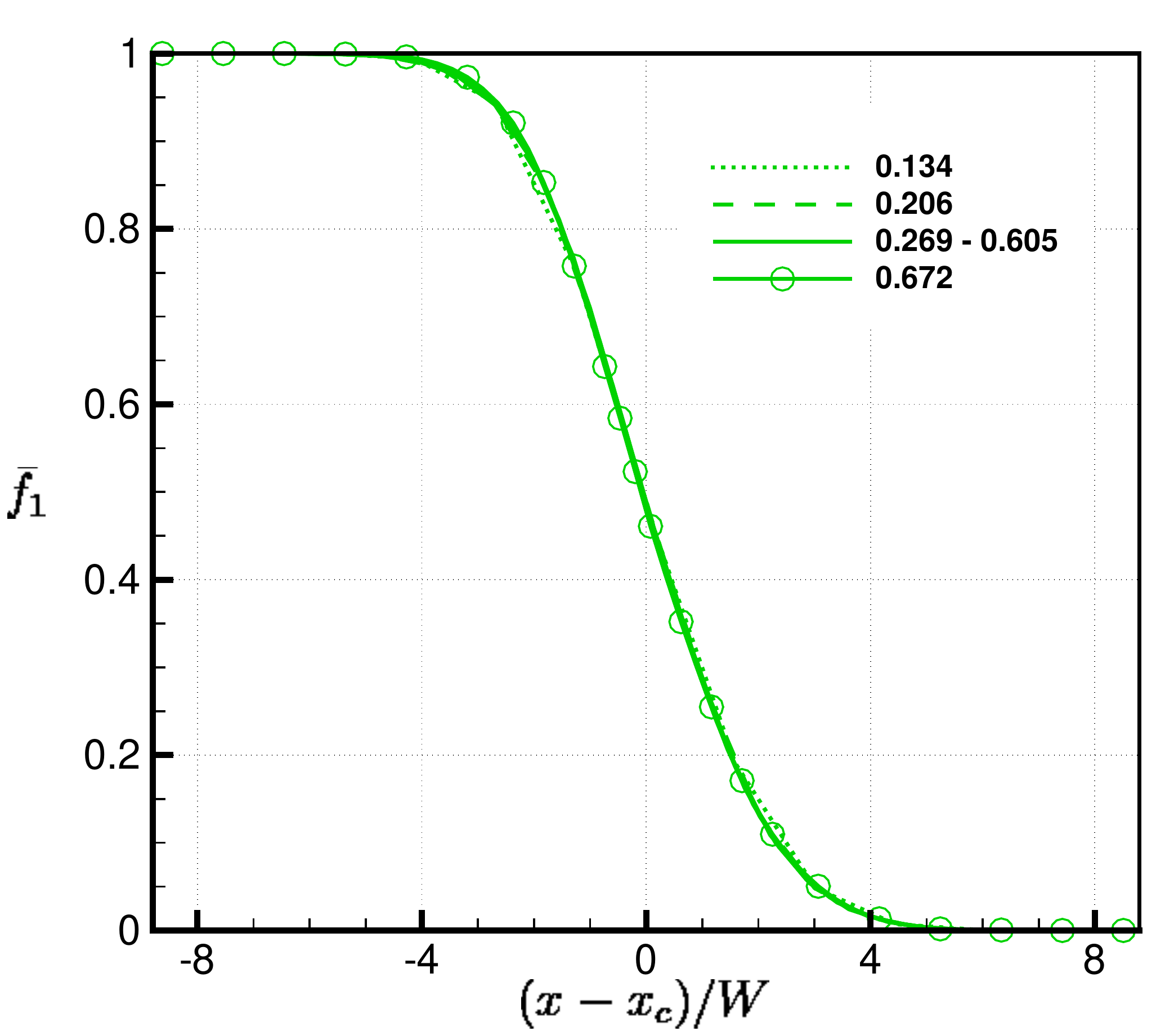}
	\includegraphics[width=0.49\textwidth]{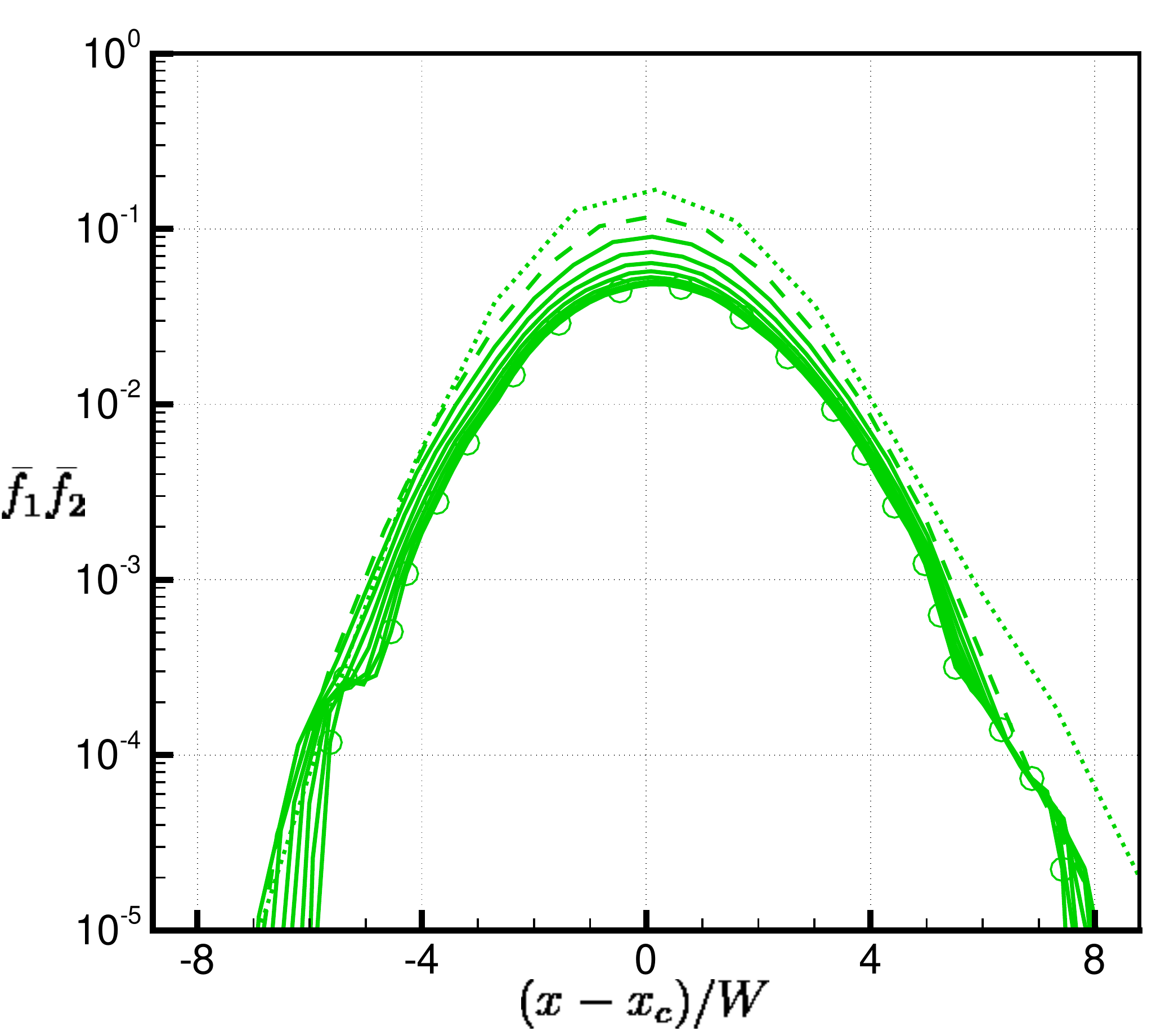}
	\includegraphics[width=0.49\textwidth]{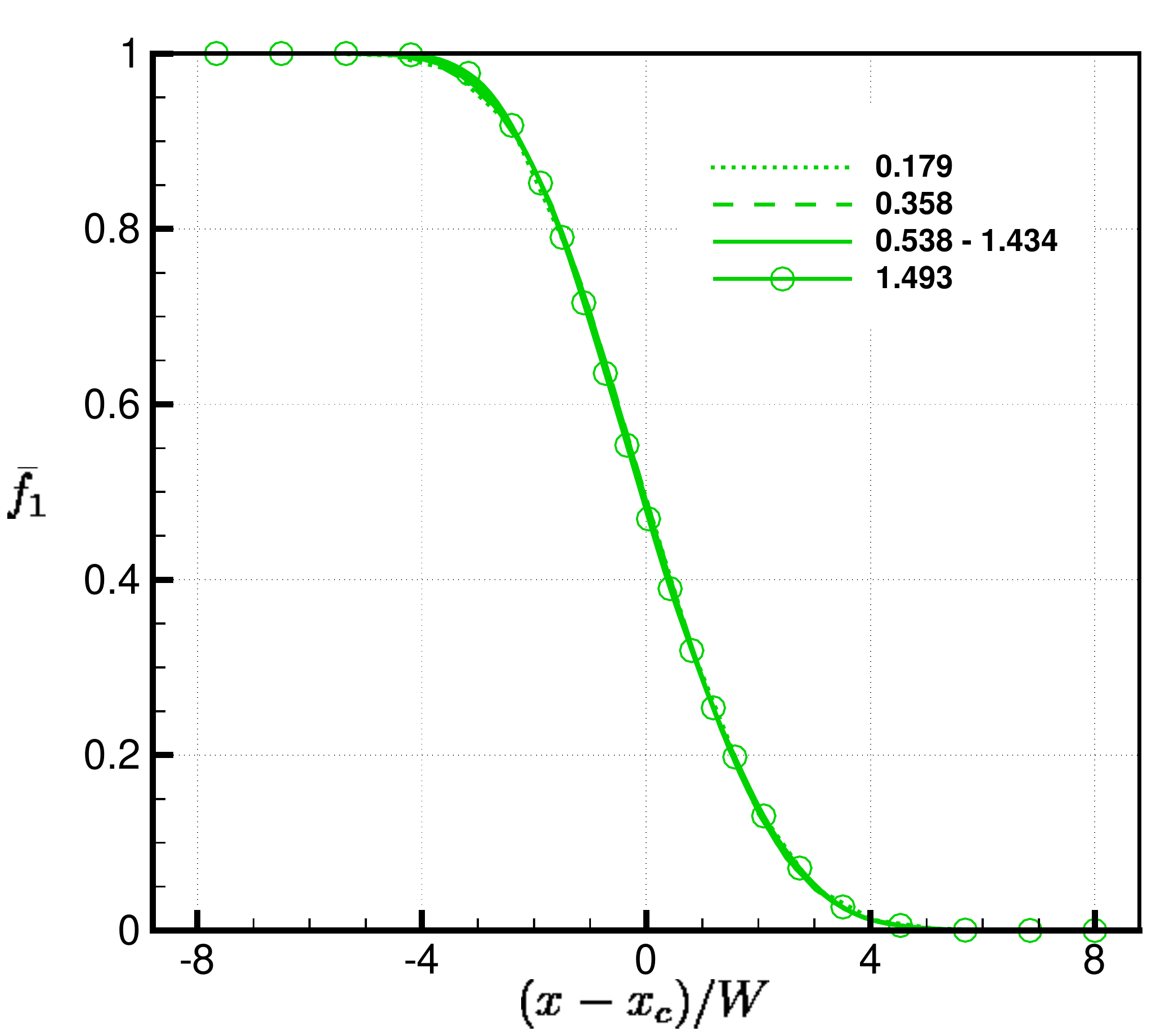}
	\includegraphics[width=0.49\textwidth]{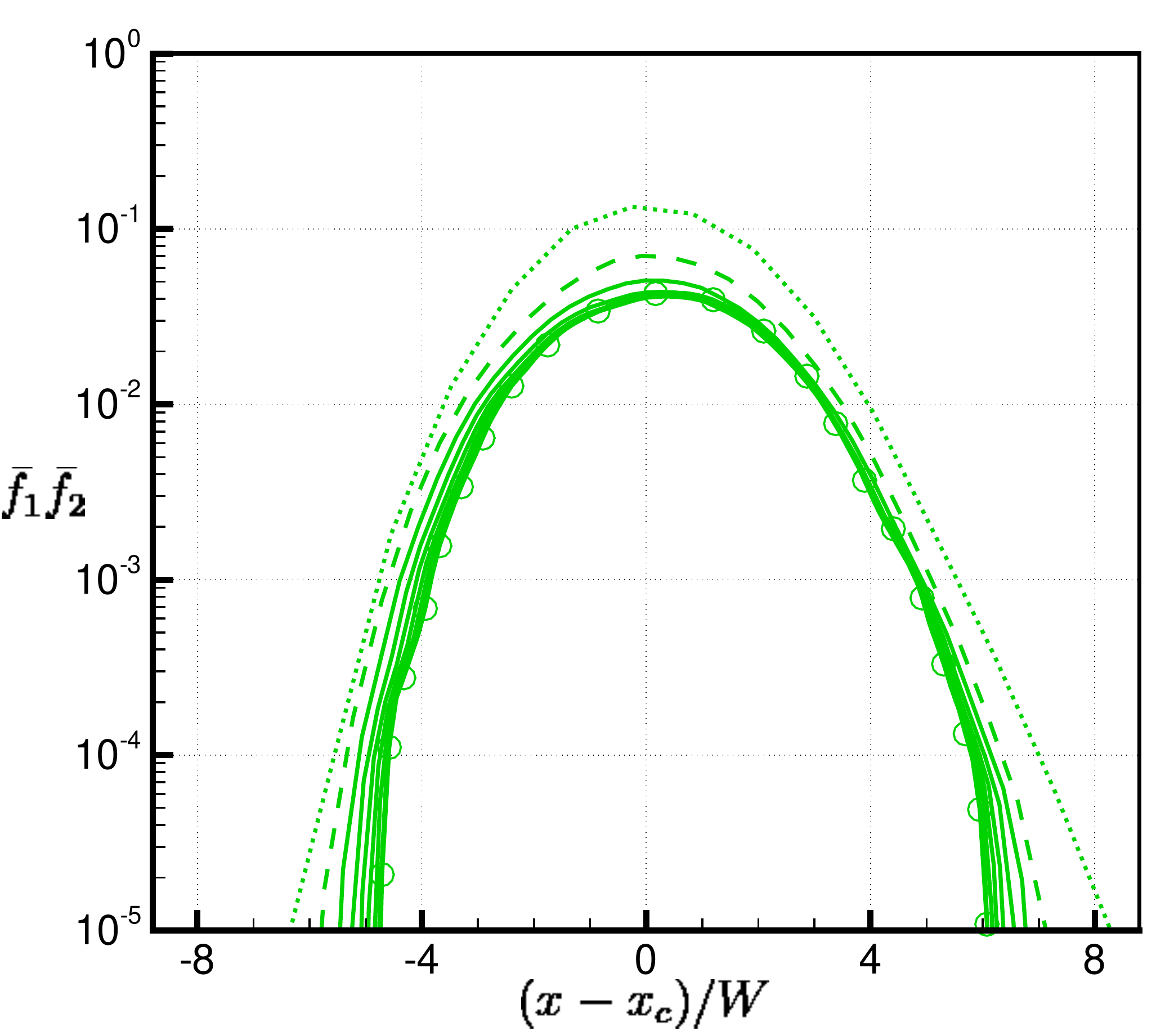}
	\includegraphics[width=0.49\textwidth]{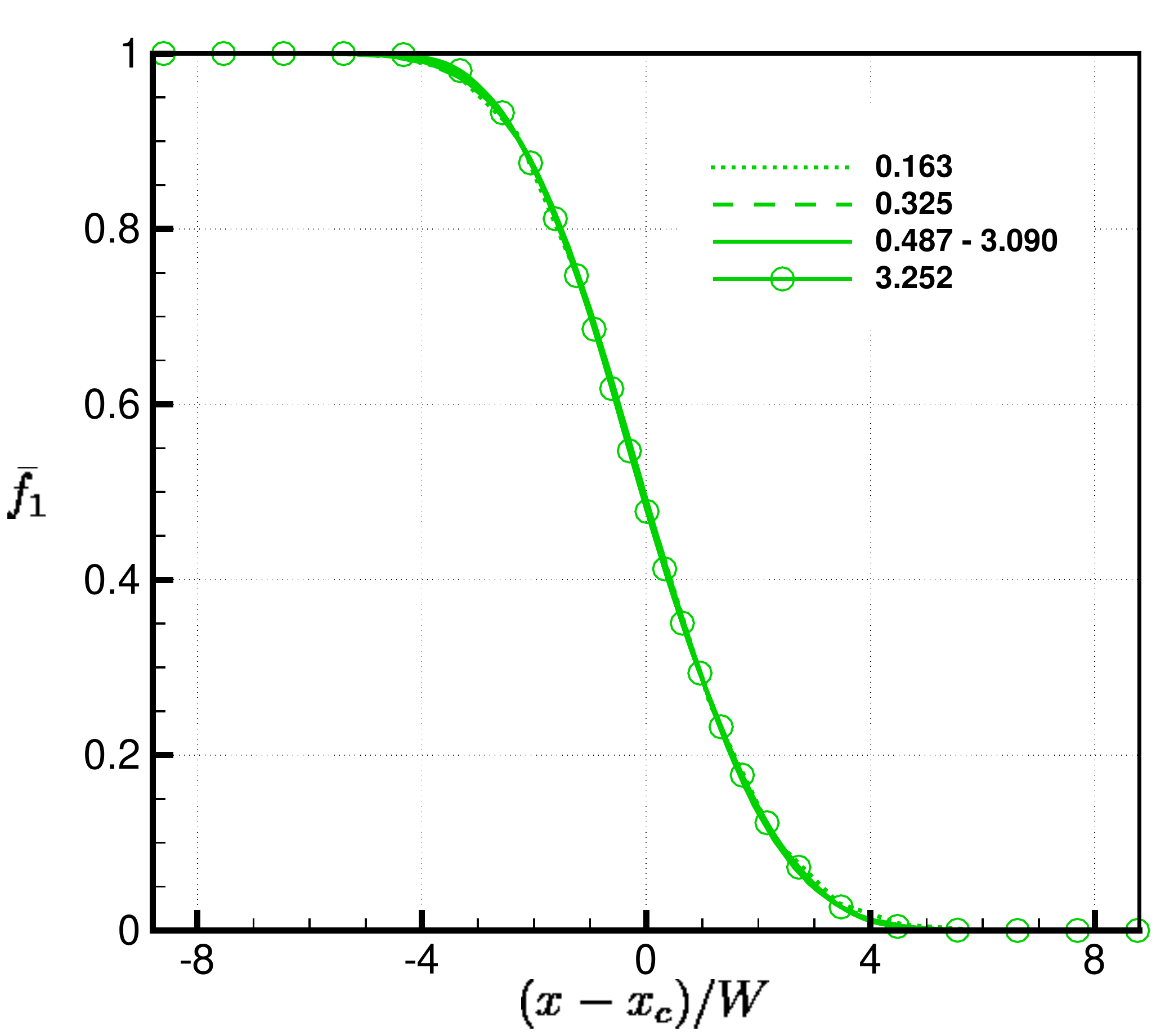}
	\includegraphics[width=0.49\textwidth]{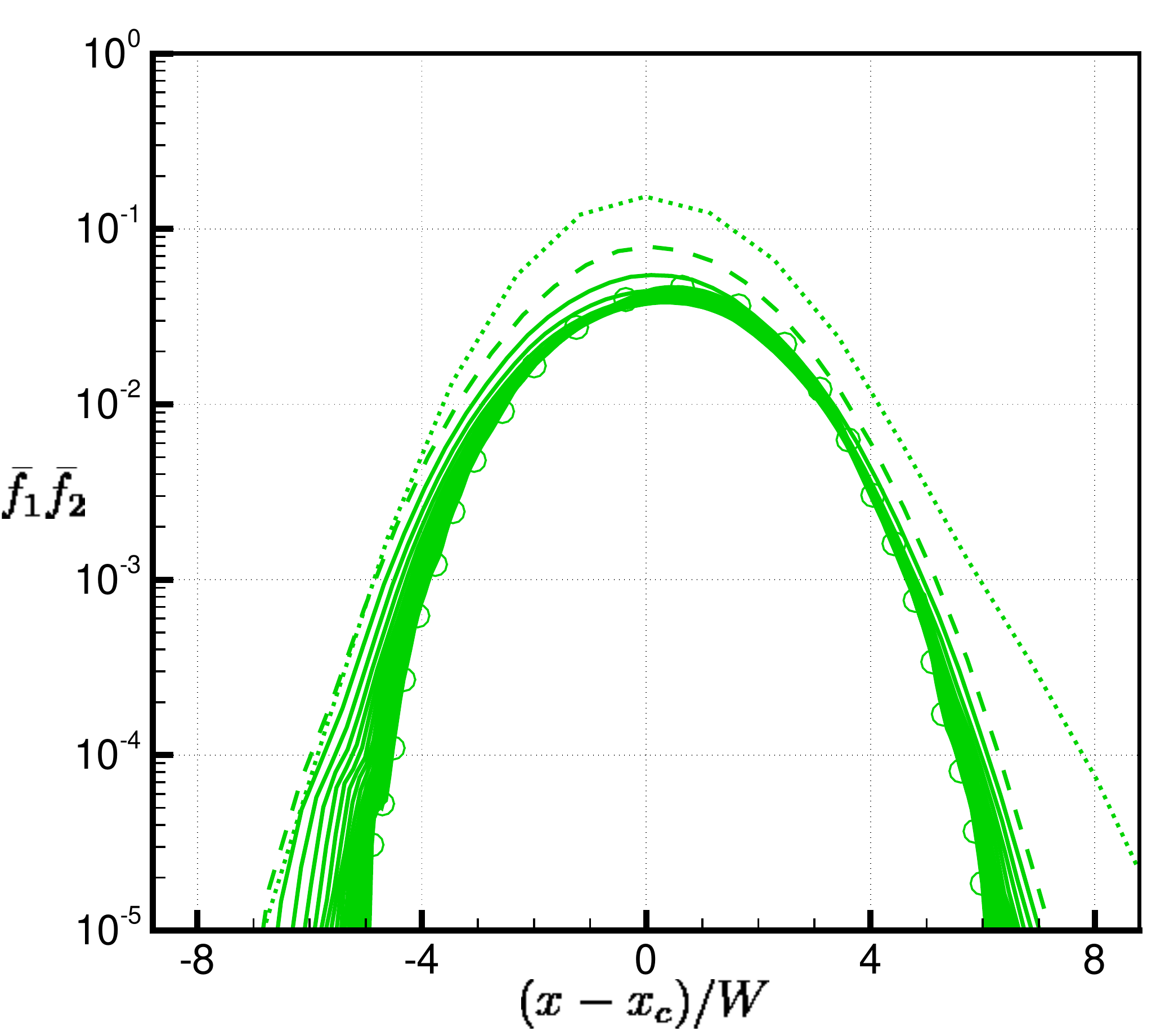}
	\caption{Plane averaged volume fraction profiles for $m=-3$. {Shown are data for $R=16$ (top), $R=32$ (middle) and $R=64$}. Dimensionless times are given in the legend.}
	\label{fig:km3}
\end{figure}

The same profiles are shown for the $m=-2$ case in Fig. \ref{fig:km2} and for the $m=-3$ case in Fig. \ref{fig:km3}. For the medium and {large bandwidths} in the $m=-2$ case, the collapse of the data when scaled by $W$ {is the best of all the cases}, even at the fringes of the layer. For the {largest bandwidth} the ratios $H_b/W$ and $H_s/W$ are 3.3 and 4.2 respectively. For the $m=-3$ case the overall collapse in the data is less good, {most notably on the bubble side}, although still acceptable when considering that the scaled bubble and spike heights are $H_b/W=3.7$ and $H_s/W=4.1$ based on the {largest bandwidth case}. Again, the fact that the $\langle f_1\rangle\langle 1-f_1\rangle$ profiles are {narrowing} in time {suggests that the degree to which mix at the boundaries of the layer influences the integral width is decreasing}. Another notable trend is that the scaled bubble heights increase with decreasing $m$ while the scaled spike heights decrease. This trend is also observed in the width of the $\langle f_1\rangle\langle 1-f_1\rangle$ profiles. 

\begin{figure}
	\centering
	\begin{subfigure}{0.49\textwidth}
		\includegraphics[width=\textwidth]{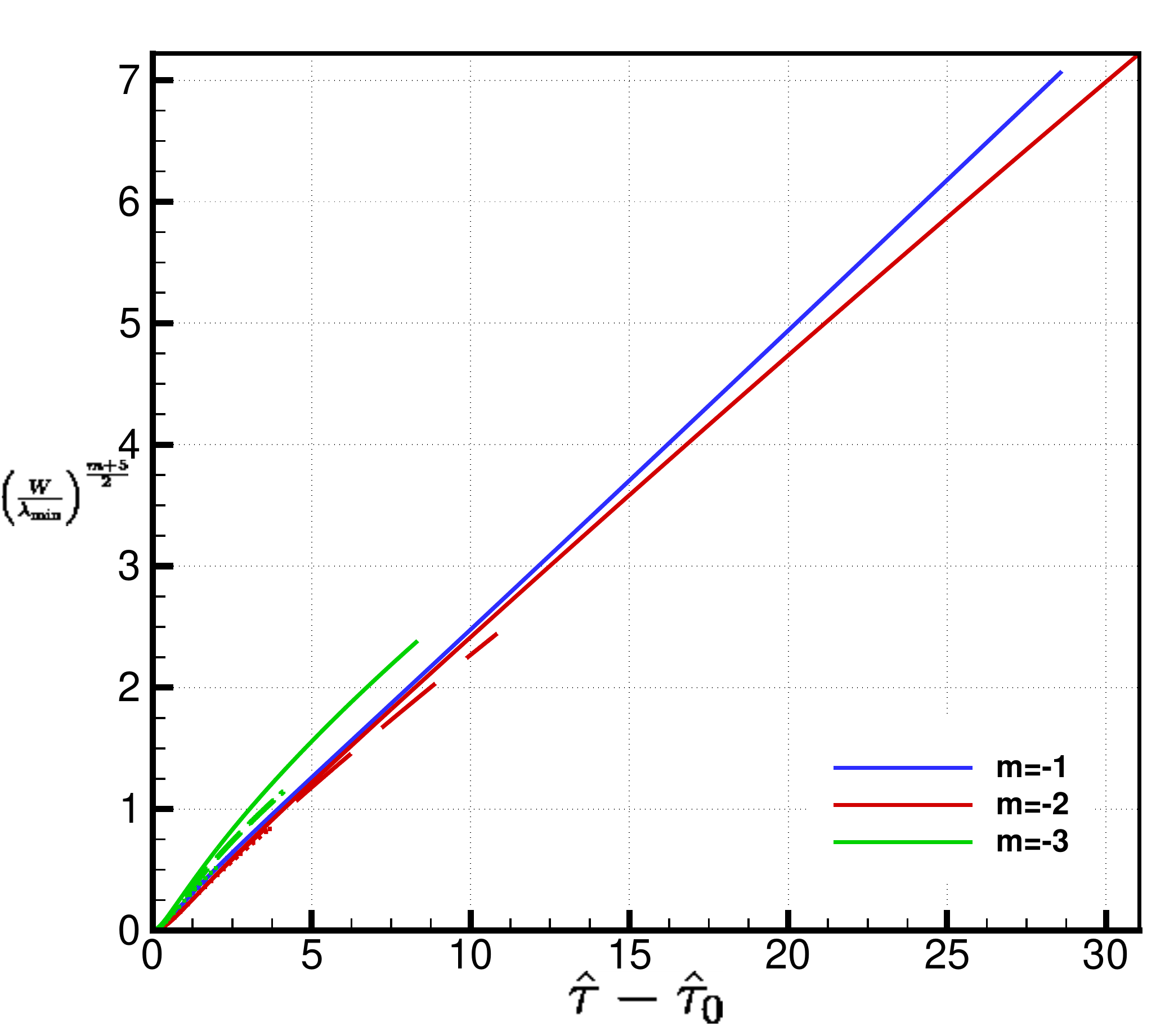}
		\subcaption{Saturation time.}
	\end{subfigure}
	\begin{subfigure}{0.49\textwidth}
		\includegraphics[width=\textwidth]{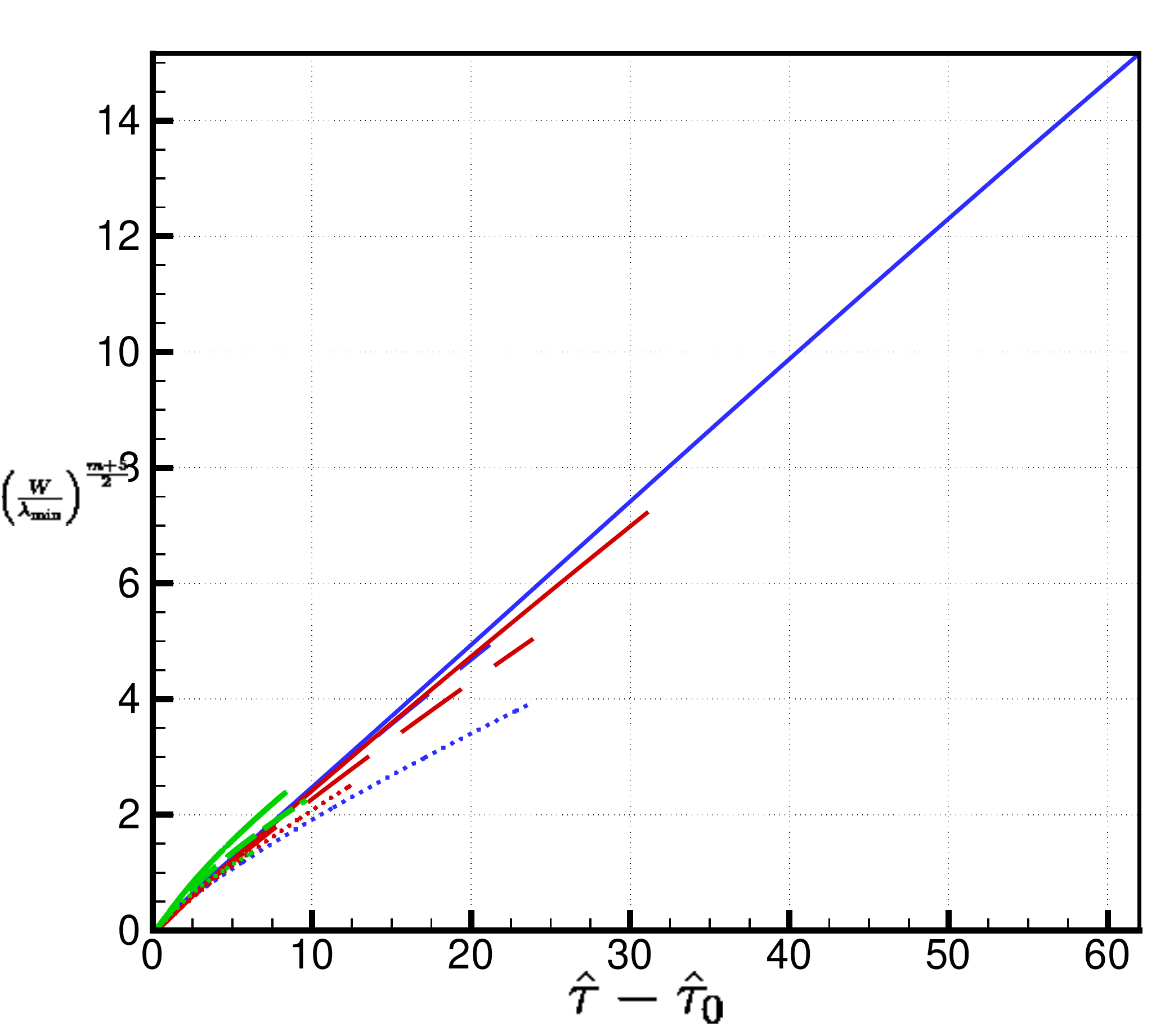}
		\subcaption{Extended time.}
	\end{subfigure}
	\caption{\label{fig:W-tau-late} Alternative non-dimensionalisation assuming self-similar growth. Dotted lines represent the {smallest bandwidth}, dashed lines the medium {bandwidth} and solid lines the {largest bandwidth}.}
\end{figure}

Based on the just-saturated mode analysis given in Sec. \ref{subsec:simulations}, as well as the observations of self-similarity made in this section, {another} non-dimensionalisation is proposed with the aim of collapsing the data at late time across different values of $m$. Under the assumption that the layer is growing self-similarly and is dominated by linear growth, just-saturated mode analysis yields the relation given in Eqn. \ref{eqn:just-saturated}. Non-dimensionalising $W$ by $\lambda_{min}$ gives 
\begin{equation}
\frac{W}{\lambda_{min}}\propto\left(\frac{B\sqrt{C}A^+\Delta u t}{\lambda_{min}^\frac{m+5}{2}}\right)^{\frac{2}{m+5}}=\hat{\tau}^{\frac{2}{m+5}}.
\label{eqn:nondim2}
\end{equation}
Therefore plotting $(W/\lambda_{min})^{(m+5)/2}$ vs. $\hat{\tau}$ should yield a linear relationship, provided $W$ is growing at the theoretical rate of $\theta=2/(m+5)$. This is shown in Fig. \ref{fig:W-tau-late}, with the data plotted up until saturation time, as well as until the very end of all simulations to explore the late time behaviour. As expected, since the {$R=64$}, $m=-1$ case obtains the theoretical growth rate, a linear relationship is obtained for this case over the majority of the simulation time. In all other cases, departure from the theoretical growth rate is reflected in Fig. \ref{fig:W-tau-late} as departure from a linear relationship. The {$R=64$}, $m=-1$ case can also be used to estimate the constant of proportionality for Eqn. \ref{eqn:nondim2}. Performing linear regression over the interval for which $\theta=0.5$, i.e. from $\tau=10$ to $\tau=30$ (or equivalently $\hat{\tau}=7.58$ to $\hat{\tau}=22.7$), gives the following line of best fit
\begin{equation}
\left(\frac{W}{\lambda_{min}}\right)^\frac{m+5}{2} = 0.25(\hat{\tau}-\hat{\tau}_0)+0.01.
\label{eqn:fit}
\end{equation}
If the intercept, which is only important at very early time, is ignored then this gives the constant of proportionality for Eqn. \ref{eqn:nondim2} to be $0.25^{2/(m+5)}$. 

\subsection{Turbulent Kinetic Energy}
\label{subsec:tke}
\begin{figure}
	\centering
	\includegraphics[width=0.98\textwidth]{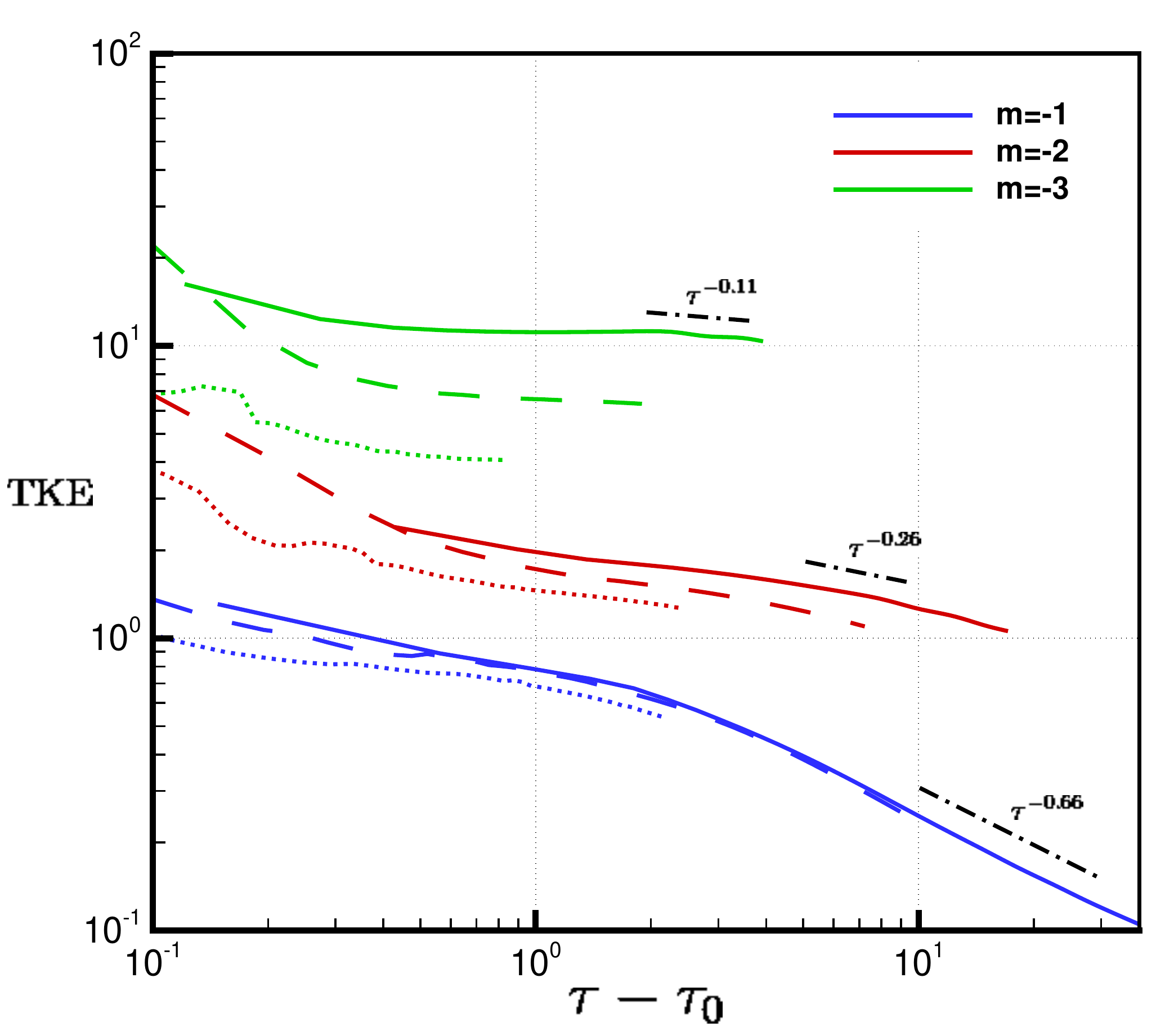} 
	\includegraphics[width=0.49\textwidth]{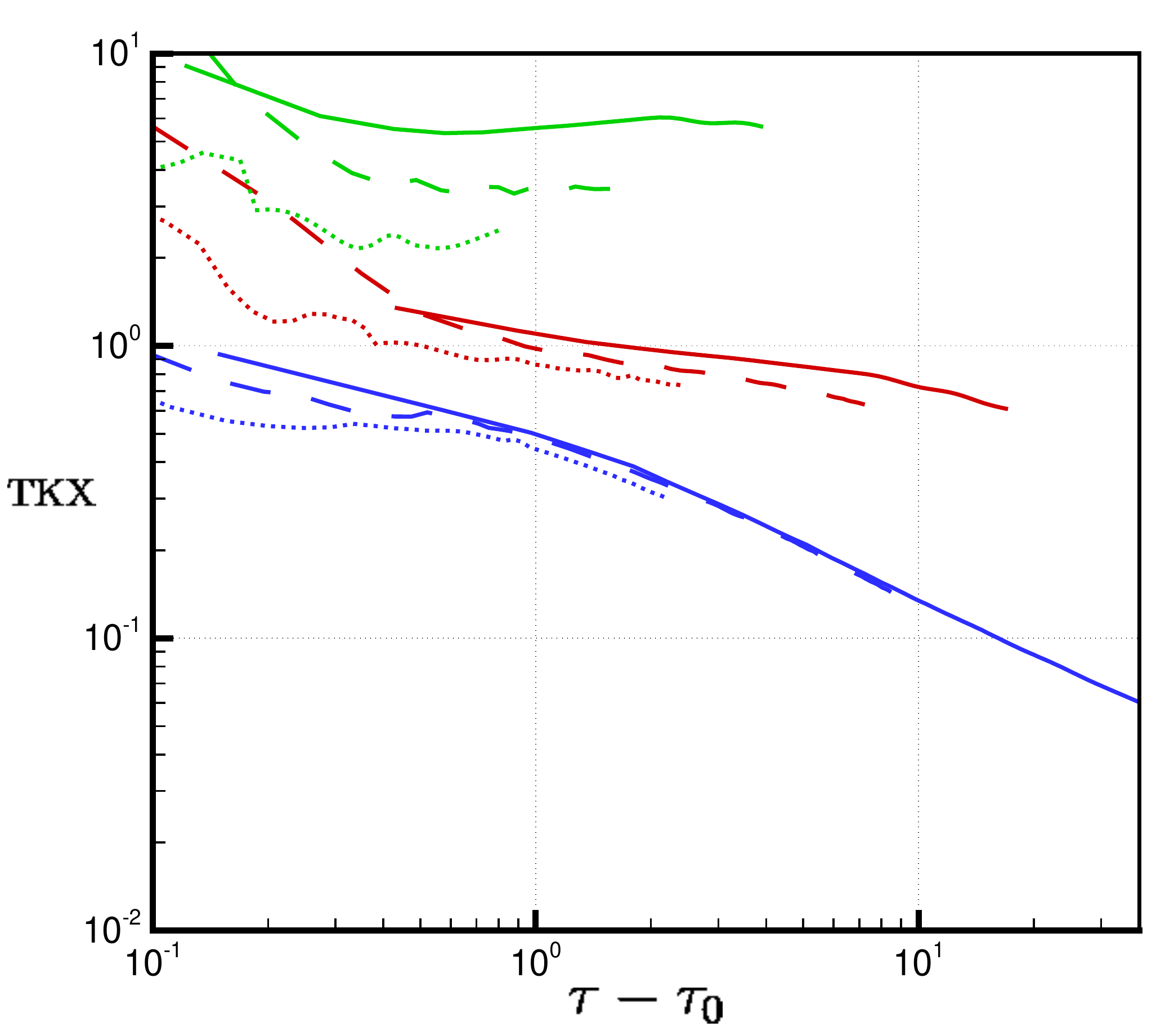}
	\includegraphics[width=0.49\textwidth]{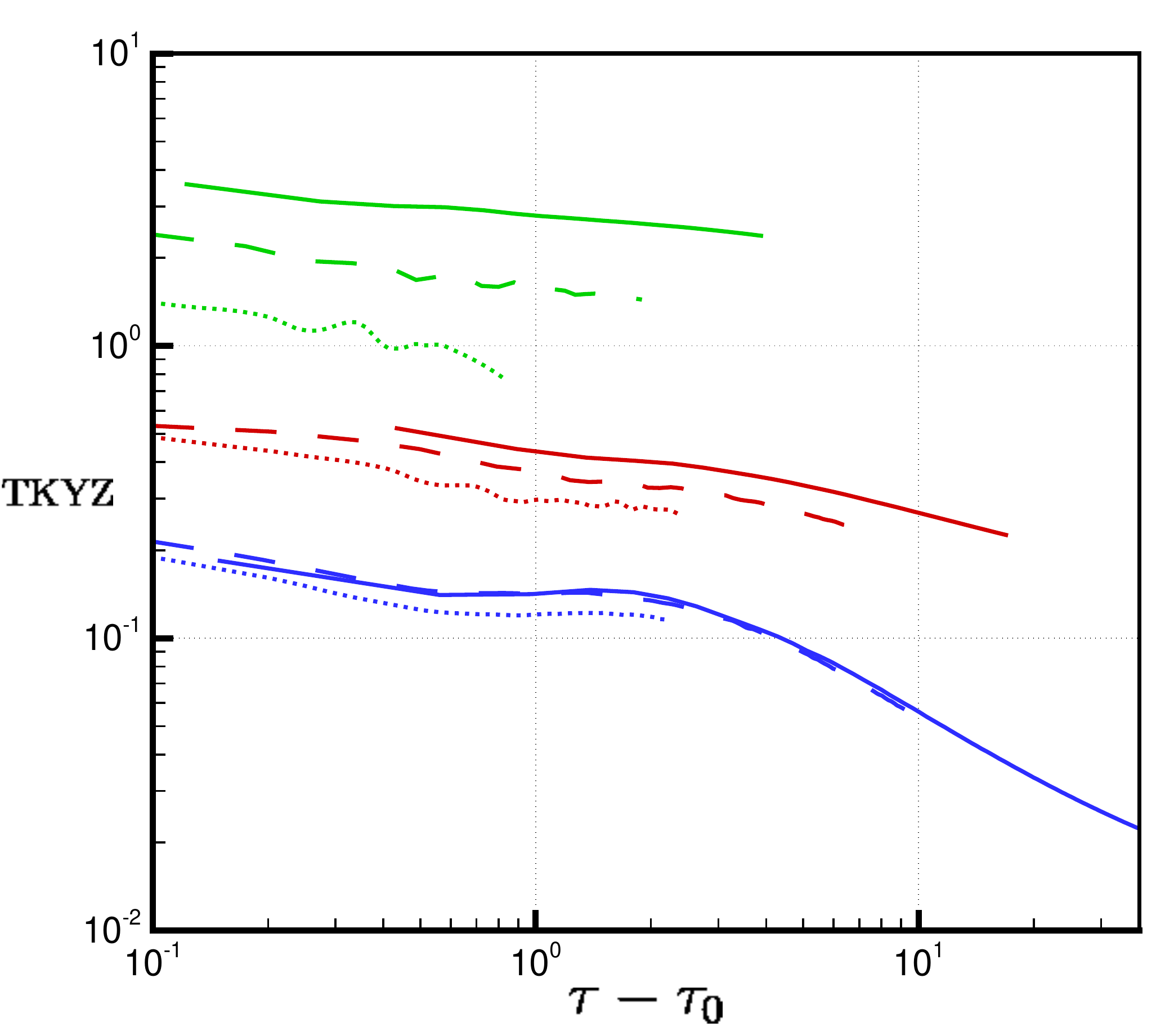}
	\caption{\label{fig:TKE} Total fluctuating kinetic energy vs. dimensionless time.  Dotted lines represent the {smallest bandwidth}, dashed lines the medium {bandwidth} and solid lines the {largest bandwidth}.} 
\end{figure}

The total fluctuating kinetic energy, presented here in terms of Favre averages, is defined as
\begin{equation}
\mathrm{TKE} = \iiint \frac{1}{2}\rho u_i^{\prime\prime}u_i^{\prime\prime}\:\mathrm{d}x\:\mathrm{d}y\:\mathrm{d}z, 
\end{equation}
where $\psi^{\prime\prime}=\psi-\widetilde{\psi}$ indicates a fluctuating quantity and $\widetilde{\psi}=\overline{\rho\psi}/\overline{\rho}$ is a Favre average. The $x$, $y$ and $z$ components of TKE are denoted by TKX, TKY and TKZ respectively. Since TKE is a large scale quantity, it is expected to converge in ILES provided there is a reasonable separation of the energetic scales from the dissipative scales \cite{Groom2019}. Fig. \ref{fig:TKE} shows the evolution of TKE and its components in time. {Since isotropy is expected in the transverse directions, the TKY and TKZ components are averaged and a single quantity, referred to as TKYZ, is presented for the transverse total fluctuating kinetic energy.} The data are presented in dimensionless form and are non-dimensionalised by {the initial growth rate i.e. by $\overline{\rho^+}\dot{W_0}^2\lambda_{min}L^2$}, where $\overline{\rho^+}=3.51$ is the mean post-shock density {and $L$ and $\lambda_{min}$ characteristic lengthscales in the homogeneous and inhomogeneous directions respectively}. {This non-dimensionalisation is useful for determining the degree to which the results are converged with respect to the infinite bandwidth limit. A reasonable collapse is observed for the $m=-1$ cases, indicating that the results for largest bandwidth are representative of the infinite bandwidth limit, at least up until the saturation time of the longest wavelength. The collapse is not as good for the $m=-2$ cases, while for the $m=-3$ cases the data are not converged at all with respect to the infinite bandwidth limit (note that as outlined in Sec. \ref{subsec:simulations} they are still considered converged for a given bandwidth). This suggests that even larger bandwidths are needed in order to obtain results that are representative of this limit. Fig. \ref{fig:TKE} also shows that in dimensionless terms, the $m=-3$ perturbations have the most kinetic energy deposited by the shock wave for a given bandwidth since all modes in the perturbation have the same growth rate, whereas for the $m=-2$ and $m=-1$ perturbations the growth rates are smaller for larger wavelengths.}

In the $m=-1$ cases, the TKE is decaying throughout the entirety of the simulation, and at early time there is a transfer of energy to the transverse directions due to the shorter wavelengths becoming nonlinear. The TKE in the $m=-2$ cases also decays throughout the simulation, however there is a less noticeable transfer of energy to the transverse directions; more energy is contained in the longer wavelengths which take a longer time to saturate and become nonlinear. An interesting phenomenon is observed in the $m=-3$ cases where at early time, starting from about the inversion time of the longest wavelength, the TKE is approximately constant, indicating zero dissipation. However, during this period there is a transfer of energy from the transverse directions to the $x$ direction. At `late' time the TKE beings to decay, mainly driven by a decay in TKX. 

The scaling of TKE over the period of dimensionless time corresponding to constant instantaneous $\theta$ is also given in Fig. \ref{fig:TKE}. In the self-similar regime TKE is expected to scale as $t^{-n}$ for some constant $n$, the value of which can be determined from the slope of these scalings. For $m=-1$, $m=-2$ and $m=-3$ the decay rates of TKE during this period are $n=0.66$, $n=0.26$ and $n=0.11$ respectively. An argument based on dimensional analysis was given in Thornber et al. \cite{Thornber2010} for the value of $n$ in terms of $\theta$. Using either the empirical relation $\epsilon\propto u^3/l$ (and equating $l$ with the integral width) or by assuming that the mean velocity in the mixing layer is proportional to the growth rate of the mixing layer itself gives $q_k\propto t^{2\theta-2}$, where $q_k$ is the mean fluctuating kinetic energy. Since the TKE is proportional to the width of the mixing layer multiplied by the mean fluctuating kinetic energy, this gives $\mathrm{TKE}\propto Wq_k\propto t^{3\theta-2}$. This predicted value of $n=2-3\theta$ has been found to be in good agreement with the measured decay rate of TKE in multiple studies of narrowband RMI \cite{Thornber2010,Thornber2017}, however for the present set of broadband cases the measured decay rates do not agree with this theoretical prediction. {This is true even for the $m=-1$ case, which is converged with respect to the infinite bandwidth limit.} Furthermore, for $\theta > 2/3$ this analysis predicts that TKE will increase in time, which is also not observed in Fig. \ref{fig:TKE}. 

\begin{figure}
	\centering
	\includegraphics[width=0.98\textwidth]{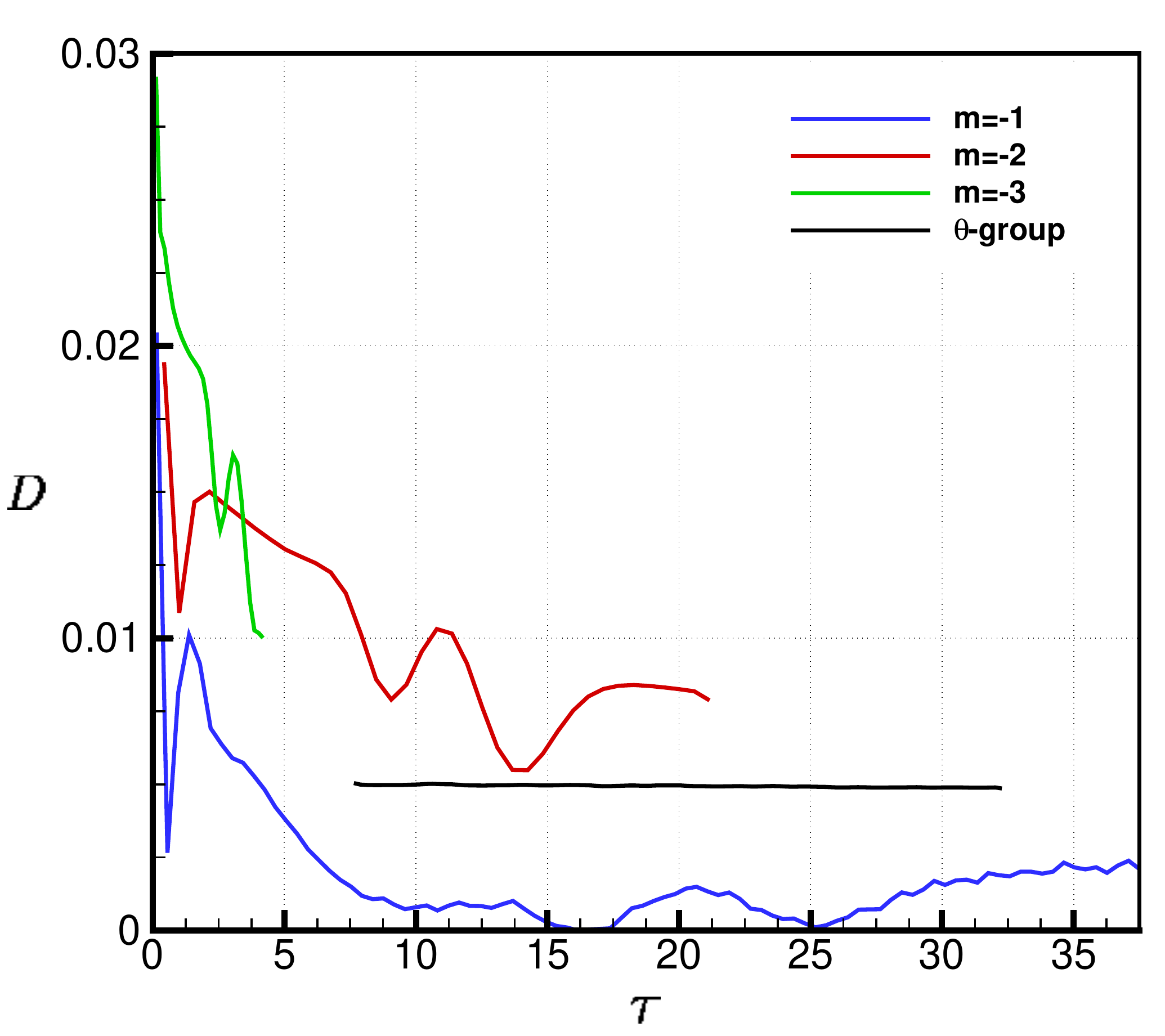} 
	\caption{\label{fig:D-tau} Plot of the RHS of Eqn. \ref{eqn:D} vs. dimensionless time for each of the {largest bandwidth} cases, as well as a portion of the quarter-scale $\theta$-group data.} 
\end{figure}

An explanation for why the TKE in the broadband case does not scale as $t^{3\theta-2}$ can be found by considering the assumptions behind the original derivation given in \cite{Thornber2010}. Starting with the relationship $\epsilon\propto u^3/l$ but retaining the constant of proportionality and equating $l$ with the integral width $W$ gives
\begin{equation*}
\epsilon=\frac{\mathrm{d} q_k}{\mathrm{d} t} = C_\epsilon\frac{u^3}{l}\propto C_\epsilon\frac{q_k^{3/2}}{W}.
\end{equation*}
Since $\mathrm{TKE}\propto Wq_k$, then it follows that
\begin{equation*}
\frac{\mathrm{d} \mathrm{TKE}}{\mathrm{d} t}\propto\frac{\mathrm{d} W}{\mathrm{d} t}\frac{\mathrm{TKE}}{W}+C_\epsilon\left(\frac{\mathrm{TKE}}{W}\right)^{3/2},
\end{equation*}
and therefore
\begin{equation}
C_\epsilon\propto\left(\frac{\mathrm{d} \mathrm{TKE}}{\mathrm{d} t}-\frac{\mathrm{d} W}{\mathrm{d} t}\frac{\mathrm{TKE}}{W}\right)\left(\frac{\mathrm{TKE}}{W}\right)^{-3/2}=D. 
\label{eqn:D}
\end{equation}
A dimensional analysis can be performed on Eqn. \ref{eqn:D}, assuming $\mathrm{TKE}\sim t^{-n}$ and $W\sim t^\theta$, which gives $C_\epsilon\sim t^{n/2-3\theta/2-1}$ and hence if $C_\epsilon$ is constant this implies $n=2-3\theta$ as before. More importantly however, all of the terms on the RHS of Eqn. \ref{eqn:D} are available, allowing for an assessment of the assumption that $C_\epsilon$ is constant. These terms are plotted in Fig. \ref{fig:D-tau} for each of the {largest bandwidth} cases, along with data for the narrowband case at late time taken from the $\theta$-group collaboration \cite{Thornber2017}. Note that the $y$-axis is not meaningful (since the data plotted is merely proportional to $C_\epsilon$) and has been scaled so that the minimum of the data is zero. The $\theta$-group data $x$-axis has also been shifted to make the figure more compact. It is clear that in the narrowband case, for which $\mathrm{TKE}$ scales as $t^{3\theta-2}$ at late time, the assumption that $C_\epsilon$ is constant is justified.  However, for the broadband cases $C_\epsilon$ is not constant and therefore the result $\mathrm{TKE}\propto t^{3\theta-2}$ does not hold. The decay rate in the $m=-1$ case is closest to the predicted value ($n=0.5$), which agrees with the observation that $C_\epsilon$ is plateauing over the period which this decay rate was measured. In theory, if the functional form of $C_\epsilon$ was known, then a new estimate for the decay rate $n$ in the broadband case could be derived. This is analogous to the recently proposed modification of a buoyancy-drag model \cite{Youngs2019}, and points to a means of adapting the dissipation terms commonly employed in Reynolds-averaged Navier--Stokes (RANS) models. 

\subsubsection{Spectra}
\label{subsec:spectra}
\begin{figure}
	\centering
	\includegraphics[width=0.49\textwidth]{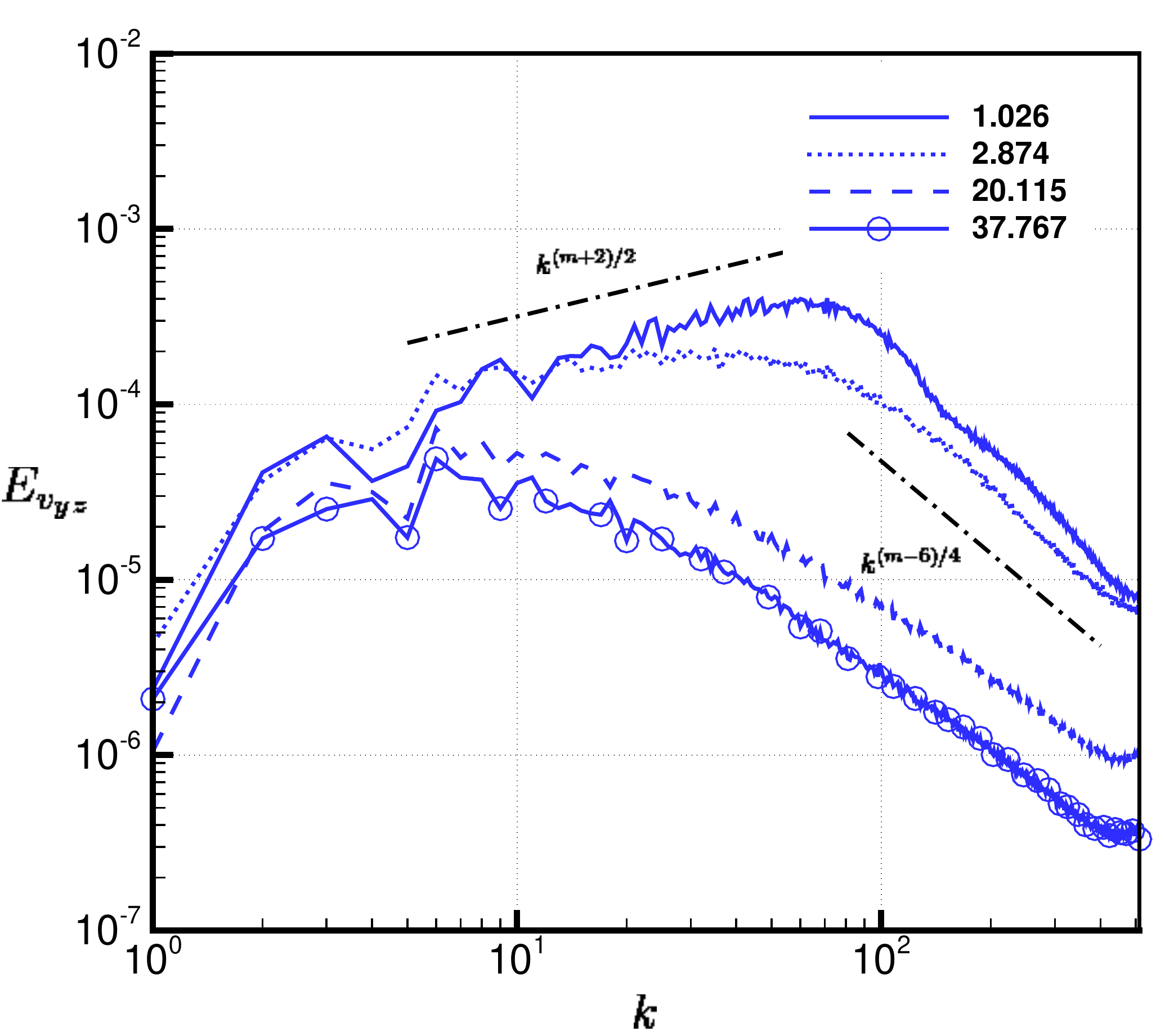} 
	\includegraphics[width=0.49\textwidth]{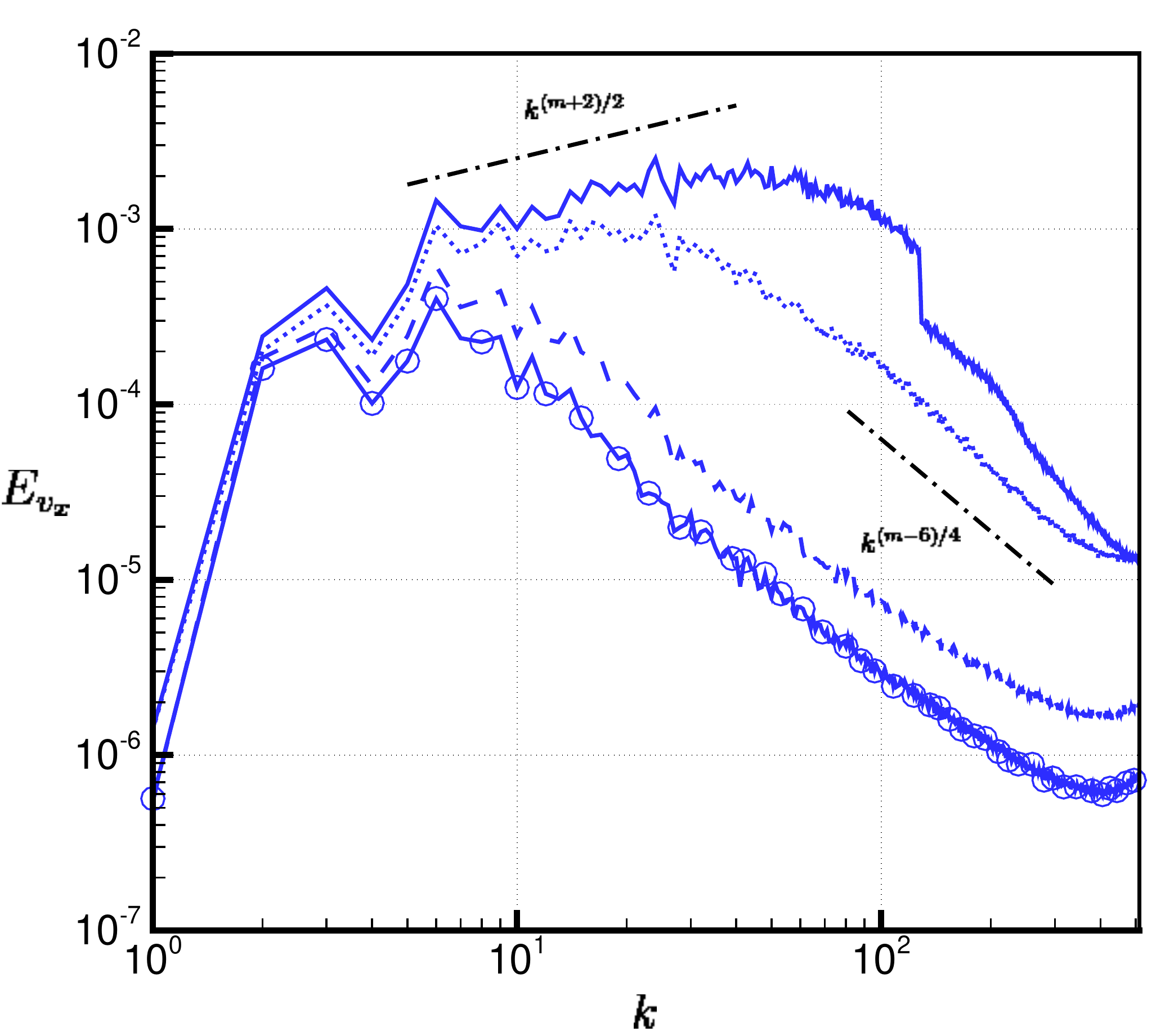} 
	\includegraphics[width=0.49\textwidth]{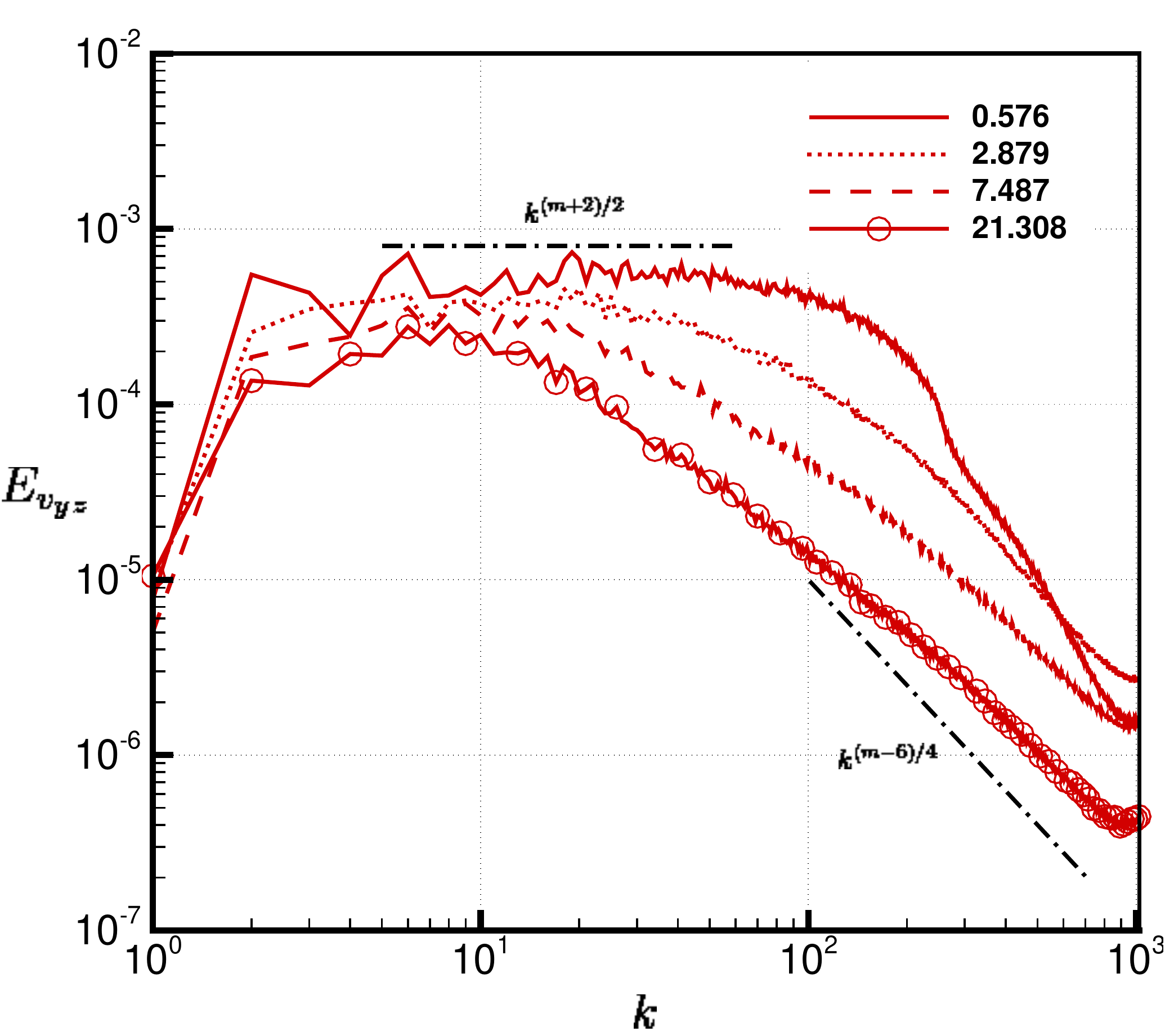} 
	\includegraphics[width=0.49\textwidth]{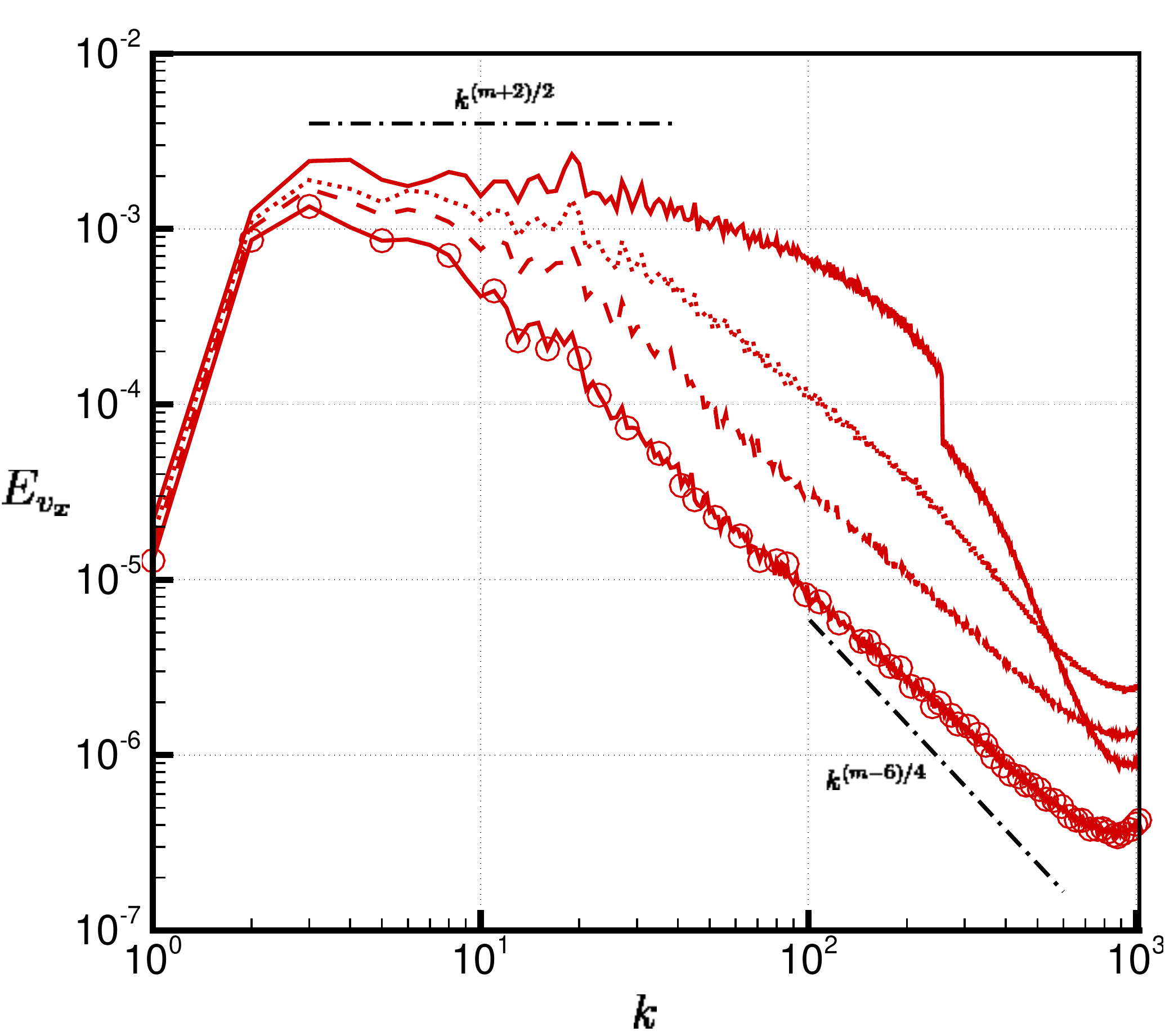}	
	\includegraphics[width=0.49\textwidth]{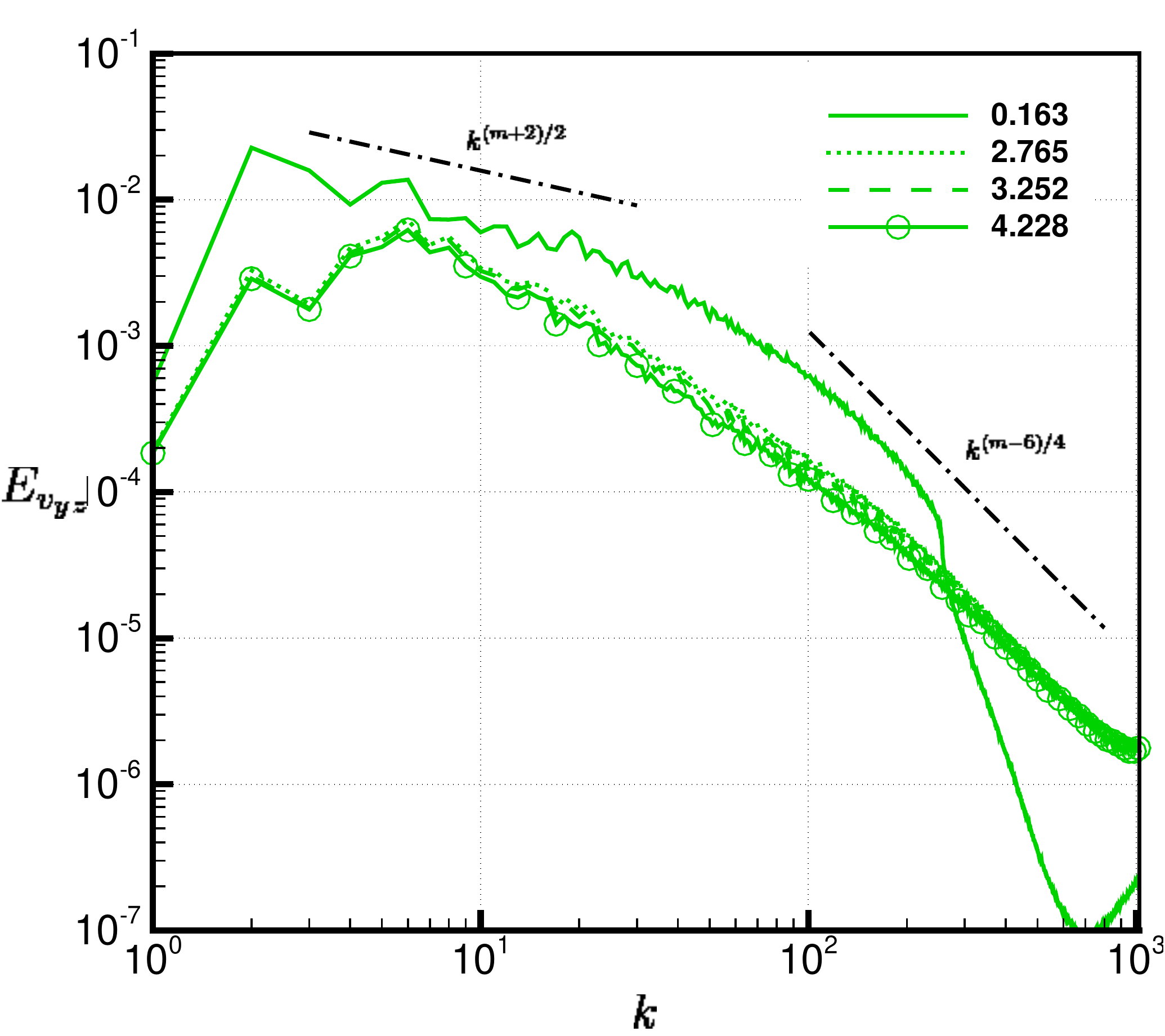} 
	\includegraphics[width=0.49\textwidth]{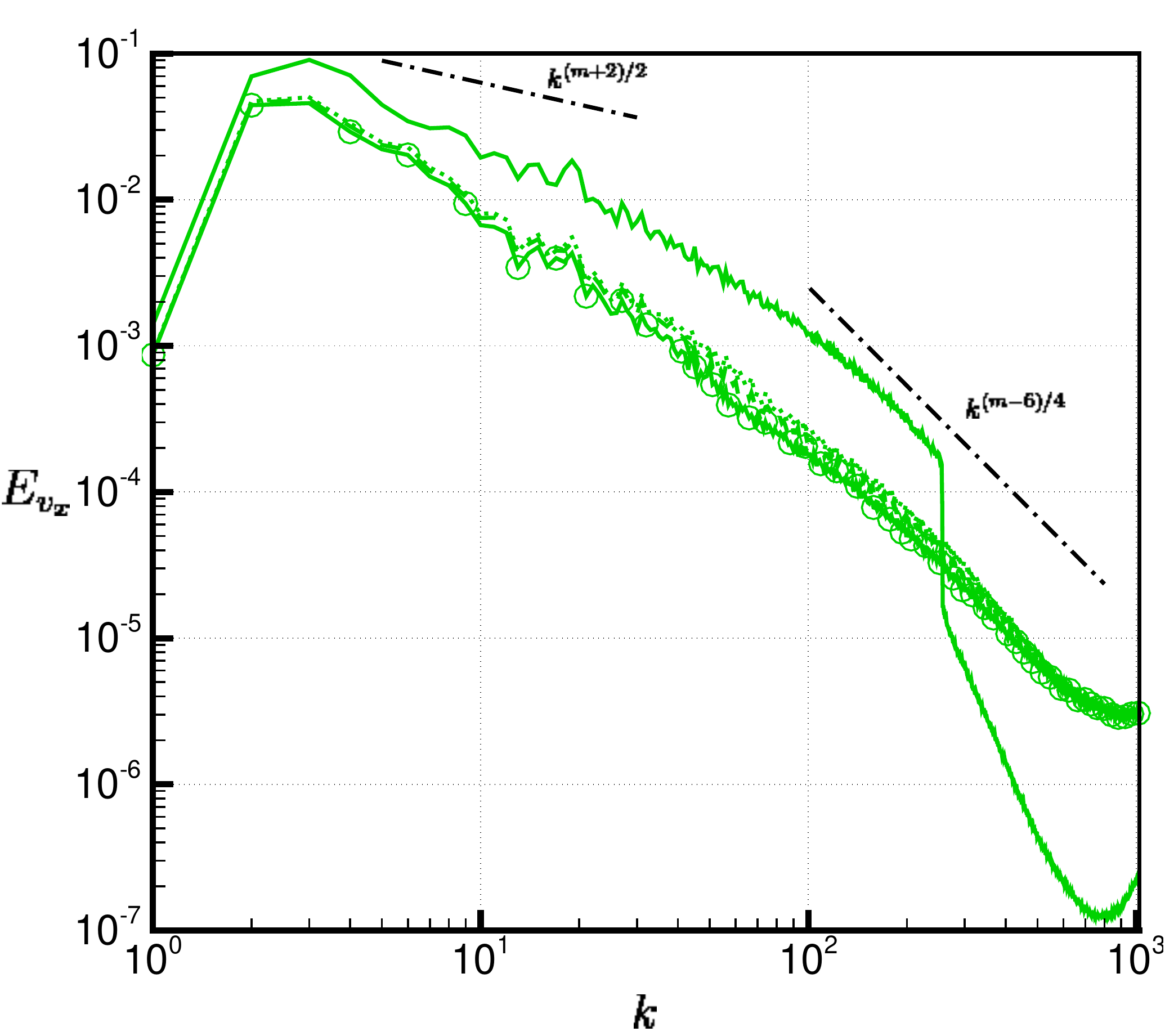}	
	\caption{\label{fig:KEV} Variable-density turbulent kinetic energy spectra for each of the {largest bandwidth} cases. Shown are data for $m=-1$ (top), $m=-2$ (middle) and $m=-3$ (bottom). Dimensionless times are given in the legend.}
\end{figure}

The distribution of turbulent kinetic energy in wavenumber space may be analysed by computing the radial power spectrum over the $y$--$z$ plane located at the mixing layer centre $x_c$. Variable-density spectra of both the normal and transverse velocity components are calculated as
\begin{equation}
E_{v_i}(k)=\widehat{\psi_i}^\dagger\widehat{\psi_i},
\label{eqn:E2D}
\end{equation} 
where $\psi_i=\sqrt{\rho}u^{\prime\prime}_i$. In Eqn. \ref{eqn:E2D}, $k=\sqrt{k_y+k_z}$ is the radial wavenumber in the $y$-$z$ plane at the mixing layer centre, $\widehat{(\ldots)}$ denotes the 2D Fourier transform taken over the plane and $\widehat{(\ldots)}^\dagger$ is the complex conjugate of this transform. Since isotropy is expected in the transverse directions, a single transverse energy spectrum is defined as $E_{v_{yz}}=(E_{v_y}+E_{v_z})/2$. The energy spectra of the transverse and normal velocity components are shown in Fig. \ref{fig:KEV} for each of the  {largest bandwidth} cases at four different dimensionless times, {with the data non-dimensionalised by $\overline{\rho^+}\dot{W_0}^2$}. Each dimensionless time plotted corresponds to; same physical time (t=0.005), same dimensionless time ($\tau=2.87$), dimensionless time at which $W/\lambda_{min}=2$ and dimensionless time at saturation of the longest wavelength respectively.

At the earliest time plotted, there is a visible discontinuity located at the highest wavenumber in the initial perturbation, particularly for the $E_{v_x}$ spectra, due to there being insufficient time for a significant amount of energy to cascade to scales smaller than $8\Delta x$.  The theoretical scalings in wavenumber space are also shown in Fig. \ref{fig:KEV}. Following Youngs \cite{Youngs2004}, a scaling of $E_v\sim k^{(m+2)/2}$ is expected at early time. This scaling is most easily visible for wavenumbers $10\lesssim k\lesssim50$ at the earliest time shown in Fig. \ref{fig:KEV}, as the highest wavenumbers in the initial perturbation break down rapidly, while the statistics of the lowest wavenumbers are not sufficient to produce a smooth line. In general, the spectra of the transverse velocity components follow this early time scaling more closely than those of the normal velocity component. {The $k^{(m+2)/2}$ scaling at early time also shows that, right after shock passage, the peak kinetic energy is located at $k_{max}$ for the $m=-1$ cases, at $k_{min}$ in the $m-3$ cases while the $m=-2$ cases have a uniform distribution of kinetic energy across all wavenumbers.}

At late time, the analysis of Zhou \cite{Zhou2001} was modified by Thornber et al. \cite{Thornber2010} to take into account the effects of the initial perturbation spectrum. This gave an expected scaling of $E_v\sim k^{(m-6)/4}$ provided $\tau_{\tiny{RM}}<\tau_{\tiny{HDT}}$, where $\tau_{\tiny{RM}}$ and $\tau_{\tiny{HDT}}$ are the characteristic eddy turnover times of Richtmyer--Meshkov and homogeneous decaying turbulence respectively, otherwise the spectra should revert to a $k^{-5/3}$ scaling \cite{Thornber2010}. The $k^{(m-6)/4}$ scaling is shown in Fig. \ref{fig:KEV} at high wavenumbers and in general the agreement with the data is mixed. In the $m=-1$ case, the transverse spectra scale as $k^{-1.47}$ at late time, which is close to the scaling observed in the narrowband case \cite{Thornber2016}. The early and intermediate time transverse spectra do suggest however that a $k^{-7/4}$ scaling is briefly obtained at higher wavenumbers. This is also observed in the normal spectra, which also retain this scaling ($k^{-1.73}$) at later times. A similar trend is observed in the $m=-2$ case, where again the transverse spectra scale as $k^{-1.47}$ at late time while at earlier times a scaling close to $k^{-2}$ is observed at high wavenumbers. The $k^{-2}$ scaling is also seen briefly in the normal spectra, which at late time are tending towards a $k^{-5/3}$ scaling. Finally in the $m=-3$ case, the transverse spectra follow a scaling of $k^{-1.53}$ at intermediate wavenumbers and a scaling close to $k^{-9/4}$ at higher wavenumbers (above the smallest initial wavelength). Meanwhile the normal spectra scale as $k^{-5/3}$ at intermediate wavenumbers and as $k^{-9/4}$ at higher wavenumbers. The fact that very little TKE is dissipated in the $m=-3$ simulation can also be seen here, with the spectra at later times collapsing almost perfectly on top of each other.

\section{Conclusions}
\label{sec:conclusion}
This paper has investigated the influence of different broadband perturbations on the evolution of a turbulent mixing layer induced by Richtmyer--Meshkov instability through a series of carefully designed numerical simulations. In particular, the effects of varying the bandwidth $R$ and spectral exponent $m$ of the initial perturbation have been analysed for three different values of both $m$ and $R$. For a given bandwidth, the total standard deviation of the perturbation was held constant for all values of $m$, and the initial amplitudes of all modes were linear (or weakly nonlinear in the $m=-1$ case). Upon non-dimensionalisation, a good collapse of the data was obtained for various integral measures such as the integral width $W$ and molecular mixing fraction $\Theta$. Both nonlinear regression and direct calculation through derivatives of $W$ were used to extract the growth rate exponent $\theta$. For the {largest bandwidth} cases, $\theta$ was found to be $0.5$, $0.63$ and $0.75$ for $m=-1$, $m=-2$ and $m=-3$ respectively, while the values of $\Theta$ at the latest times considered were $0.56$, $0.39$ and $0.20$. The degree to which the layer is evolving self-similarly was assessed using plane-averaged volume fraction profiles, with all cases showing a good collapse when scaled by a single length scale $W$. The temporal evolution of total fluctuating kinetic energy was also presented, along with the observed decay rates for each case and an argument for why they do not match the predicted value of $n=2-3\theta$. Finally, the scaling of turbulent kinetic energy was analysed in spectral space and was shown to follow the theoretical scaling of $k^{(m+2)/2}$ at low wavenumbers. At high wavenumbers the spectra of the transverse velocity components tend towards a $k^{-3/2}$ scaling at late time, while the normal velocity spectra approach a $k^{-5/3}$ scaling. {In general, the results provide a good validation for analysing the mixing layer in terms linear growth rates of individual modes in the perturbation, as well as highlighting where this analysis fails to accurately capture the behaviour of the mixing layer.}

This work highlights multiple avenues for further investigation. Given the promising comparisons that were drawn between the present simulations and experiments with broadband perturbations, a full study that aims to more closely match experimental conditions would likely provide a lot of useful insight. Another option is to perform a complete analysis of the budgets of transport equations for quantities such as turbulent kinetic energy to inform RANS modelling, as was recently done for narrowband RMI \cite{Thornber2019}. Finally, it would be useful to extend the just-saturated mode theory presented here to include the effects of viscosity and diffusivity on the growth rates of each mode. This would require direct numerical simulations (DNS) to be performed in order to validate the modifications, which would follow a similar approach to recent DNS of the narrowband case \cite{Groom2019}.





\section{Acknowledgements}
This work is dedicated to honouring the exemplary scientific career of David L. Youngs. The authors would like to acknowledge the computational resources at the National Computational Infrastructure provided through the National Computational Merit Allocation Scheme, as well as the Sydney Informatics Hub and the University of Sydney's high performance computing cluster Artemis, which were employed for all cases presented here.


\bibliographystyle{elsarticle-num} 
\bibliography{bibliography}




\end{document}